\documentclass{aa}
\usepackage{ulem}
\usepackage{soul}
\usepackage{xcolor}
\usepackage{array}
\usepackage{float}
\usepackage{tabularx}
\usepackage{booktabs}
\usepackage{amsfonts}
\usepackage{amssymb}
\usepackage{amsmath}
\usepackage{natbib}
\usepackage{wrapfig}
\usepackage{graphicx}
\usepackage{makecell}
\usepackage{caption}
\usepackage{subfigure}
\usepackage{longtable}
\usepackage{threeparttable}
\usepackage{footnote}
\usepackage[export]{adjustbox}
\graphicspath{ {./images/} }
\usepackage{txfonts}

\usepackage[colorlinks=true, linkcolor=blue, citecolor=blue, urlcolor=blue]{hyperref}
\bibpunct[, ]{(}{)}{;}{a}{}{,}

\newenvironment{acknowledgments}
{
    \section*{Acknowledgments}
}
{
}

\begin{document} 
    \title{Deep learning interpretability analysis for carbon star identification in $Gaia$ DR3}
   
    \author{Shuo Ye
          \inst{1,2,3}
          \and
          Wen-Yuan Cui
          \inst{2,3}
          \fnmsep\thanks{Corresponding author, email:cuiwenyuan@hebtu.edu.cn} 
          \and
          Yin-Bi Li
          \inst{1}
          \fnmsep\thanks{Corresponding author, 
          email:ybli@bao.ac.cn}
          \and
          A-Li Luo
          \inst{1,4,5}
          \fnmsep\thanks{Corresponding author, 
          email:lal@nao.cas.cn}
          \and
          Hugh R. A. Jones
          \inst{6}
          }

  \institute{
       Department of Physics, Hebei Normal University, Shijiazhuang 050024, China
       \and
       Guo Shoujing Institute for Astronomy, Hebei Normal University, Shijiazhuang 050024, China
       \and
       CAS Key Laboratory of Optical Astronomy, National Astronomical Observatories, Beijing 100101, China
       \and 
       University of Chinese Academy of Sciences, Beijing 100049, China
       \and
       Nanjing Institute of Astronomy and Optical Technology, University of Chinese Academy of Sciences, Nanjing 211135, China
       \and
       School of Physics, Astronomy and Mathematics, University of Hertfordshire, College Lane, Hatfield AL10 9AB, UK
    }

   \date{\today}
   
   \abstract
   {A large fraction of Asymptotic Giant Branch (AGB) stars develop carbon-rich atmospheres during their evolution. Based on their color and luminosity, these carbon stars can be easily distinguished from many other kinds of stars. However, numerous G, K, and M giants also occupy the same region as carbon stars on the HR diagram. Despite this, their spectra exhibit differences, especially in the prominent CN molecular bands.}
   {We aim to distinguish carbon stars from other kinds of stars using $Gaia$'s XP spectra, while providing attributional interpretations of key features necessary for identification, and even discovering additional new spectral key features.}
   {We propose a classification model named `GaiaNet', an improved one-dimensional convolutional neural network specifically designed for handling $Gaia$'s XP spectra. We utilized the SHAP interpretability model to determine SHAP values for each feature in a spectrum, enabling us to explain the output of the `GaiaNet' model and provide further meaningful analysis.}
   {Compared to four traditional machine-learning methods, the `GaiaNet' model exhibits an average classification accuracy improvement of approximately 0.3\% on the validation set, with the highest accuracy reaching 100\%. Utilizing the SHAP model, we present a clear spectroscopic heatmap highlighting molecular band absorption features primarily distributed around CN$_{\text{773.3}}$ and CN$_{\text{895.0}}$, and summarize five key feature regions for carbon star identification. Upon applying the trained classification model to the CSTAR sample with $Gaia$ `xp\_sampled\_mean' spectra, we obtained 451 new candidate carbon stars as a by-product.}
   {Our algorithm is capable of discerning subtle feature differences from low-resolution spectra of $Gaia$, thereby assisting us in effectively identifying carbon stars with typically higher temperatures and weaker CN features, while providing compelling attributive explanations. The interpretability analysis of deep learning holds significant potential in spectral identification.}
   

   \maketitle
%

\section{Introduction}
    Carbon stars, first recognized by \citet{secchi1869schreiben}, exhibit an inversion of the C/O ratio (C/O > 1). Compared to other stars, carbon stars possess carbon-enriched atmospheres. The enrichment can originate from mass transfer in binary systems or the activation of the third dredge-up (TDU) process during their AGB evolution phase \citep{2023A&A...674A..39G}. The carbon stars arise in binary systems likely inherited atmospheric carbon through mass transfer from its AGB companion, which now is a white dwarf \citep{2002ApJ...579..817A}. These carbon stars are known as `extrinsic' carbon stars. However, the carbon enrichment of cool and luminous carbon stars (N-type) is caused by the pollution of nuclear helium fusion products transported from the inner to the outer layers during their AGB phase \citep{2023A&A...674A..39G}. This process transports carbon from the interior region to the stellar surface, leaving the outer layers rich in carbon-containing molecules and dust particles. Thus, their spectra show C$_{\text{2}}$ and CN molecular bands stronger than usual in stars cooler than 3800 K (i.e. $G_\mathrm{BP} - G_\mathrm{RP}\geq2$; \citeauthor{2023A&A...674A..39G} \citeyear{2023A&A...674A..39G}). These carbon stars are called `intrinsic' carbon stars.
    
    The spectra of carbon stars exhibit absorption characteristics due to carbon-containing compounds of CH, C$_{\text{2}}$, and CN. The presence and intensity of these spectral features provide valuable insights into the atmospheric conditions and chemical processes within these stars. Because they tend to be in a late evolution stage of stellar mass loss, carbon stars are important contributors to the interstellar medium and serve as a good references for studying a variety of physical processes affecting the end of the life of low mass stars \citep{2023A&A...674A..39G}. 

    Traditionally, identifying and classifying carbon stars relied on parametric measurements and manual checking of their spectra. \citet{ji2016carbon} identified 894 carbon stars from LAMOST DR2 by measuring multiple line indices from the stellar spectra. \citet{2022A&A...664A..45A} reported the identification of 2660 new carbon star candidates through 2MASS photometry, $Gaia$ astrometry, and their location in the $Gaia$–2MASS diagram. \citet{2023ApJS..266....4L} distinguished carbon stars from M-type giants by selecting spectral indices satisfying the criterion [$\rm{CaH3}-0.8\times\rm{CaH2} -0.1$]$ < 0$. \citet{2023A&A...674A..39G} screened carbon stars by measuring the carbon molecular band head strength, resulting in a final selection of 15,740 high-quality `golden sample' carbon stars. \citet{2023A&A...674A..15L} classified 546,468 carbon star candidates using a method guided by narrow-band photometry \citep{1982AJ.....87.1739P}. Their SOS module examines this feature in an automated way by computing the pseudo-wavelength difference between the two highest peaks in each $Gaia$ DR3 RP spectrum, taking the median value of the results, and storing it in the parameter of median\_delta\_wl\_rp. They assumed all stars with median\_delta\_wl\_rp$ > 7$ to be C-rich stars. \citet{2024ApJS..271...12L} identified 3546 carbon stars through line indices and near-infrared color–color diagrams. Through visual inspection of these spectra, they further subclassify them into C–H, C–R, C–N, and Ba stars.

    To date, many machine-learning methods have been used to identify carbon stars based on their spectra. \citet{2014SCPMA..57..176S} applied the label propagation algorithm to search for 260 new carbon stars from SDSS (Sloan Digital Sky Survey) DR8.
    \citet{2015RAA....15.1671S} applied the efficient manifold ranking algorithm to search for 183 carbon stars from the LAMOST pilot survey. \citet{li2018carbon} identified 2651 carbon stars in the spectra of more than seven million stars using an efficient machine-learning algorithm for LAMOST (Large Sky Area Multi-Object Fiber Spectroscopic Telescope) DR4. \citet{2023MNRAS.521.2745S} investigated the use of unsupervised learning algorithms to classify the chemistry of long-period variables \citep{2023A&A...674A..15L} from $Gaia$ DR3’s BP/RP spectra (also called XP spectra; \citeauthor{2021A&A...652A..86C} \citeyear{2021A&A...652A..86C}) into O-rich and C-rich groups. They also employed a supervised approach to separate O-rich and C-rich sources using broadband optical and infrared photometry. In all, they tagged a total of 23,737 C-rich classifications based on the BP/RP spectra and identified a small population
    of C-rich stars in the Galactic bar-bulge region.  
    
    In recent years, the application of interpretable analysis techniques in astronomy has shown great potential. \citet{2019ApJS..242...13Q} used a random forest (RF) algorithm deriving rank features, then picked out 15,269 Am candidates from the early-type stars of LAMOST DR5. \citet{2022MNRAS.512.1710H} successfully distinguished between Red Giant Branch (RGB) and Red Clump (RC) stars using the XGBoost algorithm \citep{chen2016xgboost}, and used the SHAP interpretable model \citep{NIPS2017_7062} to obtain the top features that the XGBoost selected. \citet{2022ApJS..259...63S} used machine learning algorithms to search for class-one and class-two chemical peculiars (CP1 and CP2). Finally they presented a catalog of 6917 CP1 and 1652 CP2 new candidate sources using XGBoost followed by the visual investigation, and listed the spectral features for separating CP1 from CP2 using SHAP.

    Before this work, searching for carbon stars based on $Gaia$ spectra was mainly to obtain the matching giant star spectra by random forest classifier \citep{breiman2001random}, and then to screen out carbon stars with strong CN molecular bands by measuring the molecular band head strength \citep{2023A&A...674A..39G}, most of which belong to the N-type. Although this method can easily pick out the carbon stars with obvious spectral features, it misses carbon stars that exhibit relatively weak molecular bands. They are mixed with non-carbon stars in the band head strength diagrams, which makes it difficult to distinguish them using this method. Our proposed algorithm can be used as an improvement of the above method, to identify such carbon stars from their spectra in a quantitative manner.

    In this work, we explored the significant potential of deep learning model enhanced with interpretability analysis algorithm (SHAP) for identifying carbon stars  using their $Gaia$ DR3 XP spectra. In Section \ref{sec:data}, we  cover the source of the data, the preparation of the training set, and data processing methods. In Section \ref{sec:method}, we introduce the proposed deep learning classification model and important model parameters, as well as the SHAP interpretability model, and the evaluation index. In Section \ref{sec:result}, we  show how well the model performs with a  validation set, how it compares to other classification models, demonstrating the effectiveness of model interpretation and key features, as well as the newly discovered carbon star candidates. In Section \ref{sec:analysis}, we  compare the results with others,  analyze and interpret the spectral features. Finally, a general summary is provided in Section \ref{sec:conclusion}, and candidates are given in Section \ref{sec:appendix}.

\section{Data}\label{sec:data}
\subsection{CSTAR sample of $Gaia$ DR3}
    $Gaia$ Data Release 3 ($Gaia$ DR3) has been released in 2022, and its mean BP/RP spectral data was released in May of the same year, including four folders on the archive: rvs\_mean\_spectrum, xp\_continuous\_mean\_spectrum, xp\_sampled\_mean\_spectrum and xp\_summary. The spectra we adopt are from xp\_sampled\_mean\_spectrum, which has 34,468,373 BP/RP externally calibrated sampled mean spectra. 
    All mean spectra were sampled to the same set of absolute wavelength positions, viz. 343 values from 336 to 1020 nm with a step of 2\,nm, i.e. corresponding to a wavelength range of 3360 to 10200\,\AA\ in steps of 20\,\AA. To standardize the data scale range, we normalized the raw spectral data as 
    \begin{equation} 
        x^{*}=\frac{x-x_{\min }}{x_{\max }-x_{\min }} \text{,}
    \end{equation}
    where $x$ represents the original spectrum, the $x_\mathrm{min}$ and $x_\mathrm{max}$ respectively indicate the minimum and maximum value of the $x$, and the normalised spectrum is denoted as $x^{*}$.
    
    Min-max normalization is a data standardization method that maps the data to a 0-1 range in equal proportion, thus changing the distribution of the raw data. It only scales the spectrum without altering its original structural representation. It also accelerates the convergence rate of the deep learning model's loss function, thereby enhancing the efficiency and speed of achieving the optimal solution, which might otherwise be hard to converge.
    
    \citet{2023A&A...674A..39G} (hereafter referred to as C2023) released a batch of `golden samples', which include the `golden sample' of carbon stars from `CSTAR'. Their ESP-ELS module attempted to flag these suspected carbon stars. The module is based on a random forest classifier trained on the synthetic BP and RP spectra and a sample of Galactic carbon stars \citep{abia2020properties} obtained from the $Gaia$ low-resolution spectra ($\mathrm{R} = \lambda/\delta\lambda \approx 25-100$; \citeauthor{2021A&A...652A..86C} \citeyear{2021A&A...652A..86C}). In total, 386,936 targets received the `CSTAR' tag. i.e., the potential carbon star candidates. While most of these stars are M stars rather than carbon stars, only smaller fraction of the sample exhibit significant C$_{\text{2}}$ and CN molecular bands. Most of the candidate carbon stars have $G_\mathrm{BP}$ - $G_\mathrm{RP}$ > 2 mag and have colors consistent with M stars \citep{2023A&A...674A..39G}.

    As shown in Table \ref{tab:Molecular Band}, C2023 considered four molecular band heads in order to further screen for reliable carbon stars.
    \begin{table}[tb]
        \renewcommand{\arraystretch}{1.2}
        \centering
        \caption{The four molecular band head positions used to identify carbon stars.}
        \vspace{1.5pt}
        \label{tab:Molecular Band}
        \resizebox{\linewidth}{!}{%
            \begin{tabular}{c|c|c|c|c}
                \hline Molecular band  &   Strength &  $\lambda_1$ [nm] &   $\lambda_2$ [nm]   &   $\lambda_3$ [nm]\\
                \hline
                C$_{\text{2}}$    &   R$_{\text{482.3}}$    &   462.2345    &   482.3455    &   505.3195\\ 
                C$_{\text{2}}$   &    R$_{\text{527.1}}$     &   505.3195    &   527.1080    &   546.5995\\
                CN   &    R$_{\text{773.3}}$     &   716.5865   &   773.2905    &   810.7805\\
                CN   &    R$_{\text{895.0}}$     &   806.8910    &   894.9855    &   936.6820\\
                \hline
            \end{tabular}%
        }
    \end{table}
    Finally, they selected 15,740 golden sample carbon stars based on the strongest CN molecular band features. Most of these golden sample carbon stars are AGB stars of which the enriched carbon was produced and then dredged up from its interior region, i.e., the cold and bright N-type carbon stars. The calculation formula for the four molecular band head strengths is as follows:
    \begin{equation} \label{molecular band head strength}
            R_{\lambda_{2}}=\frac{f\left(\lambda_{2}\right)}{g_{\lambda_{1}, \lambda_{3}}\left(\lambda_{2}\right)} \text{,}
    \end{equation}
    where $f\left(\lambda_{2}\right)$ is the flux measured at the top of the band head of the molecular band, and $g_{\lambda_{1}, \lambda_{3}}$ is the value linearly interpolated between wavelengths $\lambda_1$ and $\lambda_3$ \citep{2023A&A...674A..39G}.  
    
    The reason C2023 did not select C$_{\text{2}}$ molecular band strengths as filtering criterion may be twofold: firstly, the C$_{\text{2}}$ features in the spectra are significantly weaker compared to CN features; secondly, more than half of the golden sample carbon stars lack prominent C$_{\text{2}}$ features in their spectra after visual inspection. Therefore, C$_{\text{2}}$ features lack clear distinction, as can also be seen in the band head strengths diagram of Figure \ref{fig:strengths}.
  
    
\subsection{Training data}\label{sec:train_data}
    After cross-matching the source\_id of CSTAR with the spectral data  of the xp\_sampled\_mean\_spectrum library in $Gaia$ DR3, we obtained 83,028 spectra. These spectra will be used to verify the feasibility of our algorithm, which we refer to as CSTAR\_XPM (CSTAR sample with xp\_sampled\_mean spectra). Of these, 8288 belong to the 15,740 golden sample of carbon stars and 74,740 are non-golden sample. We respectively refer to them as CSTAR\_XPM\_G (for the golden sample of CSTAR\_XPM) and CSTAR\_XPM\_nonG (for the non-golden sample of CSTAR\_XPM). 

\begin{figure*}[htbp]
      \centering
      \subfigure[$T_\mathrm{eff}$ and $\log{g}$ of CSTAR\_XPM]{
        \includegraphics[width=0.32\linewidth]{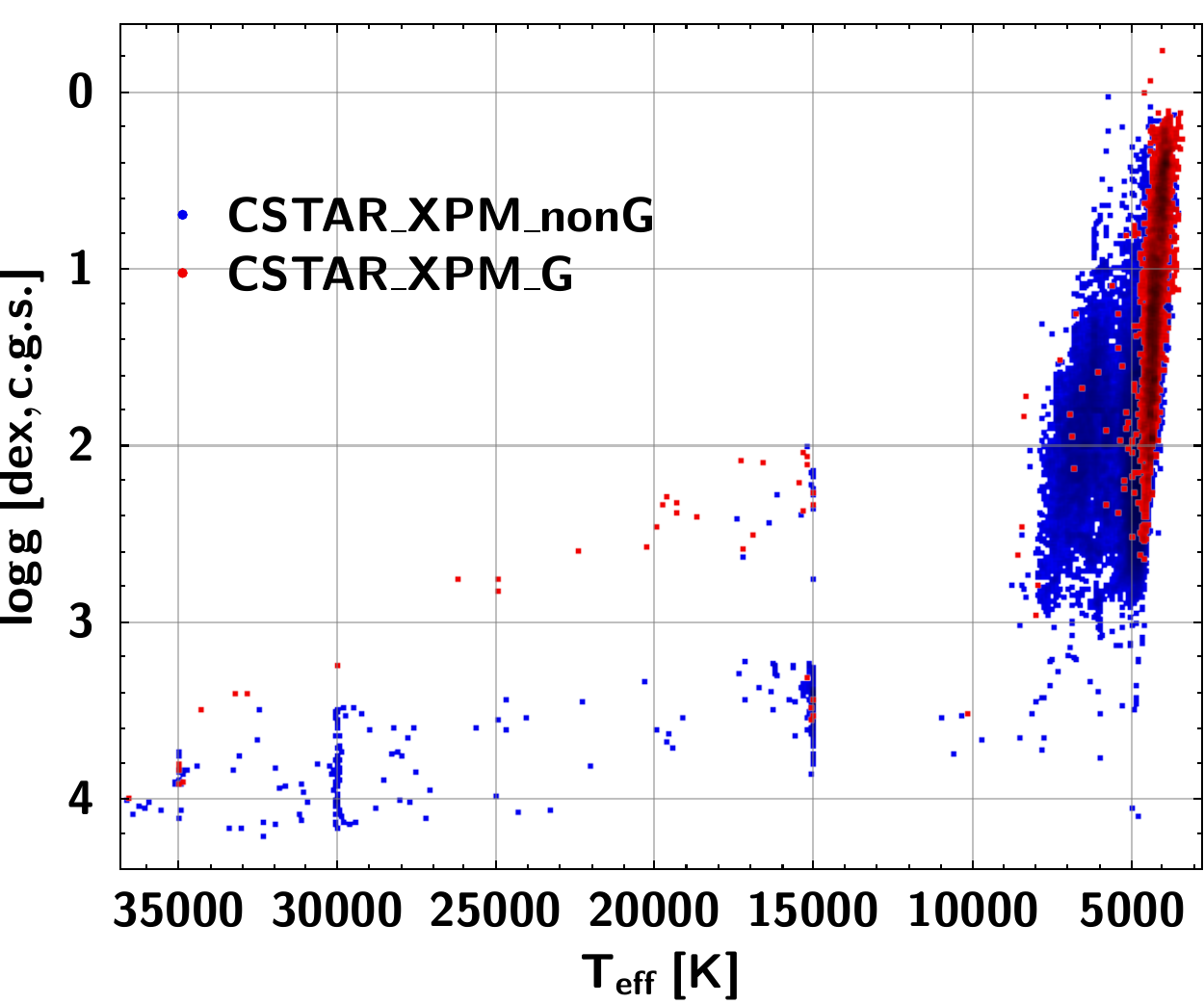}
        \label{fig:kiel}
      }
      \hfill
      \subfigure[Random sample and CSTAR\_XPM]{
        \includegraphics[width=0.32\linewidth]{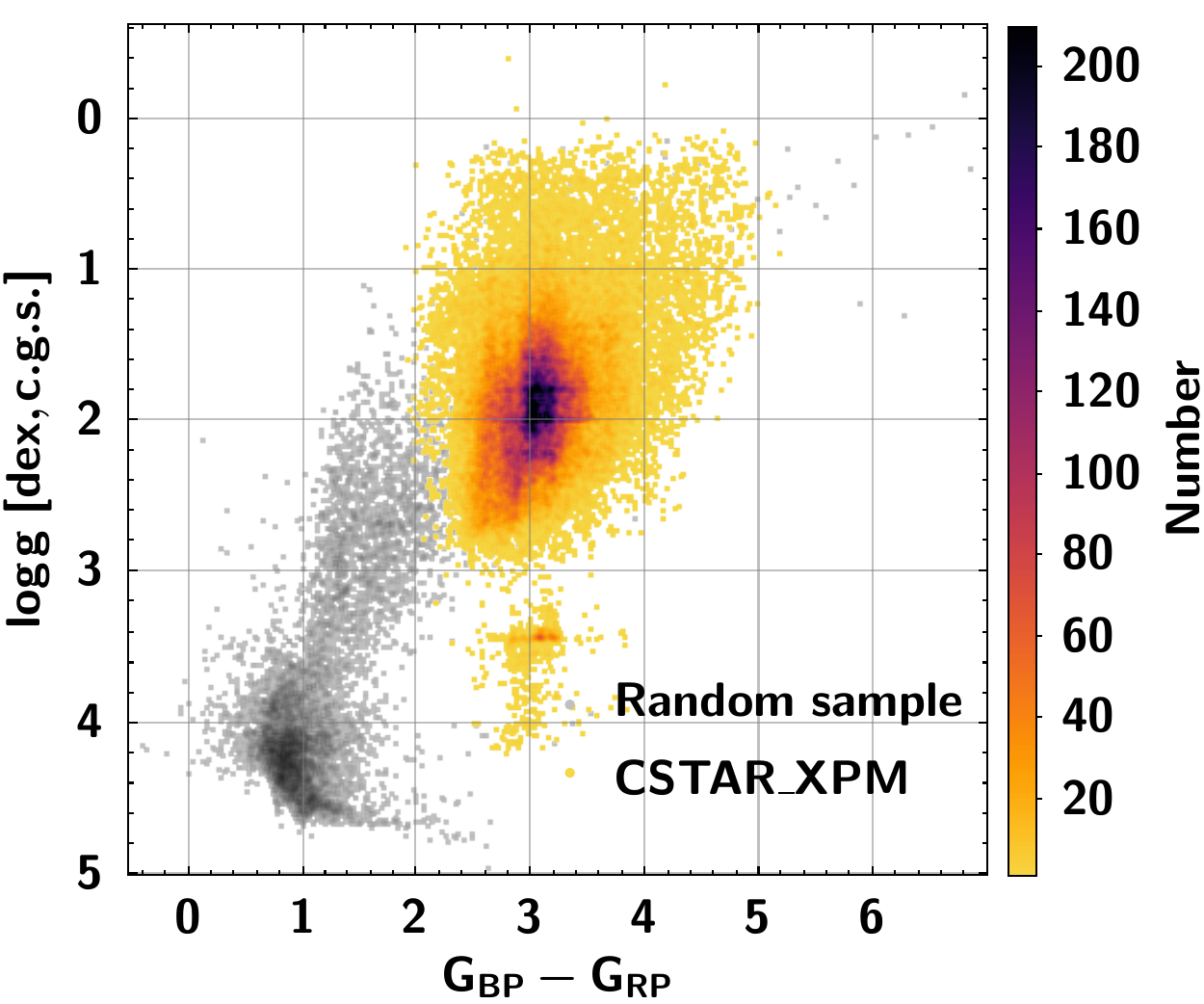}
        \label{fig:bprp_logg1}
      }
      \hfill
      \subfigure[Gold and Non-gold from CSTAR\_XPM]{
        \includegraphics[width=0.32\linewidth]{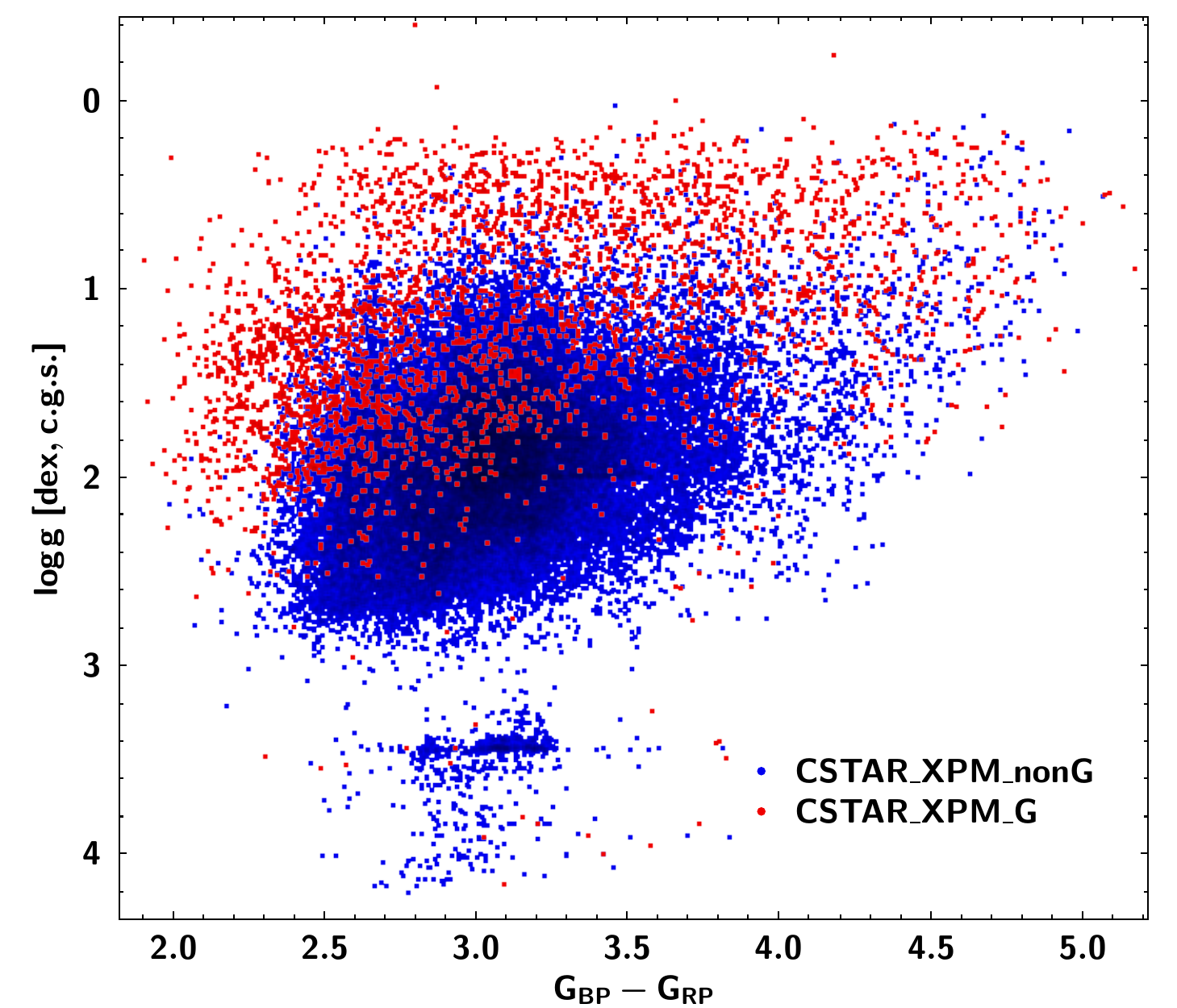}
        \label{fig:bprp_logg2}
      }
       \caption{Panel (a): the $T_\mathrm{eff}$ and $\log{g}$ diagram of CSTAR\_XPM; Panel (b): The spatial location and density distribution of the CSTAR\_XPM and random sample in the $G_{\text{BP}}-G_{\text{RP}}$ and $\log{g}$ plane. The grey random sample is randomly selected from xp\_sampled\_mean. It can be observed that these stars of CSTAR\_XPM are all located in the giant star branch, with most of them having 2 < $G_\mathrm{BP}$ - $G_\mathrm{RP}$ < 5 mag; The distribution of stars labelled as golden sample carbon stars (red) and non-golden sample carbon stars (blue) in CSTAR is shown in the right panel (c).}
      \label{fig:bprp_logg}
    \end{figure*}

    Considering that the $T_\mathrm{eff}$ provided by $Gaia$ DR3 may not be reliable, in Figure \ref{fig:bprp_logg}, we also plotted the $G_{\text{BP}}-G_{\text{RP}}$ (not dereddened) and $\log{g}$ diagrams of all CSTAR\_XPM in order to get a basic idea of the physical properties and evolutionary state of these sources. Since a large proportion of the sources are missing $T_\mathrm{eff}$ and $\log{g}$, we plotted 3245 CSTAR\_XPM\_G sources, 47,290 CSTAR\_XPM\_nonG sources, and 6763 random sample sources.
    It is clear that the CSTAR\_XPM\_nonG and CSTAR\_XPM\_G are spatially intermingled in the distribution and are difficult to distinguish from each other.
    
    From the panel (\ref{fig:kiel}) and (\ref{fig:bprp_logg2}) of Figure \ref{fig:bprp_logg}, we noted that a few targets (19) have a $\log{g}$ greater than 3 and $T_\mathrm{eff}$ hotter than 6000 K. However, their $G_{\text{BP}}-G_{\text{RP}}$ is around 3, and their spectra are typical of AGB carbon stars (showing strong CN bands), which is quite unexpected. Although the possibility of them being extrinsic carbon stars (such as dwarf carbon stars) cannot be ruled out, C2023 mentioned that the $T_\mathrm{eff}$ of these targets tends to be overestimated. Therefore, it is more likely that these are due to stellar parameter determination errors.
    
\begin{figure*}[htbp]
        \centering
        \includegraphics[width=0.95\textwidth]{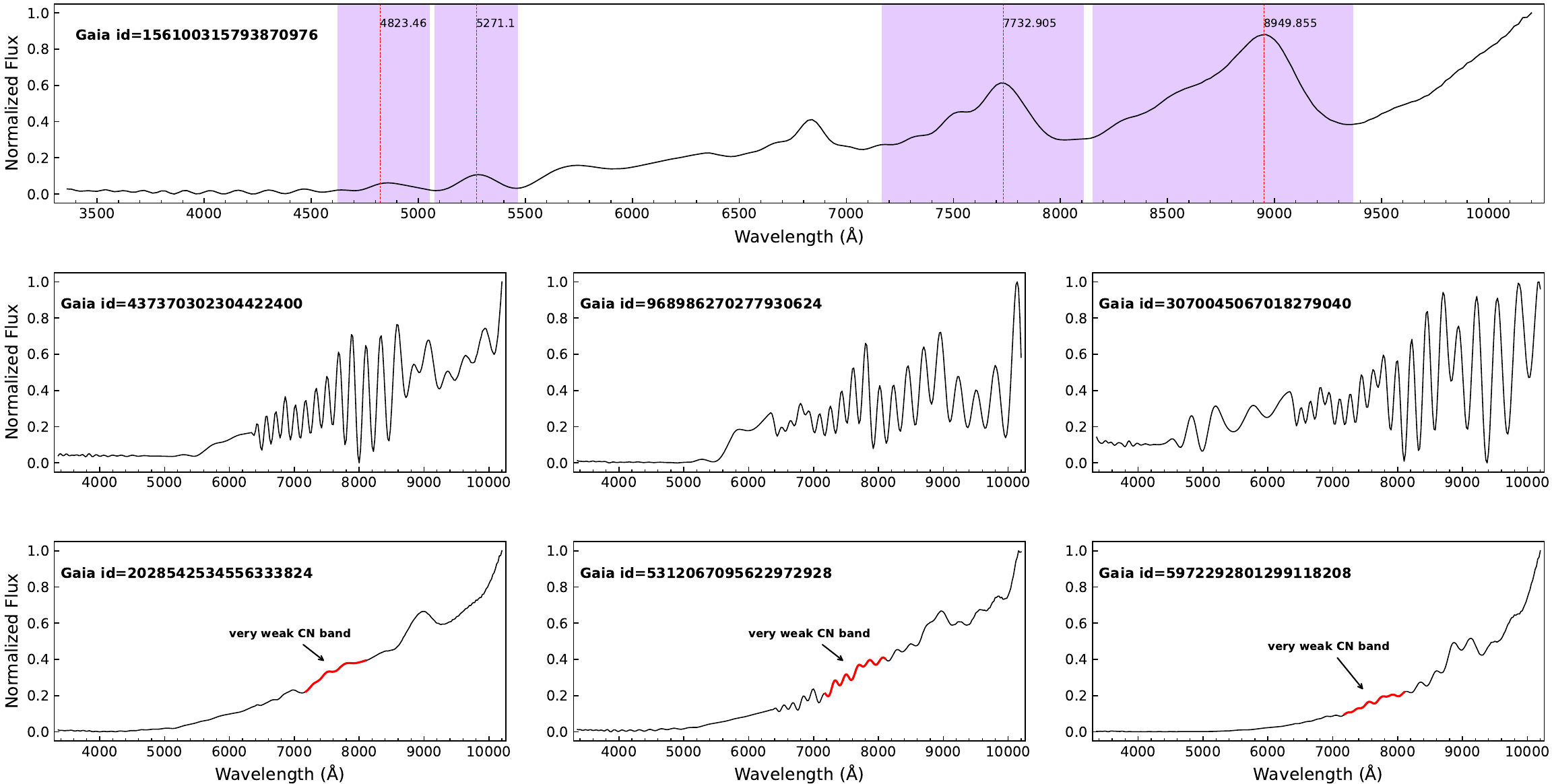}
        \caption{These images are normalized XP spectra of CSTAR\_XPM\_G. The upper panel shows a standard carbon star with four molecular band ranges marked in purple, and the red dotted line indicating the top of the band head. The middle panel shows three problematic spectra, while the bottom panel presents three spectra with very weak CN$_{\text{773.3}}$ molecular band features.}
        \label{fig:spectra1}
    \end{figure*}

    In order to further ensure the purity of the carbon star sample, we visually checked all CSTAR\_XPM\_G spectra. We found that some spectra exhibit significant contamination, which may be due to the signal-to-noise (S/N) problem, and some spectra exhibit insignificant CN characteristics. These spectra would have troubled the purity of our training set. Therefore, we need to eliminate these problematic spectra. We show the XP spectra of a standard carbon star and some of the deleted problematic stars in Figure \ref{fig:spectra1}. After screening we selected 8176 carbon stars as the positive training sample, which showed significant CN molecular band characteristics.
 
\begin{figure}[ht]
        \centering
        \includegraphics[width=0.5\textwidth]{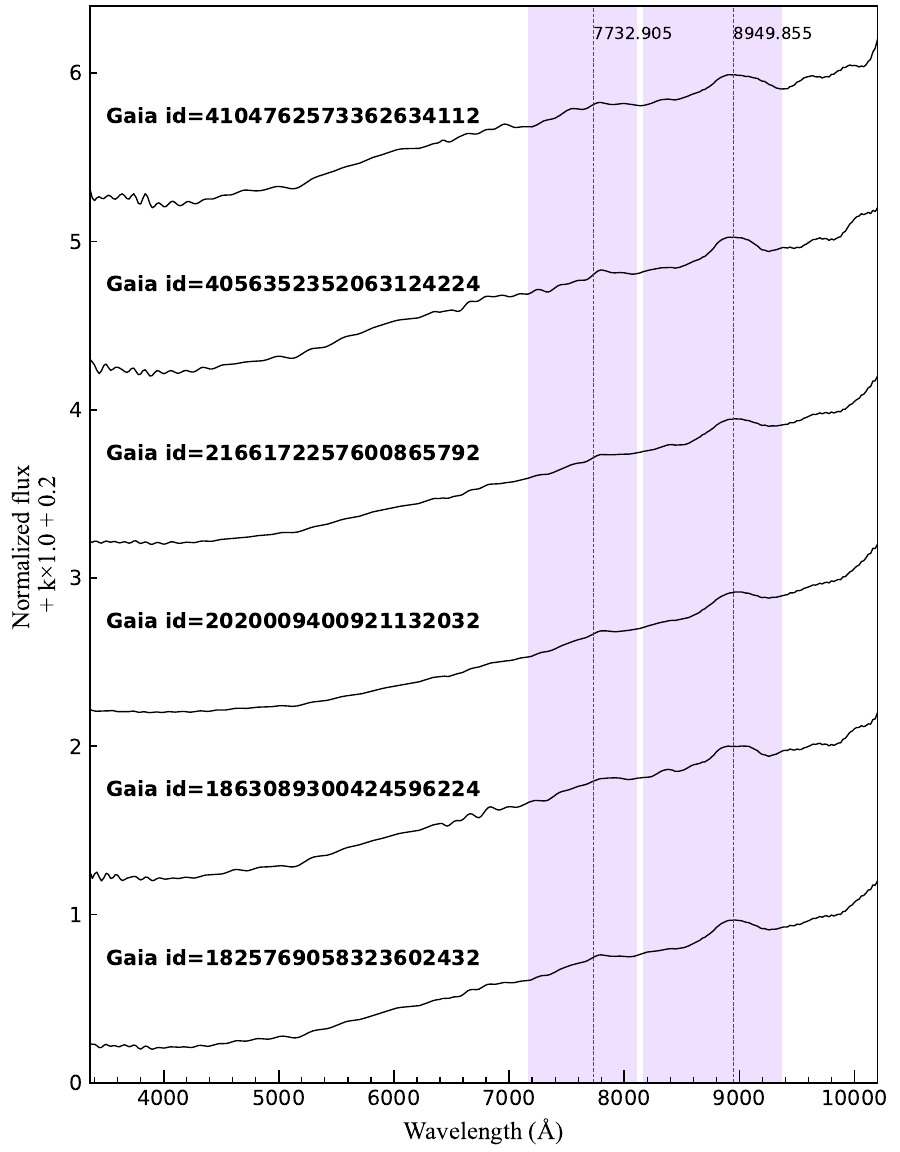}
        \caption{Six representative spectra were selected from the 885 objects removed. The two prominent CN bands (purple areas) are located at the 7165-8108\,\AA\ and 8160-9367\,\AA\ ranges. The red dashed lines mark the top of the band head. The spectra are normalised and offset from one another by $k\times1.0 + 0.2$ for clarity (where $k$ is an integer that varies from 0 to 6 from the bottom to the top spectrum). }
        \label{fig:spectra2}
    \end{figure}

    We then randomly selected 9,000 spectra out of 74,740 CSTAR\_XPM\_nonG as "negative sample", the majority of which are M-type stars \citep{2023A&A...674A..39G}. First, we made sure that none of the 9,000 spectra are included in several known common carbon star lists (Solar Neighbourhood carbon stars: \citeauthor{abia2020properties} \citeyear{abia2020properties}; Galactic carbon stars: \citeauthor{2001BaltA..10....1A} \citeyear{2001BaltA..10....1A}; carbon stars in the Large Magellanic Cloud (LMC): \citeauthor{2001A&A...369..932K} \citeyear{2001A&A...369..932K}; carbon stars in the Small Magellanic Cloud (SMC): \citeauthor{1995A&AS..113..539M} \citeyear{1995A&AS..113..539M}), and that none of them were labeled C* by SIMBAD or listed as `Carbon' by LAMOST's pipeline. 
    Second, here we must mention our training method. We randomly trained several models based on the above dataset. We identified false positive (FP) spectra resulting from obvious model misclassification, where the true category was 0 (other stars), but the model predicted category 1 (carbon star). We iteratively added these spectra to our "negative sample" dataset, finding that this data enhancement method significantly improved the model's generalization capability. 
    Finally, during the visual examination of the "negative sample", we found that many spectra exhibited relatively obvious and prominent molecular bands, particularly the two strong CN molecular bands. Therefore, in order to further ensure the purity of the "negative sample" and ensure that the model could learn the correct features, we calculated the band head strength for all spectra according to the method provided in C2023. We screened out all the suspicious potential carbon stars under the two conditions of R$_{\text{773.3}}\geq 1.03$ and R$_{\text{895.0}}\geq 1.10$, which are almost the weakest CN molecular band strength conditions. 
    We obtained 2255 spectra in total, and then carefully identified the weak CN band head by eye to exclude the interference of possible M and K type giant stars. The reason is that the absorption of M and K giants at the TiO5 (7126-7135\,\AA; \citeauthor{2023ApJS..266....4L} \citeyear{2023ApJS..266....4L}) position is more abundant, and may exhibit a pseudo-R$_{\text{773.3}}$ molecular band strength. Through this checking process, we eliminate 885 objects that exhibit weaker CN molecular bands relative to the cold and bright golden sample of carbon stars. We showed six representative spectra in Figure \ref{fig:spectra2}.     
    
\begin{figure}
        \centering
         \begin{minipage}{0.24\textwidth}
            \centering
            \includegraphics[width=\linewidth]{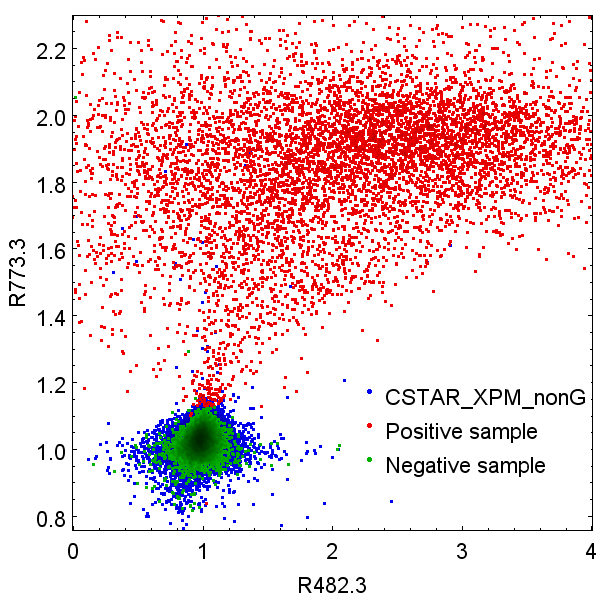}
        \end{minipage}%
        \hfill
        \begin{minipage}{0.24\textwidth}
            \centering
            \includegraphics[width=\linewidth]{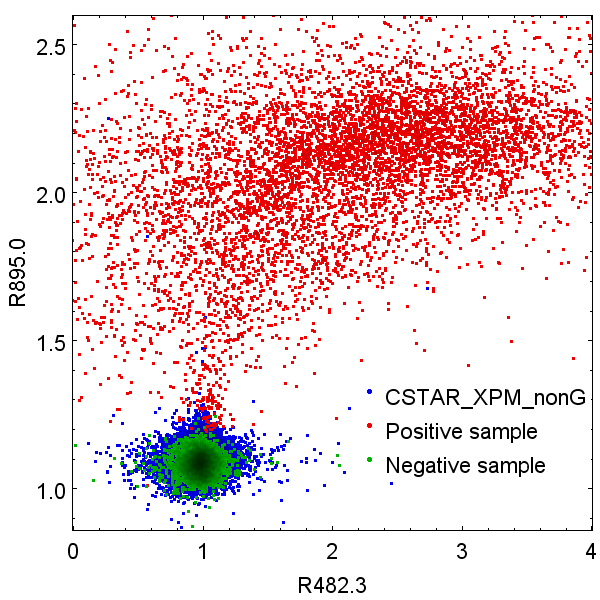}
        \end{minipage}%
        \vspace{1em}
        \begin{minipage}{0.24\textwidth}
            \centering
            \includegraphics[width=\linewidth]{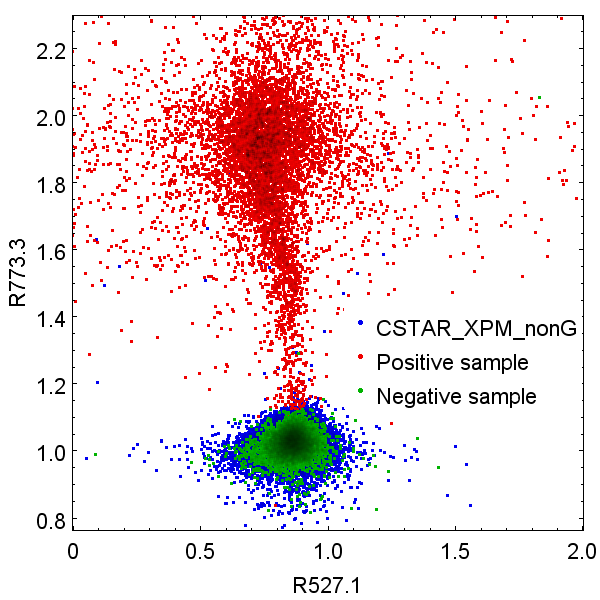}
        \end{minipage}%
        \hfill
        \begin{minipage}{0.24\textwidth}
            \centering
            \includegraphics[width=\linewidth]{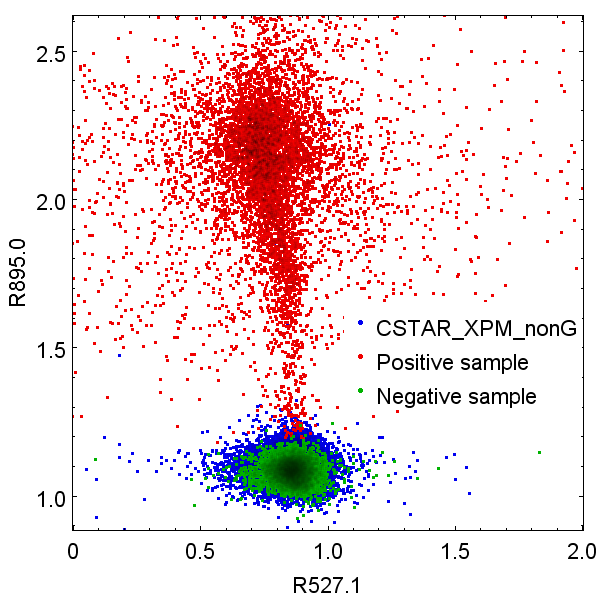}
        \end{minipage}
        \vspace{1em}
        \begin{minipage}{0.24\textwidth}
            \centering
            \includegraphics[width=\linewidth]{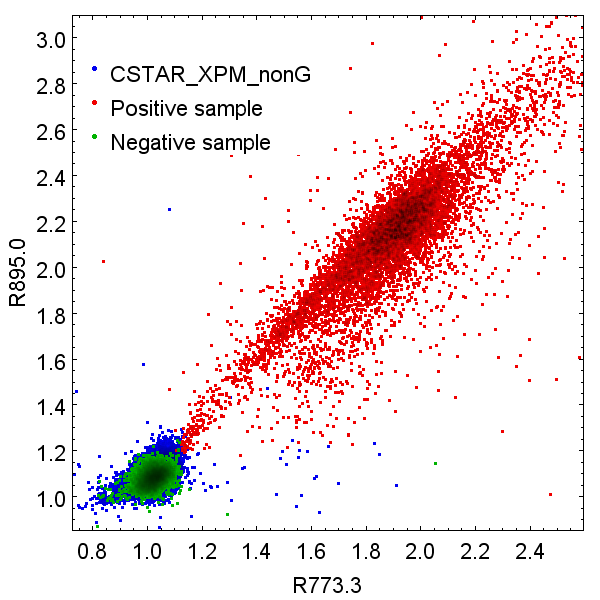}
        \end{minipage}%
        \hfill
        \begin{minipage}{0.24\textwidth}
            \centering
            \includegraphics[width=\linewidth]{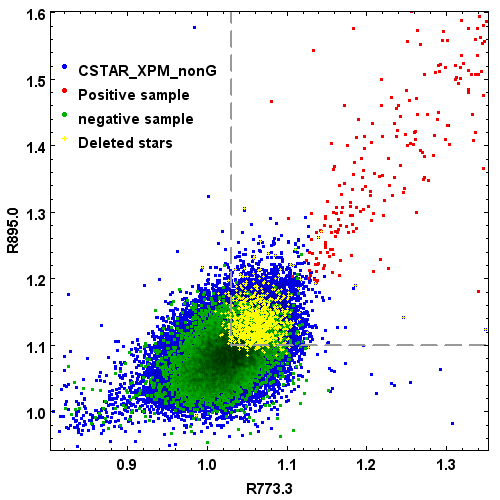}
        \end{minipage}%
        \caption{An alternative view from the C2023 sample, for the 74,740 CSTAR\_XPM\_nonG (blue) flagged by ESP-ELS; 8176 selected training positive sample stars (red) from CSTAR\_XPM\_G; 8556 selected training negative sample stars (green) from CSTAR\_XPM\_nonG and 885 deleted stars (yellow). The grey dashed line in the last subplot represents the boundary for the weakest CN strength.}
        \label{fig:strengths}
    \end{figure}

    Note that these spectra only show weak CN molecular bands, which does not mean that they are all potential carbon stars, we just try to eliminate carbon star contamination as much as possible to obtain pure giant star sample. We also searched and excluded the problematic spectra with condition R$_{\text{773.3}}$ > 1.10 or R$_{\text{895.0}}$ > 1.30, as some of the problematic spectra may have large R$_{\text{773.3}}$ or R$_{\text{895.0}}$. Then we screened the obtained spectra three times to ensure that all spectra did not show any carbon characteristics and then added some FP spectra that were apparently misclassified by the model mentioned above. Finally we kept 8556 training negative sample\label{sec:dataset}, which can be confidently identified as non-carbon stars. We plotted the molecular band head strength distribution of the sample in Figure \ref{fig:strengths}.

\begin{figure}[ht]
        \centering
        \includegraphics[width=0.5\textwidth]{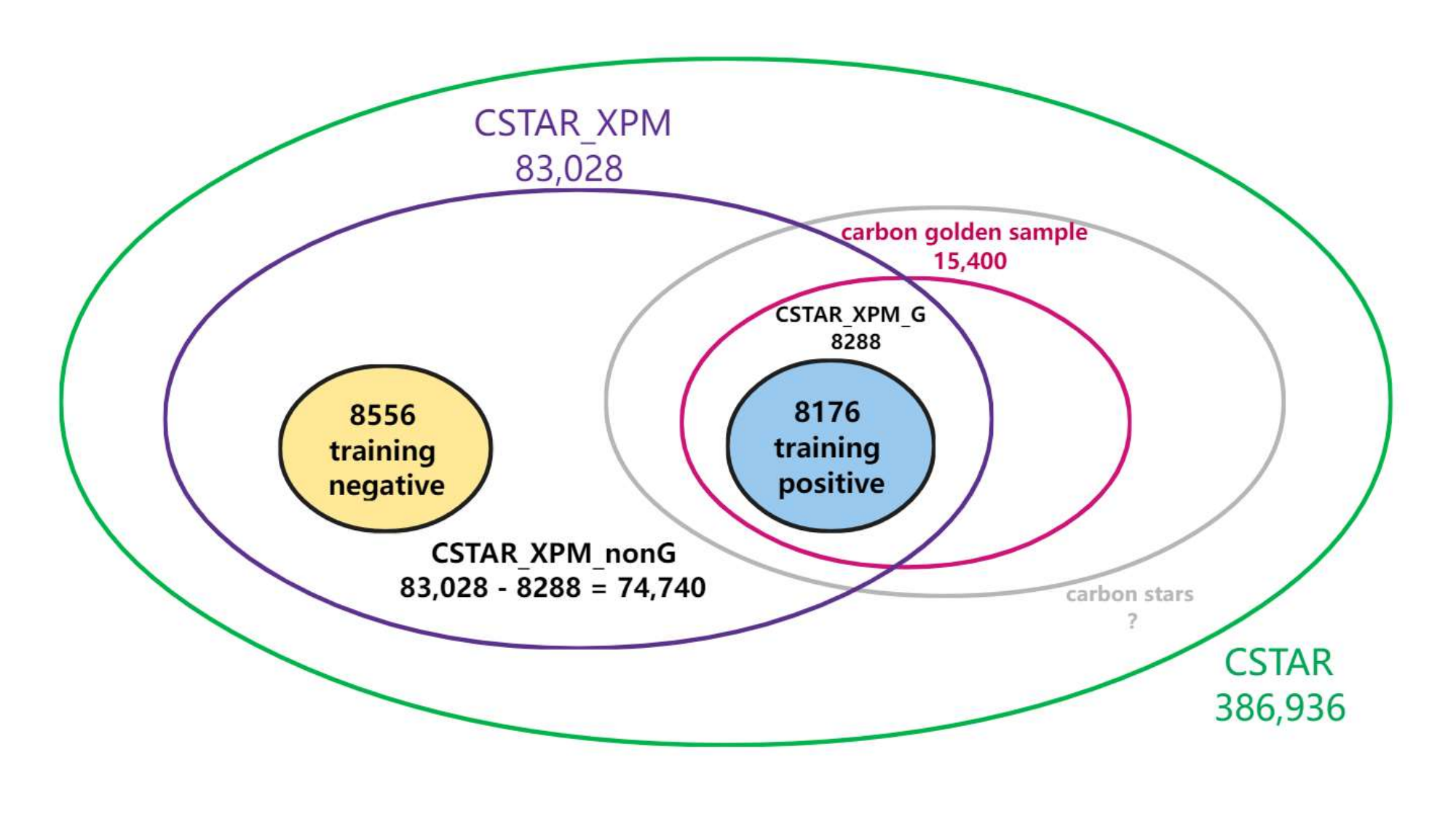}
        \caption{Venn diagram of the relationship of all the samples in CSTAR (green frame). The spectra we used is CSTAR\_XPM (purple frame, left in the figure). CSTAR\_XPM\_G is a cross between CSTAR\_XPM and golden sample of carbon stars (purplish-red frame, right in the figure), and CSTAR\_XPM\_nonG is the complement of CSTAR\_XPM and CSTAR\_XPM\_G. Negative sample (yellow) is selected from CSTAR\_XPM\_nonG, and positive sample (blue) is selected from CSTAR\_XPM\_G. The grey ellipse frame indicates the range of possible carbon stars.}
        \label{fig:venn}
    \end{figure}

    All the samples we used are summarised in Table \ref{tab:data_description}. The Venn diagram in Figure \ref{fig:venn} depicts the relationship between these samples.
    \begin{table*}[htbp]
        \caption{The sample sets used to validate our algorithm.}
        \label{tab:data_description}
        \centering
        \renewcommand{\arraystretch}{1.25}
        \begin{tabularx}{\textwidth}{c|c|c}
            \hline
            \hline
            \textbf{Sample name} & \textbf{Number of spectra} &    \textbf{Sample definition} \\
            \hline
            CSTAR\_XPM & 83,028 & CSTAR\textsuperscript{($a$)} sample with xp\_sampled\_mean spectra \\ 
            CSTAR\_XPM\_G & 8,288 & Golden sample\textsuperscript{($b$)} of CSTAR\_XPM, the vast majority of which are carbon stars \\
            CSTAR\_XPM\_nonG & \textbf{74,740} & Non-golden sample of CSTAR\_XPM, most of which are non-carbon stars \\
            \hline
            Positive sample & 8176 & Carbon stars selected from CSTAR\_XPM\_G, used for model training \\
            \hline
            Negative sample & 8556 & Non-carbon stars selected from CSTAR\_XPM\_nonG, used for model training \\
            \hline
        \end{tabularx}
        \begin{tablenotes}
            \item \textbf{Notes.}
            \item \textsuperscript{($a$)} 386,936 candidate carbon stars received the `CSTAR' tag \citep{2023A&A...674A..39G}.
            \item \textsuperscript{($b$)} 15,740 golden sample carbon stars from CSTAR.
        \end{tablenotes}
    \end{table*}
    We finally got 16,732 stars as training sample, the total number of which is approximately 20.15\% of the total number of CSTAR\_XPM sample. This training sample is used to verify the accuracy of our classification model and the explanatory performance of the interpretable model. We speculate that there may be some additional carbon stars in CSTAR\_XPM\_nonG, so we will apply the trained model to this sample.
     
\section{Method}\label{sec:method}
\subsection{GaiaNet}
     Convolutional Neural Network (CNN; \citeauthor{lecun1998gradient} \citeyear{lecun1998gradient}) is proposed by the concept of perceptual field in biology. By setting several convolutional kernels of different sizes, the model can effectively capture a wider range of global and local information, thereby enhancing the feature extraction of the input signal. The neural network model proposed in this work is named `GaiaNet'.

\subsubsection{The structure of GaiaNet}
    The model is a "light-weight" one-dimensional convolutional neural network, an improvement based on the TextCNN model proposed in \citet{kim-2014-convolutional}. The main reason for choosing this algorithm is that the flexible-sized convolutional kernel of the 1D CNN can effectively capture and extract detail and overall key features when sliding the convolution in one direction, such a working characteristic makes the model more convincing and interpretable.

    There is experimental evidence that extending the depth of a 1D convolutional model can effectively enhance the model's fitting and feature learning capabilities \citep{chen2015convolutional}. A parallel CNN structure consisting of multiple sets of convolutional kernels of different sizes can effectively capture key features of different sizes and achieve similar effects to ensemble learning. We have drawn on these advantages, which are well reflected in the improvement of our model. We first modified the size of the convolutional kernel in the input layer so that it can receive one-dimensional data, i.e., data of shape 1 × 343. We added regularization units consisting of BN (Batch normalization; \citeauthor{ioffe2015batch} \citeyear{ioffe2015batch}) and Dropout. Both methods are effective in improving the generalization of the model. For the data in each batch, the BN layer first normalizes them. The formula for BN is as following:
    \begin{equation}
        \hat{x}=\frac{x-\mu_B}{\sqrt{\sigma_B^2+\epsilon}} \text{,}
    \end{equation}
    where $x$ denotes the input data, $\mu_B$ and $\sigma_B^2$ are the mean and variance of the datas within that batch, respectively, and $\epsilon$ is a very small constant to prevent division by zero errors. The normalized data are then linearly transformed by a set of learnable parameters ($\gamma$ and $\beta$), which can be optimized by a back-propagation algorithm to obtain the final output $y$:
    \begin{equation}
        y=\gamma\hat{x}+\beta \text{.}
    \end{equation}
    The BN layer avoids drastic changes in the input data distribution caused by parameter updates in the preceding model layer and `pulls' the data back into a stable distribution space. Experimental results indicate that it stabilizes the network, effectively accelerates the training and convergence processes, and enhances the classification accuracy of the model.

    Dropout is a commonly used regularization method to avoid overfitting deep network models \citep{hinton2012improving}. Dropout reduces the risk of model overfitting by randomly discarding the outputs of neurons to reduce the complex dependencies within the network during model training. It is implemented by temporarily erasing the output of a neuron with probability $p$:
    \begin{equation}
        y_i=\begin{cases}
        \frac{x_i}{1-p}, & \text{with probability } 1-p \\
        0, & \text{with probability } p 
        \end{cases}
    \end{equation}
    where $x_i$ denotes the original output of neuron $i$ and $y_i$ denotes the output of neuron $i$ after Dropout. During the training phase, each neuron is randomly selected and turned off with a probability of $p$ and retained with a probability of $1-p$. The retained neuron needs to be scaled by dividing its output value by $1-p$ to keep the expectation of the output value constant. We add Dropout before the final fully connected layer of the model to prevent overfitting and to improve the model's generalization.

\begin{figure*}[htbp]
        \centering
        \includegraphics[width=0.95\textwidth]{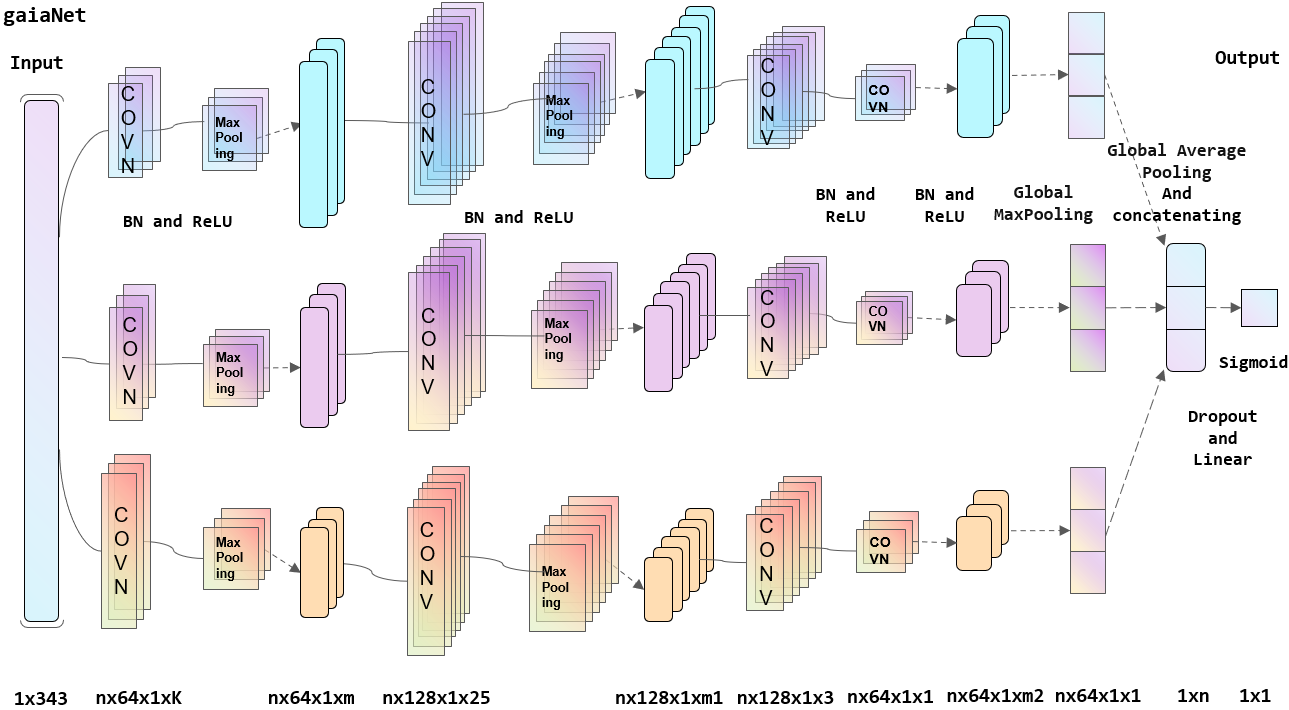}
        \caption{
        Convolutional kernels of different sizes are used to extract features from different receptive field sizes. The arrows indicate the tensors of feature maps after convolution, pooling, or linear operations. A global max-pooling layer downsamples the feature map to compress its size and captures the most sensitive features. These are then combined into a tensor, which then undergoes global average pooling to share each feature value across the entire feature map. The resulting tensor is passed through a fully connected layer, which summarizes the features to produce a high-level abstract representation. Finally, a Sigmoid layer converts the neural network's output into a probability distribution, making the predictions easier to interpret. Below the image, the corresponding convolution, pooling, and feature map shapes are labeled.
        }
        \label{fig:model_flowchart}
    \end{figure*}

    Our model is made up of 16 parallel network structures, each with different-sized convolutional input layers to be able to better capture and extract key features in different fields of view. The same deep hidden layer network structure is then used, which consists of three convolutional layers, two pooling layers, a maximum pooling layer, and a average pooling layer. The deeper hidden layer structure helps to enhance the model's non-linear fitting capability, allowing for the extraction of key non-linear features. The overall architecture of the model is derived from `textCNN', which has been improved by several iterations. We referenced the NIN network by introducing a 1×1 convolution, and a global average pooling operation \citep{lin2013network}. The Inception Module network structure from GoogLeNet is borrowed, i.e., a structure consisting of multiple parallel and smaller dense convolutions with pooling approximating sparse connections \citep{szegedy2015going}. The flow chart of the model is presented in Figure \ref{fig:model_flowchart}.
    
\begin{figure*}[htbp]
        \centering
        \includegraphics[width=0.95\textwidth]{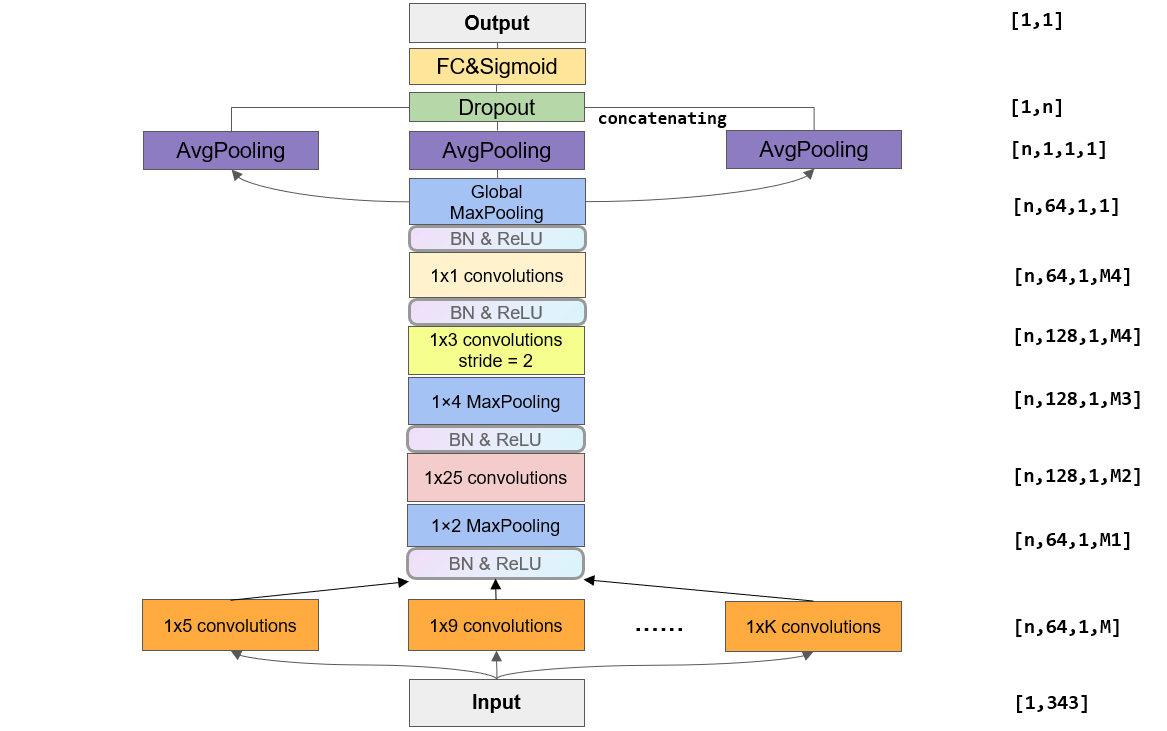}
        \caption{The input data is initially passed through convolutional input layers of different sizes and depths of hidden layers, and then through a global average pooling layer before being stitched together, and through a fully connected layer to obtain a probabilistic output of [0,1] using the Sigmoid layer.}
        \label{fig:model_structure}
    \end{figure*}

    The 1×1 convolution \citep{lin2013network} can further deepen the network, effectively reduce the number of parameters in the overall convolution layer of the model, enhance the model's non-linear fitting ability, allow for flexible dimensioning up and down of the feature map, and enable cross-channel information interaction and integration. Global averaging pooling replaces the original practice of stitching the maximum pooled feature maps directly into the fully connected layer. These feature maps are passed through the global average pooling layer and then concatenated to the linear layer to give the corresponding categories. The purpose of this is to share information from the multi-channel feature maps so that they can all contribute to the final result, which is one of the main innovations of this model. The formula is as below:
    \begin{equation}
        y_{i}=\frac{1}{H} \sum_{j=1}^{H} x_{i, j} \quad i \in[1, n] \text{.}
    \end{equation}
    The global average pooling is handled with parameters $x_{i, j}$ denotes the feature map value at the $j$-th row position of the $i$-th input tensor, and $H$ denotes the height of the tensor. The global averaging pooling layer calculates the average of all pixel points within a single tensor activation feature map. The overall structure of the model is shown in the Figure \ref{fig:model_structure}.
    
    We conducted extensive comparative experiments, and the results indicate that balancing the depth and width of the network, combined with the above improvements, can significantly enhance the model's performance.

\subsubsection{Important model parameters}
    The output layer of our model finally passes through a Sigmoid activation function \citep{lippmann1987introduction}, which is an `S'-type function that converts the real values of the output into a 0-1 probability distribution to represent the confidence level of a positive case\label{sec:confidence}. We usually set 0.5 as the threshold value, case greater than the threshold being classified as positive, and case less than or equal to the threshold being classified as negative. It is in line with the 0,1 label defined when we process secondary classification tasks. We use the  BCELoss (Binary Cross Entropy Loss; \citeauthor{krogh1991simple} \citeyear{krogh1991simple}) function to measure the difference between the model output and the true label, which can be interpreted as a maximum likelihood estimate, i.e., maximizing the probability of the observed data and finally converging the loss by an optimizer. It is a natural pairing with the Sigmoid activation function when dealing with dichotomous problems, which makes the model output easier to interpret and understand.
    The formula for calculating the loss of the output from the Sigmoid activation function using BCELoss is as below:
    \begin{equation} \label{Entropy Loss}
        \mathcal{L} = - \frac{1}{N} \sum_{i=1}^{N}\left[y_i \log(\hat{y}_i) + (1-y_i) \log(1-\hat{y}_i)\right] \text{,}
    \end{equation}
    where $N$ represents the number of input cases in a batch, $y_i$ denotes the true label of the $i$th case, and $\hat{y}_i$ signifies the predicted probability (model output value) of the $i$th case.
    The $p$-parameter of the Dropout layer is also an important parameter, which refers to the proportion of randomly `lost' neurons. The output of our several parallel neural network structures is then spliced together to obtain a vector that is finally discarded with a given $p$-ratio after the Dropout layer. This approach promotes the model's generalization, effectively avoids model overfitting, and achieves a good integrated learning effect.\label{sec:parameters}

    In addition, it is important to balance the size of the convolutional kernels. As larger kernels can effectively expand the model's receptive field and capture features in a broader context, while smaller kernels excel at capturing local details. Therefore, we need to strike a good balance between the whole and the local depending on the actual task.

    Furthermore, the batch\_size refers to the number of input cases in a batch used to update the model parameters during each iteration. A larger batch\_size can speed up convergence by reducing the frequency of parameter updates. Conversely, a smaller batch\_size often improves the model's generalization ability by introducing more variability into each batch, thereby preventing overfitting to the training set. However, the choice of batch\_size should be comprehensively considered based on factors such as dataset size, task complexity, memory limitations, and computational resources.

    The number of epochs for training is a critical hyperparameter that requires careful consideration. Excessive epochs can lead to overfitting, where the model performs well on the training set with good loss convergence but poorly on the validation set due to decreasing accuracy rates with increasing training epochs. On the other hand, insufficient epochs may result in underfitting, causing the model to inadequately learn key information and perform poorly on the actual test set. Therefore, it is crucial to select an reasonable number of epochs based on factors such as model capacity, data volume, and task complexity.

\begin{figure}[htbp]
        \centering
        \includegraphics[width=0.5\textwidth]{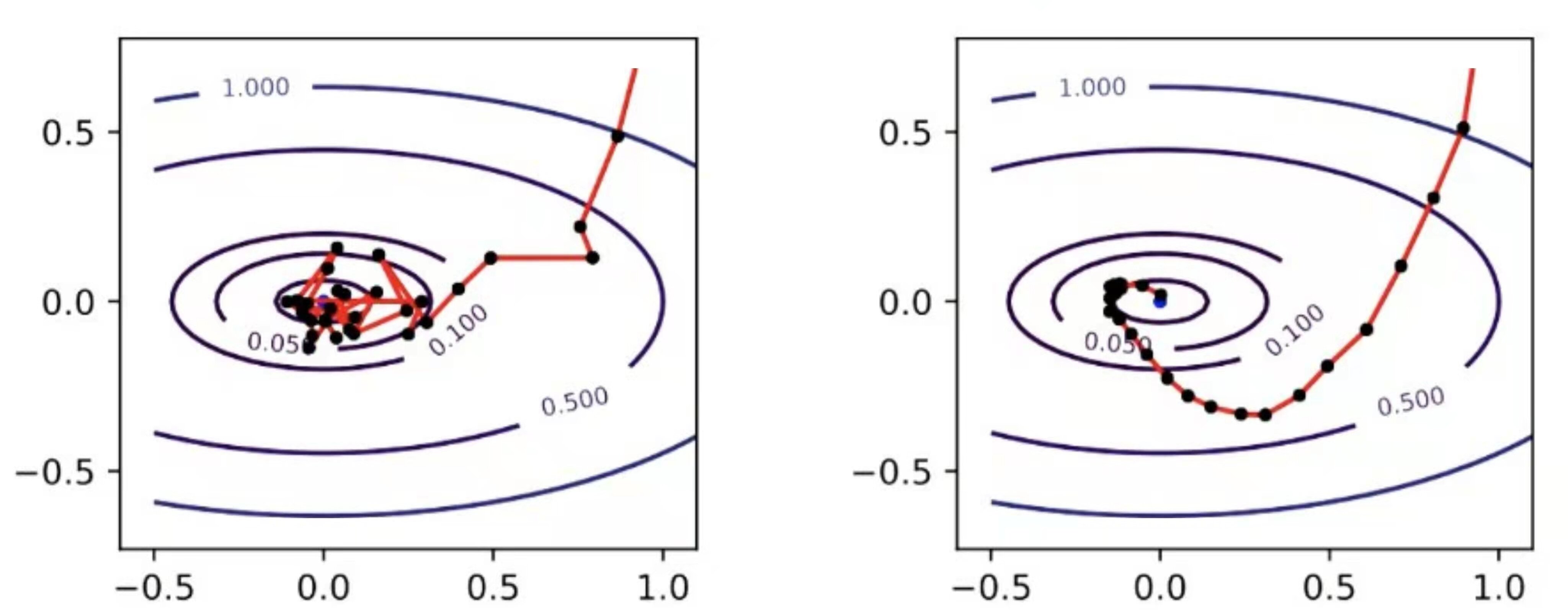}
        \caption{As shown in the figure, the contour line depicts a binary loss function. The left panel depicts the loss update path of a normal SGD, while the right panel shows the loss update path with the addition of momentum, we can see that the update path of the SGD with momentum is smoother and more stable, and converges more easily to the global optimal solution location.}
        \label{fig:momentum}
    \end{figure}

    There are several parameters that play a crucial role in the SGD (Stochastic Gradient Descent) optimization algorithm during the training phase of the model. The $lr$ (learning rate) is a factor that controls the step size of the model weights update, which represents the size of the updated weights at each iteration. A large learning rate may lead to fast convergence of the model during training, but may also lead to the model skipping the optimal solution. A smaller learning rate may result in slower convergence of the model, but it is more likely to achieve a better optimal solution. In our practical tests, we opted for a smaller learning rate to avoid missing the global optimal solution. 
    During the SGD training shown in Figure \ref{fig:momentum}, the size of each step is fixed, but with the introduction of the momentum learning algorithm, the movement of each step depends not only on the magnitude of the current gradient but also on the accumulation of past velocities. `Momentum' uses historical gradient information to adjust the direction and speed of parameter updates \citep{polyak1964some}, thus accelerating the convergence process of the SGD gradient descent and reducing oscillations during the gradient update process. It can be seen as an inertia introduced to leverage the previous gradient information in parameter updates. The core idea of momentum is to introduce an accumulated gradient history variable, which is similar to momentum in physics and records the direction and velocity of the previous gradient's motion. During each iteration of the update, the momentum algorithm considers not only the current gradient but also the trends of previous gradients. This allows the parameters to be updated with a certain `momentum' in the direction of the gradient, resulting in faster crossing of flat areas and less invalid oscillation. The parameter update formula in the momentum algorithm is shown below:
    \begin{equation}
        v_{t+1} = \mu v_{t} - \varepsilon \nabla f\left(\theta_{t}\right) \text{,}
    \end{equation}
    \begin{equation}
        \theta_{t+1} = \theta_{t} + v_{t+1} \text{,}
    \end{equation}
    where $\varepsilon$ > 0 represents the learning rate, $\mu\in[0, 1]$ is the momentum parameter, and $f\left(\theta_{t}\right)$ is the gradient at $\theta_{t}$ \citep{sutskever2013importance}.
    We found in our practical tests that it was excellent for improving model accuracy and further converging model loss.
    
    Weight decay is a regularisation technique that reduces the complexity of the model by reducing the size of the weights. It applies an L2 regularisation penalty to the weight parameters in the loss function to prevent overfitting, so that larger weights are penalized more during training, thus encouraging the model to learn a simpler weight distribution.
    
\subsection{SHAP}
    SHAP (SHapley Additive exPlanations) is an interpretable model based on the game theory principles approach to explain the output of any machine learning model. It connects optimal credit allocation with local explanations using the classic Shapley values from game theory and their related extensions. SHAP can give the extent to which each input feature affects the output of the model and offer a comprehensive and interpretable analysis of the anticipated results of the entire model \citep{NIPS2017_7062}. The Shapley value is a concept in game theory that measures the contribution of each player to the cooperative game. In machine learning, we can consider different features as "players" and the combination of their values as a "cooperative game". Using the principle of equivalent assignment of Shapley values, the contribution or importance of each input feature to the overall model output can be calculated, and each feature can be assigned a relative contribution value. And SHAP extends Shapley values to deep learning models.
    
    The prediction value of the model is interpreted as the sum of the importance of each input feature. The relation between the predicted value and importance scores is performed below:
    \begin{equation}\label{eq:marginal contribution}
    \centering
        f(x)  = g\left(x^{\prime}\right) = \phi_{0}+ \sum_{i=1}^{M}\phi(x_{i}^{\prime}) \text{.}
    \end{equation}
    Specifically, for a deep learning model $f:X \to Y$, where $X$ denotes the input features and $Y$ denotes the class or real value of the output, $M$ is the number of features. The explanation model $g\left(x^{\prime}\right)$ matches the original model $f(x)$ when $x = h_x(x^{\prime})$, where $\phi_{0} = f(h_x(0))$ represents the model output with all simplified inputs toggled off (i.e. missing). 
    SHAP defines the Shapley value for an input $x_i \in X$ as $\phi(x_{i}^{\prime})$, which represents the marginal contribution of feature $x_i$ to the model prediction.
    
    The Shapley value of each feature represents the degree of its influence on the predicted outcome of a single input case. The calculation of the Shapley value is based on combinatorial game theory, which takes into account the interactions between each feature and other features and assigns the final predicted outcome to each feature. For a specific case, the calculation of the Shapley value $\phi(x_{i}^{\prime})$ of feature $x_i$ is performed below:
    \begin{equation}
    \centering
        \phi(x_{i}^{\prime})  = \sum_{S \subseteq X \setminus x_{i}^{\prime}} \frac{|S|! (M-|S|-1)!}{M!} [f(x_S \cup {x_{i}^{\prime}})-f(x_S)] \text{,}
    \end{equation}
    where $S \subseteq X \setminus x_{i}^{\prime}$ is any possible subset of all input features excluding $x_i$. $f(x_S \cup {x_{i}^{\prime}})$ represents the output obtained by adding feature $x_i$ to the subset $S$. $f(x_S)$ represents the output obtained by using the subset $S$ for prediction, and the result of subtracting the two values represents the effect of the addition of feature $x_i$ on the model output. And the ${|S|! (M-|S|-1)!} \setminus M!$ represents the weight.
    The overall equation represents the calculated contribution for each feature subset $S$. This is the contribution from adding feature $x_i$ to the feature subset $S$. Finally, the contribution of each feature subset $S$ is summed to obtain the Shapley value of the feature $x_i$.
    
    This method allows us to compare the Shapley values of different features to determine how much they affect the output of the model. It has been flexibly applied to various deep learning models, including convolutional neural networks (CNN), recurrent neural networks (RNN; \citeauthor{werbos1988generalization} \citeyear{werbos1988generalization}), and others. We will use the SHAP in our model to interpret the key features that affect the model output. In addition, SHAP provides several visualization tools and methods, which will help us understand the model better.
    
\subsection{Evaluation index}
    In addition to the accuracy criterion, we also introduce other  evaluation metrics such as recall and precision to effectively measure the model's data mining performance during dataset validation. We define TP (true positive) as the number of cases whose categories are true and predicted categories are also positive, FN (false negative) as the number of cases whose categories are true but predicted categories are negative, FP (false positive) as the number of cases whose categories are false but predicted categories are positive, and TN (true negative) as the number of cases whose categories are false and predicted categories are also negative. 
    
    Accuracy refers to the proportion of the number of accurate cases classified by all categories to the total number of cases. We  calculate it as below:
    \begin{equation} \label{Accuracy}
        Accuracy = \frac{TP + TN}{TP + TN + FP + FN} \text{.}
    \end{equation}
    Precision refers to the proportion of the number of correctly predicted positive cases to the total number of predicted positive cases. It is calculated as below:
    \begin{equation} \label{Precision}
        Precision = \frac{TP}{TP + FP} \text{.}
    \end{equation}
    Recall refers to the proportion of correctly predicted positive cases relative to all actual positive cases. It is calculated as below:
    \begin{equation} \label{Recall}
        Recall = \frac{TP}{TP + FN} \text{.}
    \end{equation}
    We also use confusion matrix to see the performance of the model more intuitively, as it effectively illustrate the distribution across different classes. For our carbon star data mining work, the idea is to have as many TP cases as possible and as few FN cases as possible, i.e., as large a recall rate as possible, so we use this as our main validation set diagnostic.

\subsection{Resampling technique}
    The resampling technique \citep{10.1214/aos/1176344552} is an important data processing method used to solve the problem of imbalance in the categories of a data set. For example, when the number of positive cases is slightly less than the number of negative cases, resampling techniques can be used to increase the number of positive cases. This is done by randomly copying or generating new positive cases, thus making the number of positive and negative cases consistent and balanced. This can effectively prevent the models from "collapsing" into categories with large case sizes during training.
    
\section{Results}\label{sec:result}
    After validation with several sets of parameters in Section \ref{sec:parameters}, we fixed the training parameters of the model as follows: $p=0.4$, $lr=0.005$, momentum$=$0.9, weight\_decay$=$0.001.
    
\subsection{Analysis and experiments on the dataset}
    Almost all N-type carbon stars are in the AGB evolution stage, where the carbon abundances in their atmosphere were produced in the stellar interiors, and were brought to the surface by the TDU process. Almost all of them are intrinsic, thermally pulsing AGB stars, i.e., TP-AGB stars with large luminosities \citep{2002ApJ...579..817A}. This kind of carbon star typically exhibit stronger C$_{\text{2}}$ and CN molecular bands than usual stars with temperatures below 3800\,K \citep{2023A&A...674A..39G}. Of course, there are also some kinds of carbon stars with relatively high surface temperatures, whose spectral types can be K or even G, and they are often extrinsic carbon stars. However, there is a large number of giant stars with the spectral type G, K, and M, which have similar colors, surface temperatures, and luminosities to the carbon stars, making it impossible to distinguish between the carbon stars and other giants just from the HR diagrams, but now possible through $Gaia$'s spectroscopy. \citet{2023A&A...674A..15L} and \citet{2023MNRAS.521.2745S} have demonstrated its potential for separating C-rich and O-rich stars.
     
\begin{figure*}
        \centering
        \includegraphics[width=1.0\textwidth]{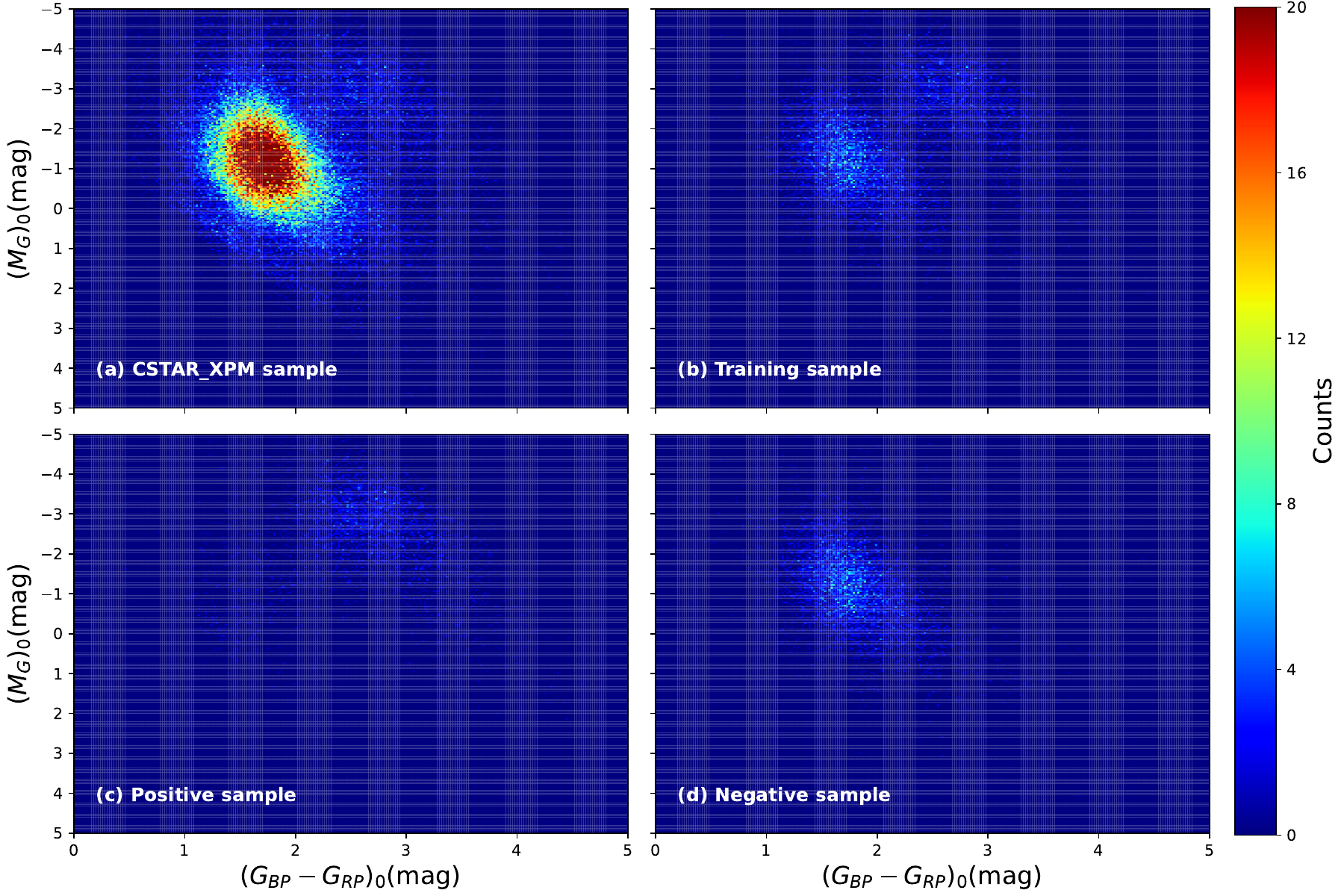}
        \caption{The distribution of CSTAR\_XPM and training sample on $(M_G)_{0}$ vs. $(G_{\text{BP}}-G_{\text{RP}})_{0}$ diagram. All the samples have been corrected for 3D extinction. Panel (a): the distribution of the CSTAR sample. Panel (b): the distribution of the training sample. Panel (c): the distribution of carbon stars from the training sample. Panel (d): the distribution of non-carbon stars from the training sample.}
        \label{fig:BPRP_MG}
    \end{figure*}

    We found that almost all of the CSTAR\_XPM\_nonG are M giants through the $G_{\text{BP}}-G_{\text{RP}}$ and $\log{g}$ diagram (see Figure \ref{fig:bprp_logg}), and their distribution overlaps with the region where the golden carbon stars located. In order to obtain reliable luminosity and color parameters, we corrected the photometric data with 3D interstellar extinction \citep{2019ApJ...887...93G}, and then a color-magnitude diagram is shown in Figure \ref{fig:BPRP_MG}.
    
    After the extinction correction, we calculate the absolute magnitudes $(M_G)_{0}$ and the intrinsic color index of $(G_{\text{BP}}-G_{\text{RP}})_{0}$ for 4002 positive sample stars, 6761 negative sample stars, and 64,761 stars of CSTAR\_XPM sample, respectively. Most of the CSTAR\_XPM stars have 1 < $(G_{\text{BP}}-G_{\text{RP}})_{0}$ < 3 and $(M_G)_{0}$ < 2 mag, and most of the carbon stars have 2 < $(G_{\text{BP}}-G_{\text{RP}})_{0}$ < 4 and $(M_G)_{0}$ < -1 mag. This suggests that the CSTAR\_XPM\_G sample may consist mostly of brighter and cooler N-type carbon stars. From the corrected color-magnitude diagram, we also found that about half of the negative sample stars have a $(G_{\text{BP}}-G_{\text{RP}})_{0}$ $\geq 1.6$, while the $(G_{\text{BP}}-G_{\text{RP}})_{0}$ of M giants should usually be larger than 1.6. We estimated that there are about half of the G and K giants in the CSTAR\_XPM\_nonG, which is corroborated by our later cross-matched results with the LAMOST DR10 spectroscopic data in Section \ref{sec:cross}. Hence, the CSTAR sample also includes a large mixture of other types of stars, rather than overwhelmingly M stars. In addition, a portion of the CSTAR\_XPM\_G has $(G_{\text{BP}}-G_{\text{RP}})_{0}$ < 2 and relatively lower luminosities. If these sources are located in the crowded regions of the Galactic plane where the dust distribution is complex and non-uniform, the true interstellar medium does not follow a simple Gaussian process \citep{2019ApJ...887...93G}. In this case, their extinction must be treated with caution. If their color index estimates are correct, as suggested by Figure 11 of \citet{2024ApJS..271...12L}, they are most likely extrinsic warm carbon stars with spectral types of G or K like Ba stars. It can be seen that there is a slightly clearer distinction between positive sample and negative sample in the color magnitudes, but some of the stars are still mixed together. However, elemental abundances are also reflected in the spectra \citep{2002ApJ...579..817A}. Specifically, the carbon stars are characterized by the strong absorption lines of carbon molecules in their spectra. This offers the possibility of a spectroscopic data-driven approach to distinguish the two directly from the stellar spectra. C2023 proposed screening carbon stars by R$_{\text{773.3}}$ and R$_{\text{895.0}}$, it could be partially contaminating and would result in the loss of some potential carbon stars through our inspection. 
    
    As mentioned in Section \ref{sec:train_data}, we carefully selected a negative sample of 8556 spectra from CSTAR\_XPM\_nonG and a positive sample of 8176 spectra from CSTAR\_XPM\_G. We equalized the number of positive and negative samples by randomly adding negative spectra using the resampling techniques. So we composed a training dataset after fine visual inspection and machine ranking to try to ensure better distinction for deep learning way and meaningful interpretable analysis.
    
\subsubsection{Model validation}
    We randomly selected 70 percent of the datasets to train the classification model, and the remaining 30\% was used as the validation set to measure the model's performance as the epochs increased. 
    
\begin{figure}
        \centering
        \includegraphics[width=0.5\textwidth]{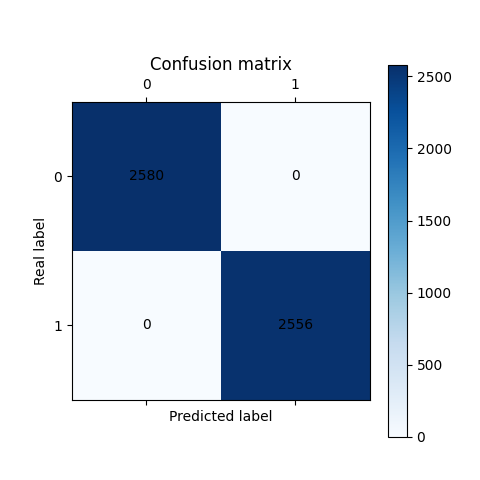}
        \caption{Label 0 represents a non-carbon star and 1 represents a carbon star. The abscissa represents the predicted category and the ordinate represents the actual category.}
        \label{fig:conf}
    \end{figure}

    The best performance of the model occurs when the epochs are 9, 16, 17, 18, and 19. The final recall rate in the validation dataset is 100 percent, the accuracy is also 100 percent. The confusion matrix is shown in Figure \ref{fig:conf}, showing that our model has extremely high accuracy and reliability for classifying all stars in the validation dataset. Our model can easily distinguish the spectra of cold and bright carbon stars with strong carbon molecular bands from the spectra of giant stars of G, K, M types. We will provide necessary comparisons and explanations in Section \ref{sec:key feature} \label{sec:valid}.
    
\begin{figure}[htbp]
      \centering
      \subfigure[The accuracy of the training set and the validation set]{
        \includegraphics[width=0.5\textwidth]{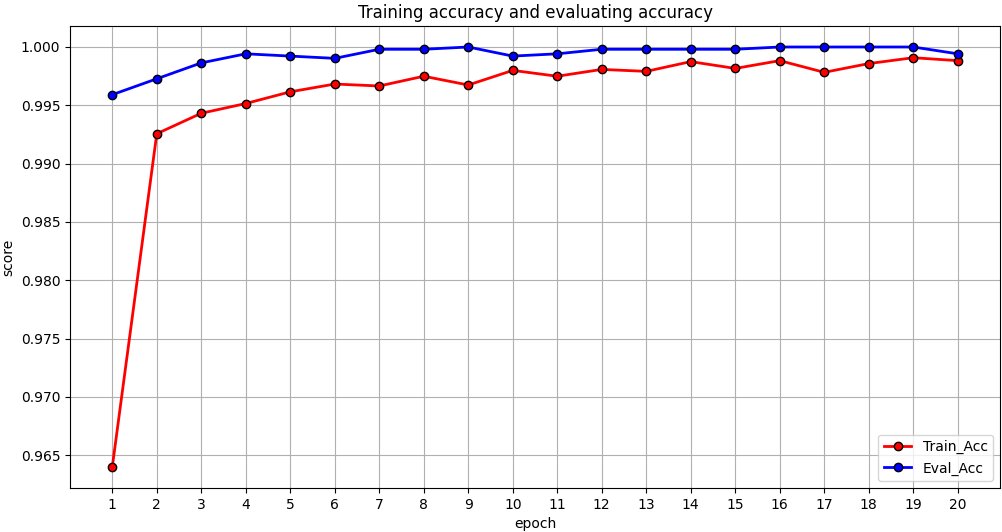}
        \label{fig:train_eval_acc}
      }
      \hfill
      \subfigure[FN and FP]{
        \includegraphics[width=0.5\textwidth]{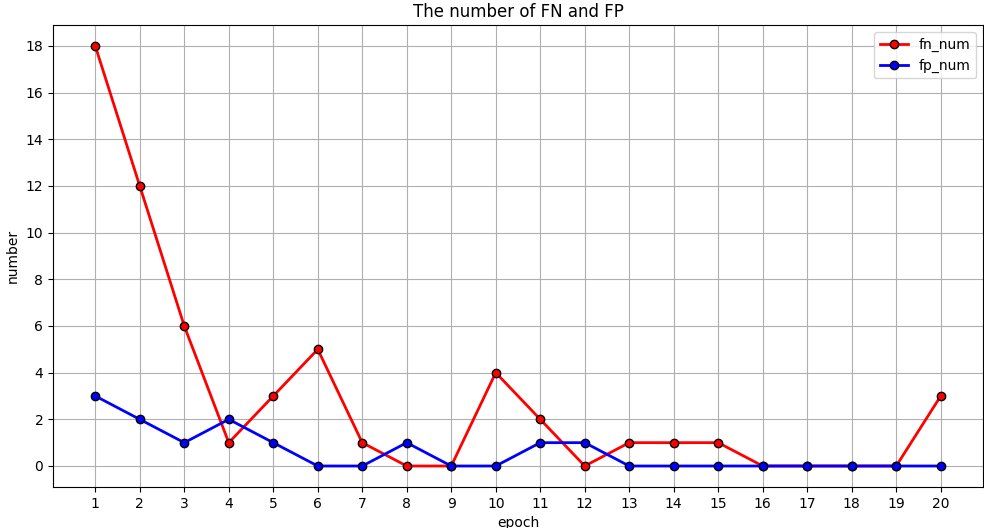}
        \label{fig:fn_fp}
      }
      \caption{Panel (a): the performance of the model on the training set and the validation set changes with the number of epochs; Panel (b): the number of misclassified spectra of the model varies with the number of epochs.}
      \label{fig:eval}
    \end{figure}

    We also plot the change in the correctness of the training and validation sets and the number of FP, and FN spectra as the training epochs increase, as shown in Figure \ref{fig:eval}. The best performance on the validation set (30\%) is 100\% accurate, with excellent convergence and no overfitting of the model.
    
\subsubsection{Comparison with other algorithms}    
    In the past decades, many machine learning includes integrated learning methods (KNN: \citeauthor{cover1967nearest} \citeyear{cover1967nearest}; SVM: \citeauthor{cortes1995support} \citeyear{cortes1995support}; Random Forest: \citeauthor{breiman2001random} \citeyear{breiman2001random}; XGBoost: \citeauthor{chen2016xgboost} \citeyear{chen2016xgboost} etc.) have been successfully applied to stellar spectral classification and physical parameter estimation. In contrast, convolutional neural networks are well known to have been used with great success in the field of image recognition, so we see the potential for the use of one-dimensional convolutional deep learning networks in spectral recognition. 

    In Table \ref{tab:Comparison}, we compiled a comparison of the accuracy (\%), recall, and error rate (\%) of several common machine learning classification models with our model on the performance of the validation set.

    \begin{table}
        \centering
        \caption{For each model, the data was split into 70\% for training, and 30\% for validation.} 
        \vspace{1.5pt}
        \begin{tabular}{llcc}
            \hline \hline \text{ Method } & \text{ Accuracy (\%) } & \text{ error (\%) } & \text{ Recall } \\
            \hline \text { Our Model\textsuperscript{1} } & \text { 100 } & 0.0 & 1.0 \\
            \hline \text { XGBoost } & \text { 99.82 } & 0.18 & 0.99794 \\
            \text { Random Forest } & \text { 99.80 } & 0.2 & 0.99678 \\
            \text { KNN } & \text { 99.70 } & 0.3 & 0.99518 \\
            \text { SVM } & \text { 99.62 } & 0.38 & 0.99282 \\
            \hline
        \end{tabular}

        \begin{tablenotes}
        \item {\textbf{Notes.}} \\
        \textsuperscript{1} The best performance in the validation set is selected when the epochs are 9, 16, 17, 18, and 19.
    \end{tablenotes}
        \label{tab:Comparison}
    \end{table}
    Recall is a very important evaluation index that effectively measures the rate of retaining positive sample. Our model retains all stars of positive sample and demonstrates approximately a 0.43\% improvement over other algorithms, resulting in an overall accuracy increase of about 0.3\%. The reason for the error rate of 0 in the validation set is likely to be the excellent overall architecture of our model, which is mainly reflected in the natural advantages of the convolutional model in feature extraction, as well as some necessary innovative improvements. On the other hand, we have conducted "expert" screening and purification of the training data to exclude erroneous spectra and ensure the high purity and reliability of the training data, combined with some necessary data enhancement methods to ensure that the model can learn the appropriate information. Though of course, The spectra of positive sample exhibit strong CN molecular bands, making them easily distinguishable from negative sample.

\subsection{Model interpretation and key features} 
    The SHAP value is used to quantify the effect of features on the model output. We consider each wavelength in a spectrum as a feature, and the corresponding flux to each wavelength as the feature value. The SHAP interpretable model can explain our deep learning model by calculating a SHAP value for each feature of a given spectrum, which indicates the importance impact or contribution of the feature to the model output (i.e. the confidence of a carbon star as mentioned in subsection  \ref{sec:confidence}). A positive SHAP value indicates that the feature contributes positively to the model’s prediction of the spectrum being a carbon star, while a negative SHAP value indicates that the feature contributes negatively to the model’s prediction of the spectrum being a carbon star. The larger the absolute value of SHAP, the greater the impact of the feature on the model output. For a spectrum, the sum of the SHAP values  of each feature can be used to approximate the output of the classification model. We can also calculate the distribution of the most important features based on the statistics of SHAP values from a large sample. 

\begin{figure*}
        \centering
        \includegraphics[width=0.95\textwidth]{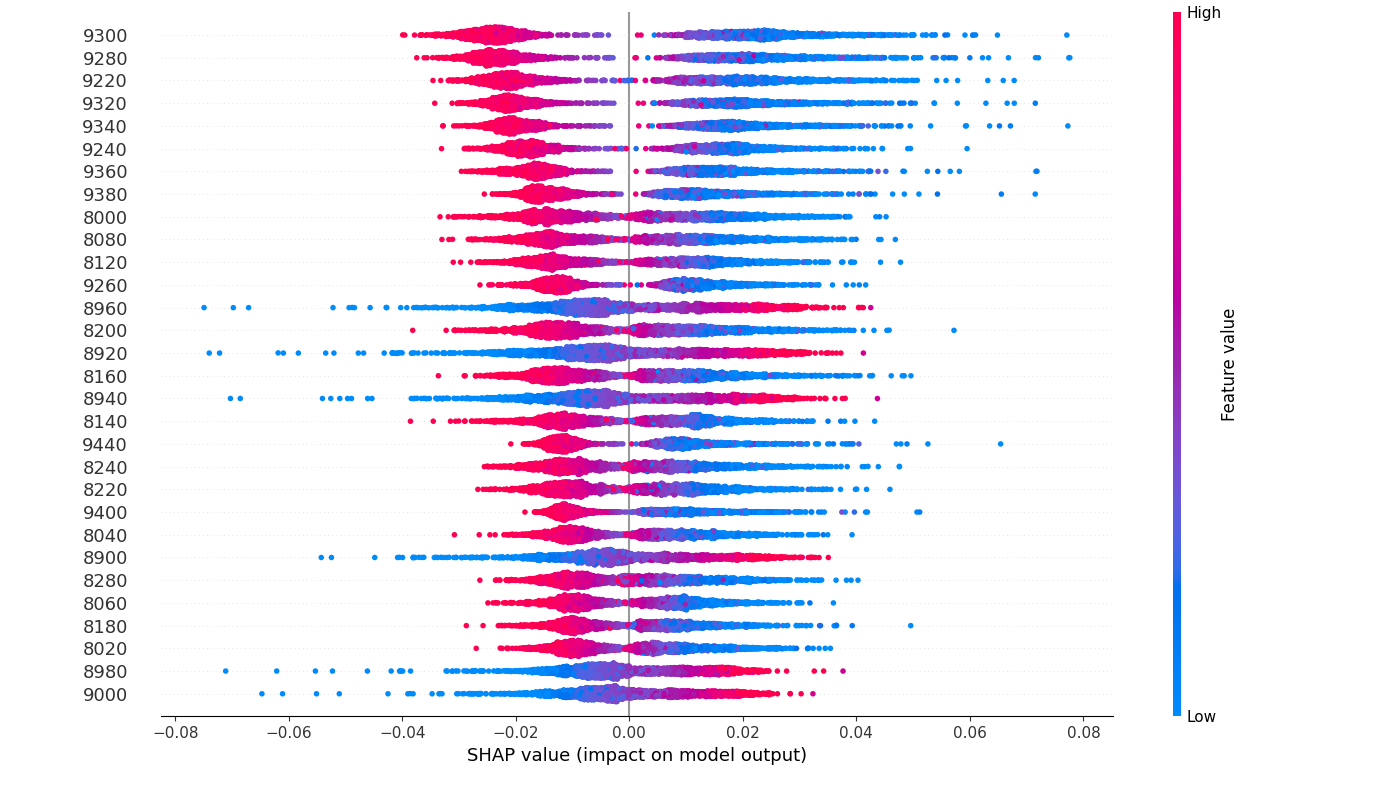}
        \caption{The beeswarm plot visualizes the link between feature values and model interpretation. Each point in the beeswarm represents a feature point from a spectrum. The horizontal axis displays SHAP values corresponding to each feature point, and the vertical axis is the top 30 most important features (i.e. wavelengths) for identifying carbon stars. From top to bottom, features contribute decreasingly to the model prediction. The color of each feature point represents the size of the feature value.}
        \label{fig:bees}
    \end{figure*}

    To get an overview of which features are most important and intuitively understand the influence of them on the model prediction results, we randomly selected 1024 carbon stars of positive sample and 1024 non-carbon stars of negative sample as background spectra of the deep explainer interpreter of SHAP,  calculated the SHAP values of these spectra, and plotted in Figure \ref{fig:bees}. The top 30 most important features were shown in this figure, which were ranking by the sum of absolute SHAP value magnitudes of them over all 2048 spectra, and uses SHAP values to show the distribution of the impacts each feature has on the model output. For a particular feature, SHAP first normalizes its values across all spectra, the normalized feature values are then mapped onto a color gradient bar. The feature point will be assigned different colors (from blue to red), the color blue represents a feature point value that is relatively small, and red means that a feature point value is relatively large.

    Therefore, the beeswarm plot can be an intuitive reflection of the influence of the feature on the output results for different values. As shown in the figure, we can see that the wavelength range 8900-9000\,\AA\ is the location of the peak arising from the molecular band head. The larger the values of these feature points are, the more positively the model predicts carbon stars. Near 8100\,\AA\ and 9300\,\AA\ are the locations of the troughs arising from CN molecular absorption bands. The smaller these feature values are, the more negatively the model predicts carbon stars. It is evident that the most striking distinguishing feature is the presence of these two troughs.

\begin{figure*}[htbp]
          \centering
          \subfigure[Carbon star]{
            \includegraphics[width=0.95\textwidth]{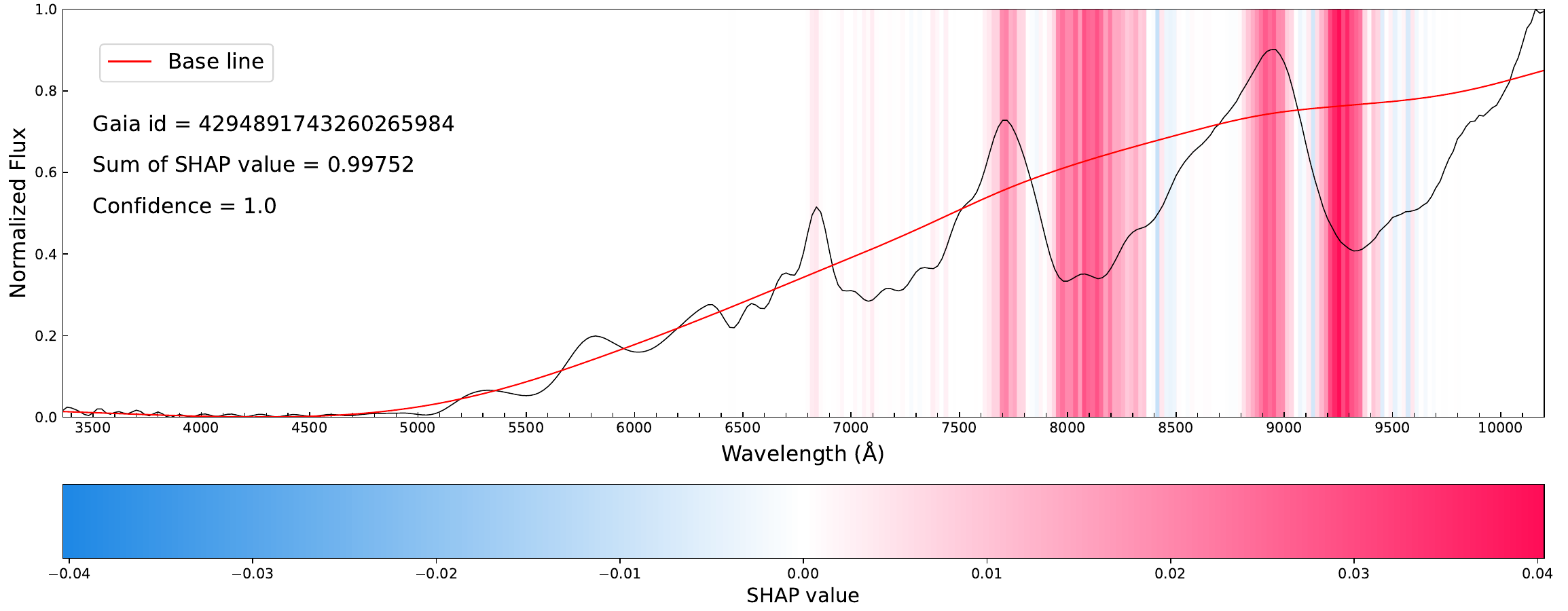}
            \label{fig:carbon_star}
          }
          \hfill
          \subfigure[Non-carbon star]{
            \includegraphics[width=0.95\textwidth]{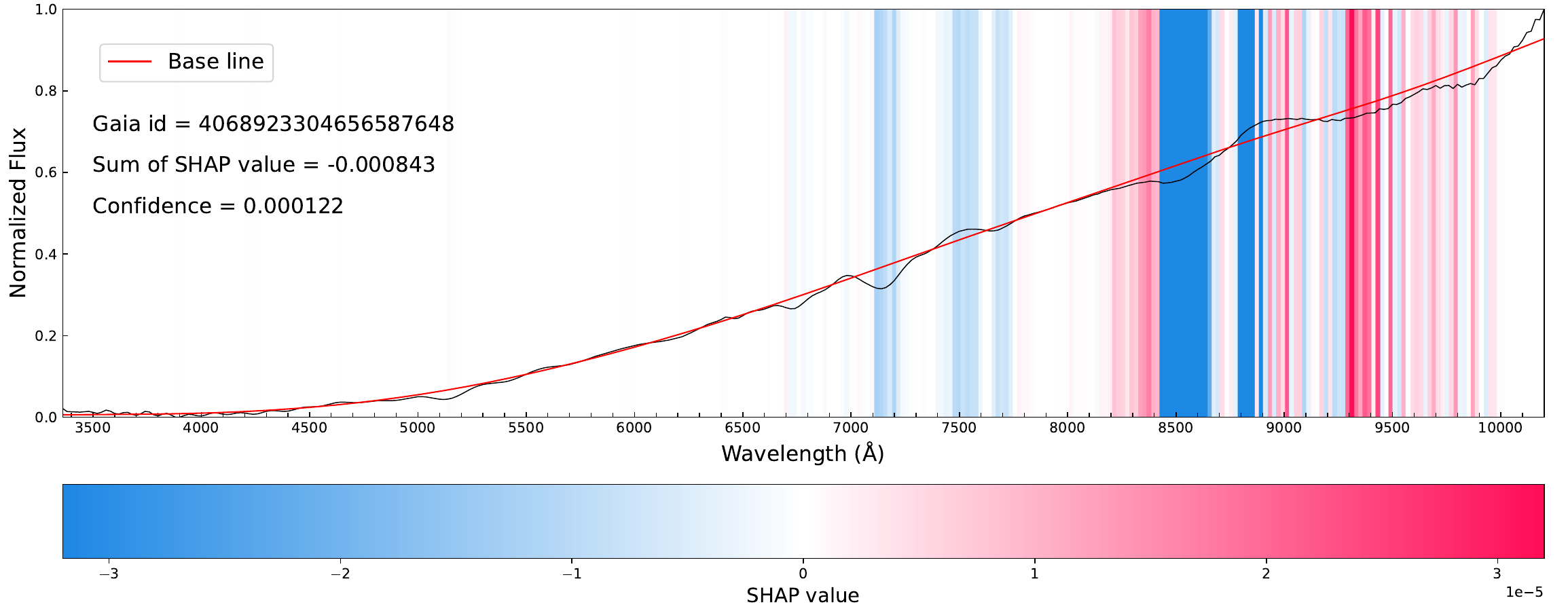}
            \label{fig:non_carbon_star}
          }
          \caption{The baseline is a background pseudo-continuum used for computing SHAP values in order to better highlight spectral features. The color represents the size of the SHAP value, and a redder feature indicates its stronger positive contribution of the SHAP model's interpretation of the carbon stars. The bluer the color, the more strongly the feature contributes negatively to the model interpretation of the carbon star. The sum of the SHAP values can be considered as the approximation of the model output, and the larger the sum of the SHAP values, the more inclined to consider the spectrum as a carbon star.}
          \label{fig:spectra}
    \end{figure*}

    To better illustrate the differences in key feature contributions within a single spectrum, the background spectrum for calculating SHAP values can be chosen as the smooth pseudo-continuum after polynomial fitting using the approach of \citet{2021ApJS..256...14Z}. It serves as a baseline and reference for individually calculating the SHAP value of each feature in each spectrum. We believe that contrasting with such a baseline can effectively highlight the trough and peak features in each spectrum. Subsequently, we can generate an intuitive and clear feature distribution heatmap for given spectrum shown in Figure \ref{fig:spectra}, which facilitates the easy location of the key spectral features concerned by our model.
    
\begin{figure*}[htbp]
          \centering
          \subfigure{
            \includegraphics[width=0.95\textwidth]{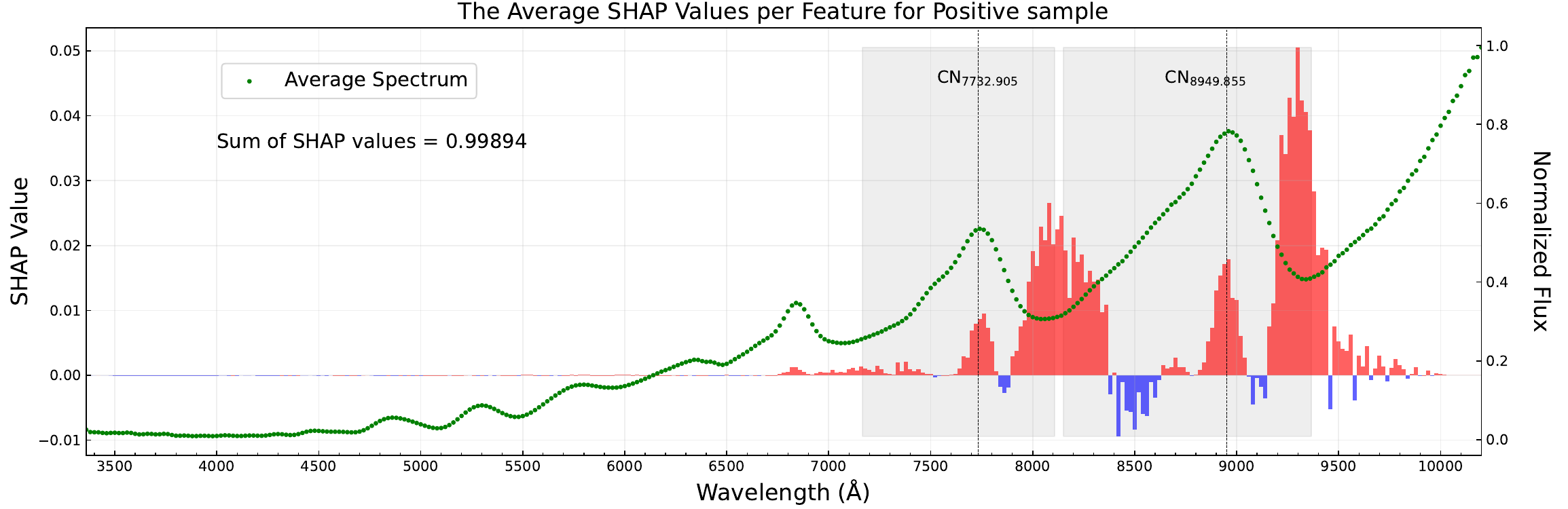}
            \label{fig:car_sum}
          }
          \hfill
          \subfigure{
            \includegraphics[width=0.95\textwidth]{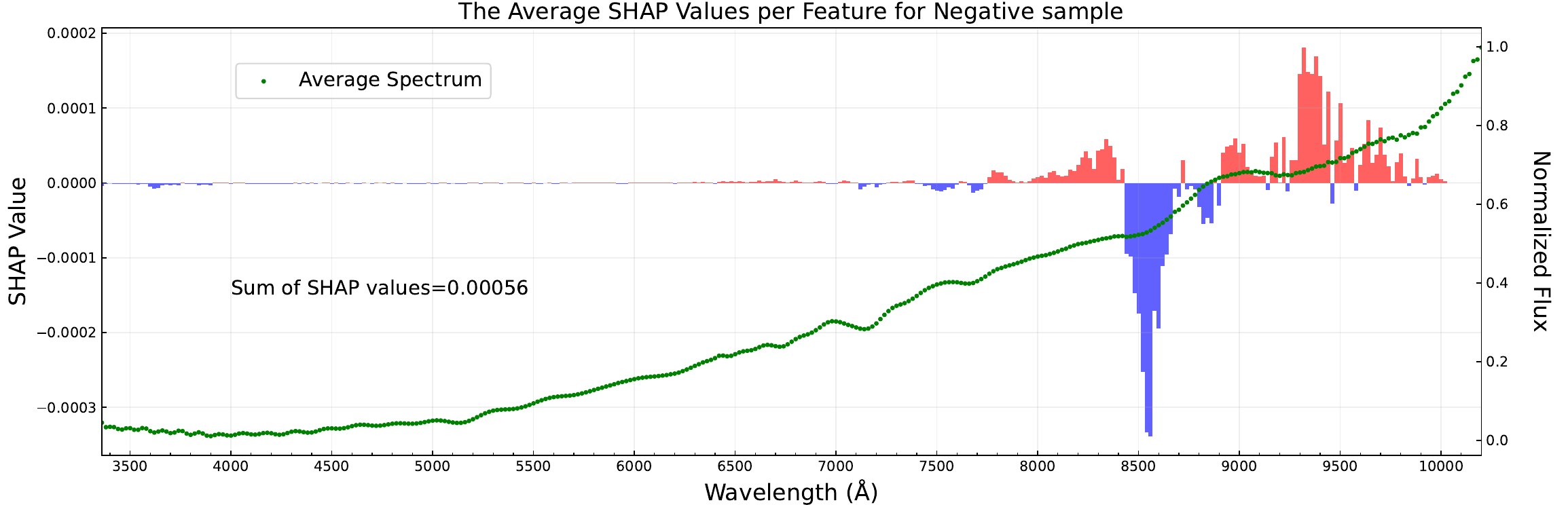}
            \label{fig:neg_sum}
          }
          \caption{The average SHAP values per feature is shown using a wavelength bin size of 20\,\AA. Red and blue represent  positive and negative SHAP values respectively. As a contrast, the average spectra of positive and negative samples are overlaid as green scatter points in the graph. In the upper panel, the two prominent CN bands are highlighted with gray shading. The dashed lines mark the positions of the tops of the two CN molecular band heads.}
          \label{fig:sum_shap_values}
    \end{figure*}

    However, if we want to get a more statistically signigicant and reach a quantified conclusion, we need statistical information on the SHAP values of a large enough sample. For this purpose, we calculate the average SHAP values of each feature for all positive and negative samples. As shown in Figure \ref{fig:sum_shap_values},  we can see that the interpretation ranges of positive features for the positive sample are consistent with the CN features. It is mainly distributed in the ranges 9140-9440\,\AA, 7900-8360\,\AA, 8820-9060\,\AA, and 7600-7820\,\AA, showing that relatively large SHAP values were assigned to both peak and trough ranges in the CN molecular bands. However, compared with the positive carbon star sample, the average SHAP values in these regions of the negative sample are much smaller. In addition, the negative sample showed a very strong negative contribution in the range 8420-8900\,\AA, which is the most significant distinguishing feature of this sample. We note that the average SHAP values of the positive sample is also negative in the 8420-8620\,\AA\ region. We suspect that this is related to the absorption feature of Ca\,II (8484-8662\,\AA), which is considered by the model to be a negative feature. We can see that the location of the model's main concern is the strong CN molecular band feature at the red end of the spectrum, which is consistent with the actual interpretation in Section \ref{sec:interpretation}. Next, we will showcase these key features.
    
    The sum of SHAP values in Figure \ref{fig:sum_shap_values} represents the sum of the average SHAP values across the feature dimensions of positive or negative sample. For the positive sample, the sum of SHAP values is 0.99894. We summarize the main feature regions of the golden sample of carbon stars, which will serve as the basis for our subsequent identification work. We calculate the sum of the SHAP values over 5 main intervals listed in Table \ref{tab:key characteristic}\label{sec:key feature}.
    \begin{table}[ht]
        \centering
        \caption{The wavelength range and SHAP value score of the 5 most important features from positive sample.}
        \vspace{1.5pt}
        \begin{tabular}{llcc} 
            \hline \text {Number} &   \text {Feature area range } & \text {Sum of SHAP values} \\
            \hline \text Feature 1   &   { 9140-9560\,\AA\ } &   0.46825 \\
            \text Feature 2  &   { 7900-8360\,\AA\ } &   0.37919 \\
            \text Feature 3  &   { 8820-9060\,\AA\ } &  0.11090 \\
            \text Feature 4  &   { 7600-7820\,\AA\ } &   0.05297 \\
            \text Feature 5  &   { 6780-7460\,\AA\ } &   0.02585 \\
            \hline
        \end{tabular}
        \label{tab:key characteristic}
    \end{table}
    Among them, the range 6780-7460\,\AA\ includes the peak characteristic of 6780-6920\,\AA\ and the absorption trough characteristic of 7000-7460\,\AA.

    The reason our model only identifies the prominent CN features in the red end of the positive sample spectra, while almost failing to detect the C$_{\text{2}}$ features in the blue end, can be attributed to two factors. Firstly, the flux values in the blue end are too small relative to the entire spectrum, making the C$_{\text{2}}$ features much less significant compared to the prominent CN features. Secondly, C2023 filtered golden sample carbon stars in relatively cold (most of them have $G_\mathrm{BP}$ - $G_\mathrm{RP}$ > 2; \citeauthor{2023A&A...674A..39G} \citeyear{2023A&A...674A..39G}) CSTAR sample based on the prominent CN features rather than the much weaker C$_{\text{2}}$ features. This results in more than half of the targets in the golden sample of carbon stars having almost no C$_{\text{2}}$ features. 

\subsection{451 new carbon star candidates}
\begin{figure}[h]
        \centering
        \includegraphics[width=0.5\textwidth]{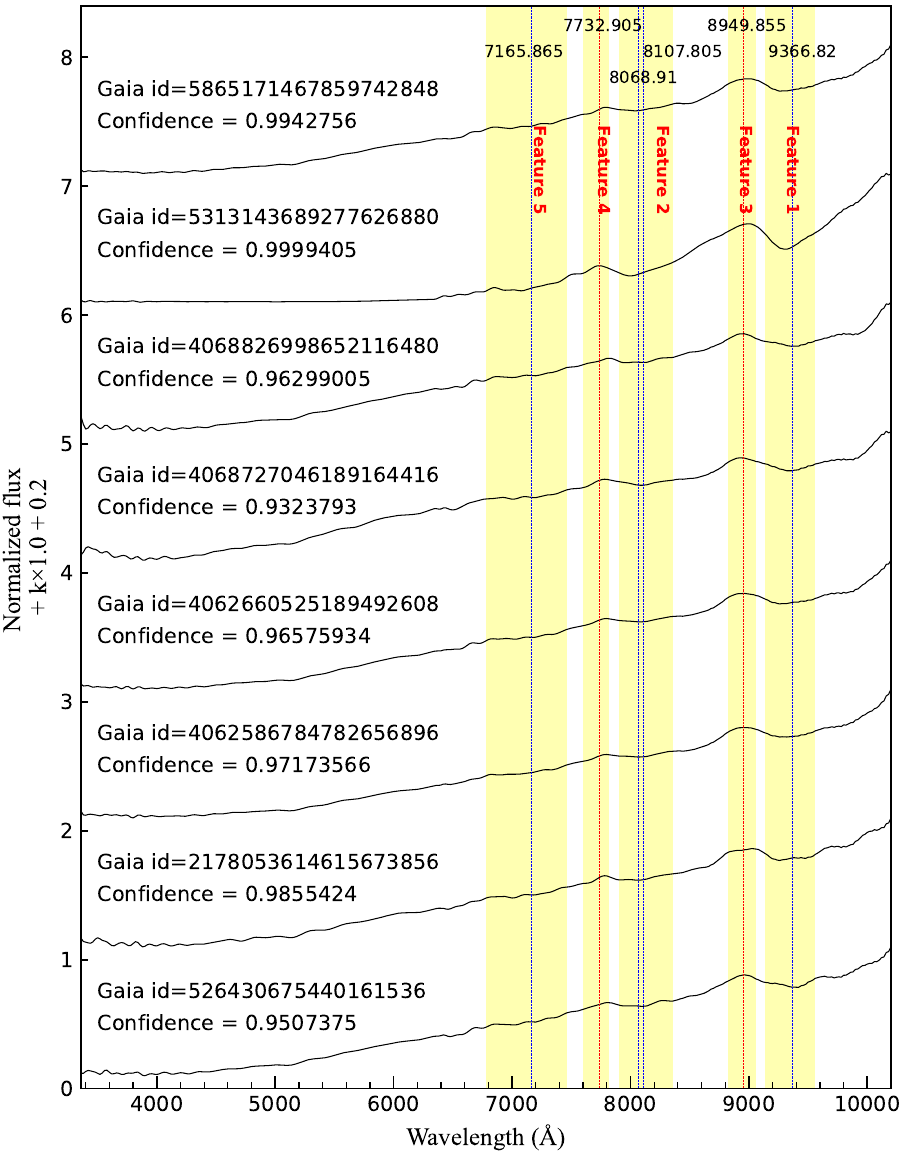}
        \caption{Eight spectra were selected from the candidates. The spectra are normalized similarly to those in Figure \ref{fig:spectra2}. The yellow areas are the key features we summarized. The blue dashed line marks the position of the CN characteristic troughs and the red marks the position of the peaks.}
        \label{fig:spectra3}
    \end{figure}

\begin{figure}[ht]
      \centering
      \begin{minipage}{0.2445\textwidth}
          \centering
          \includegraphics[width=\linewidth]{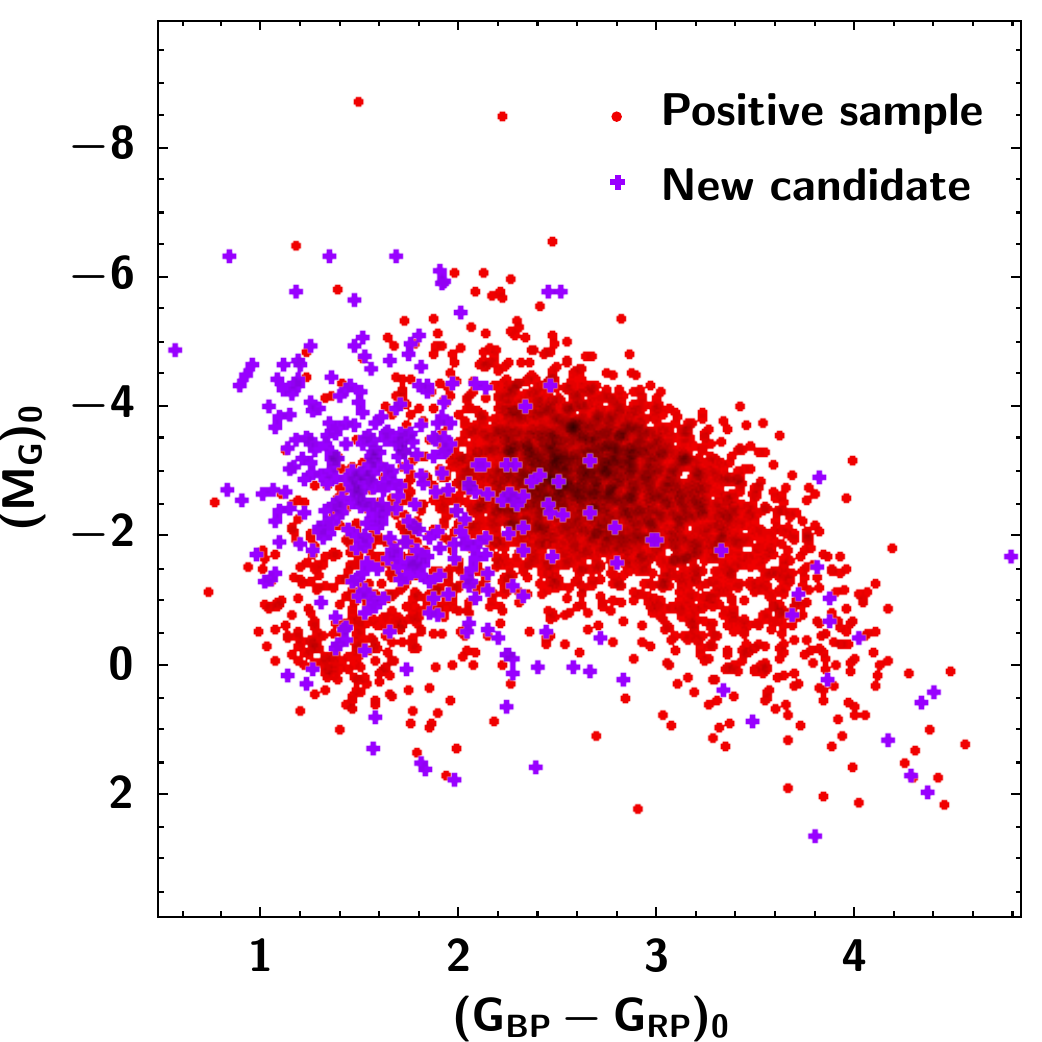} 
      \end{minipage}%
      \hfill
      \begin{minipage}{0.2445\textwidth}
          \centering
          \includegraphics[width=\linewidth]{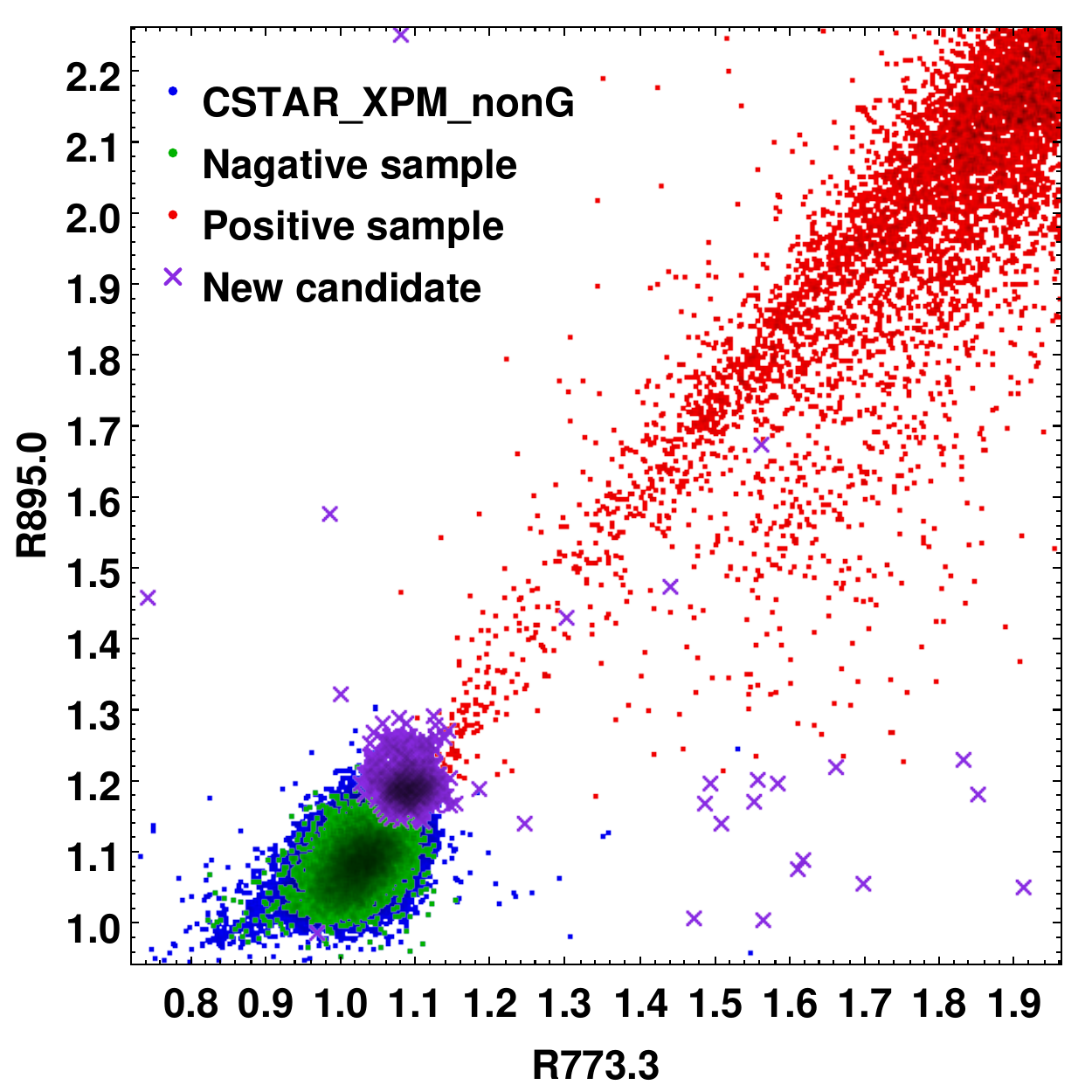} 
      \end{minipage}%
      \caption{The color-magnitude diagram (dereddened) of the new candidates is shown on the left, and the molecular band head strength of the new candidates is shown on the right.}
      \label{fig:new_candidates}
    \end{figure}

    We combined the model training and validation accuracy, as well as the changes in FP and FN in Section \ref{sec:valid}, and ultimately determined the training epoch number of the model as 16. We applied the model trained based on the dataset provided in Section \ref{sec:dataset} to the 74,740 stars of CSTAR\_XPM\_nonG, and 451 new carbon star candidates were obtained. The spectra of these stars exhibit relatively prominent CN molecular bands compared to typical giant stars but are weaker than most golden sample stars. Their spectra show reduced flux at the blue end and hardly any significant C$_{2}$ features are visible. In Figure \ref{fig:spectra3} we show some spectra of these stars.
    The color-magnitude diagram and distribution position of the CN band head strengths are shown in Figure \ref{fig:new_candidates}.
    The color-magnitude map shows that our new carbon star candidates occupy the location of the carbon star golden sample but mainly concentrate in the hotter regions. From the strength map, our new carbon star candidates mainly occupy the position of weak CN band carbon stars mentioned in C2023.

\begin{figure}[ht]
        \centering
        \includegraphics[width=0.5\textwidth]{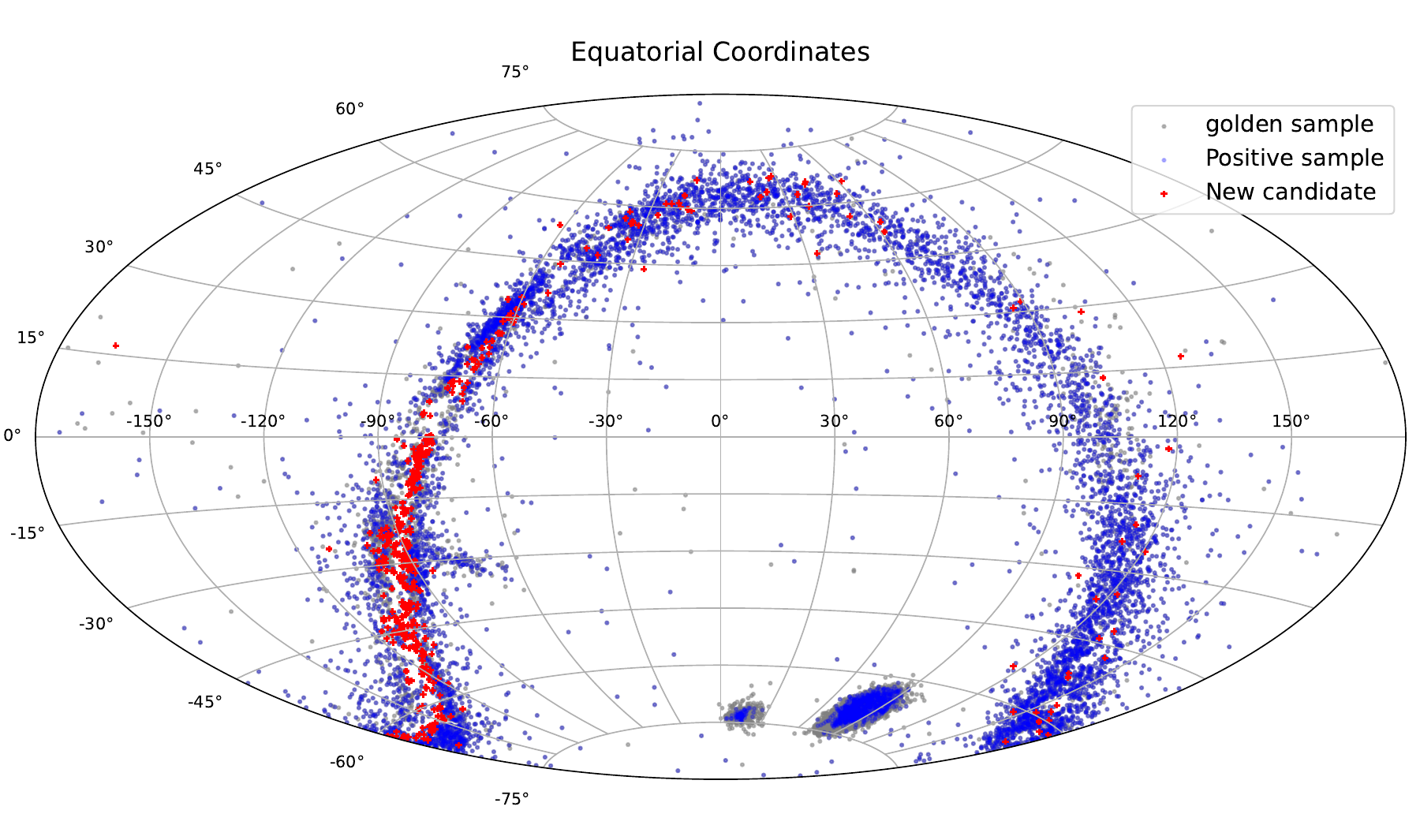}

        \vspace{1em}

        \includegraphics[width=0.5\textwidth]{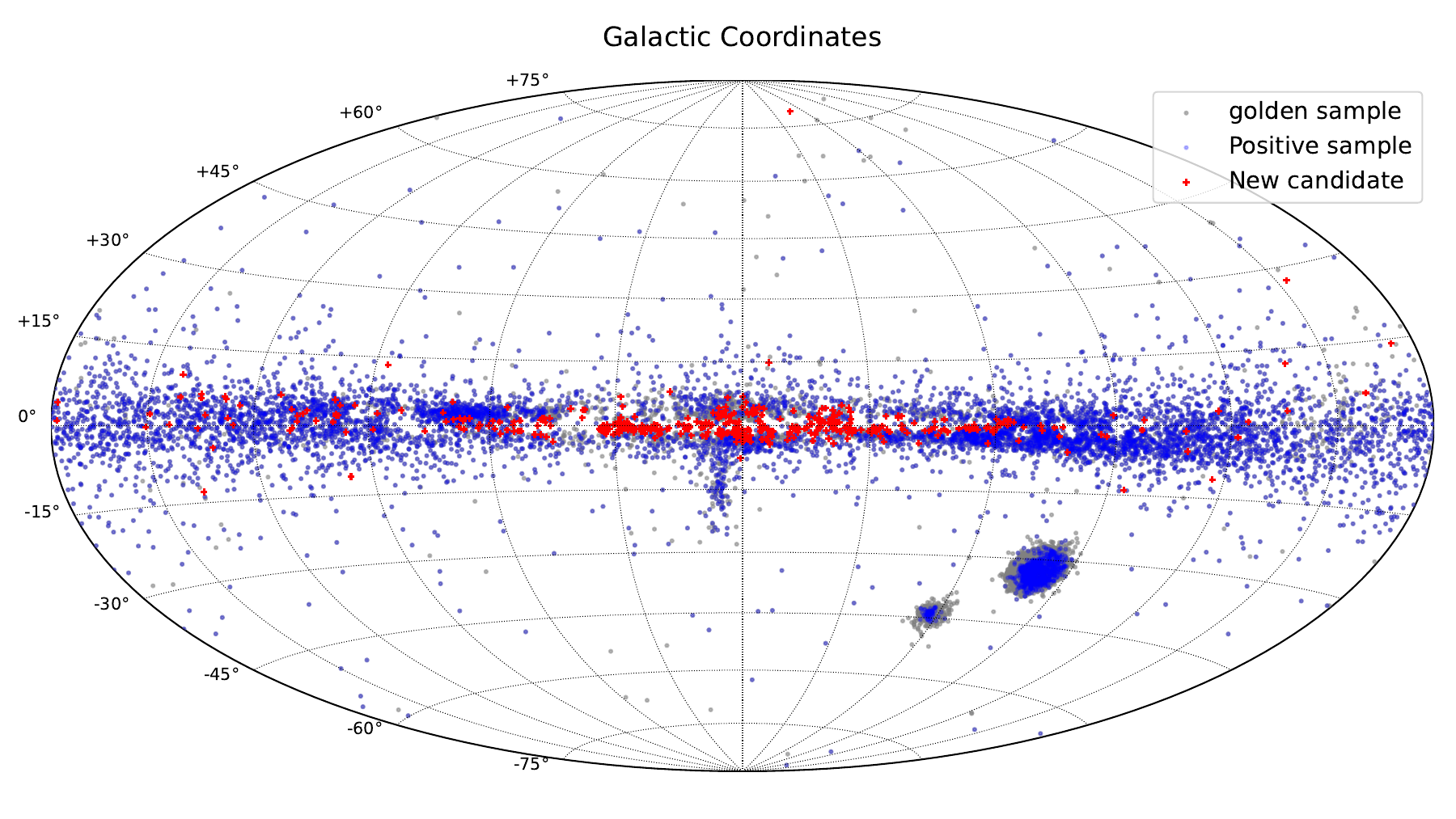}
        
        \caption{Spatial distributions of our results are drawn in equatorial and Galactic coordinates. The gray spots denote the 15,400 golden sample of carbon stars. The blue spots represent the 8176 positive sample from CSTAR\_XPM\_G. The red spots are the 451 new carbon star candidates from CSTAR\_XPM\_nonG.}
        \label{fig:coordinates}
    \end{figure}

    The spatial distribution of the candidates is shown in Figure \ref{fig:coordinates}. Since the positive sample comes from the golden sample of carbon stars, they are distributed throughout the Galactic disc, the LMC, and the SMC. However, these new candidates are primarily distributed in the dense Galactic inner disc, an area characterized by substantial amounts of interstellar dust that contribute to increased extinction. Consequently, the observed degradation of their spectral shape, particularly at the blue end, is likely caused by a lower S/N or observation in crowded and highly reddened sky regions \citep{2023A&A...674A..15L}. This also explains why many of them became hotter ($(G_{\text{BP}}-G_{\text{RP}})_{0}$ < 2) after 3D extinction correction \citep{2019ApJ...887...93G}. If their color index estimates are correct, similar to the study of C-rich stars in the Milky Way’s bar-bulge \citep{2023MNRAS.521.2745S}, these stars are likely to be the result of binary evolution. However, the possibility of single star evolution cannot be ruled out due to the possible extinction errors.
    
    After cross-matching our 451 new carbon star candidates with LAMOST DR10, four common sources were obtained, two of which were labeled as carbon stars by LAMOST's pipeline. Cross-matching with SIMBAD yielded 73 common sources, 28 of which were labeled as C* in the main\_type or other\_types attribute. Detailed results are provided in Table \ref{star_table}.
    
\section{Discussion and Analysis}\label{sec:analysis}
    Following \citet{1962ApJS....7..165W}, we plot all the CN absorption line positions  as dotted lines at the red end of the spectrum in Figures \ref{fig:conv_sp} and \ref{fig:explanations}.

\subsection{Cross-match with LAMOST DR10}
    We cross-matched the RA and Dec of all the 83,028 CSTAR\_XPM sample stars with the catalog of low-resolution spectra from LAMOST DR10,  to obtain the subclass types given by the LAMOST's pipeline \citep{2014AJ....147..101W}. There are 442 sources labeled as carbon stars, and 1131 are not. Among the sources not labeled as carbon stars, there are 351 G-type stars (31\%), 232 K-type stars (20.5\%), and 509 M-type stars (45\%), the vast majority of them come from CSTAR\_XPM\_nonG\label{sec:cross}. 

\begin{figure*}[h]
        \centering
        \includegraphics[width=1.0\textwidth]{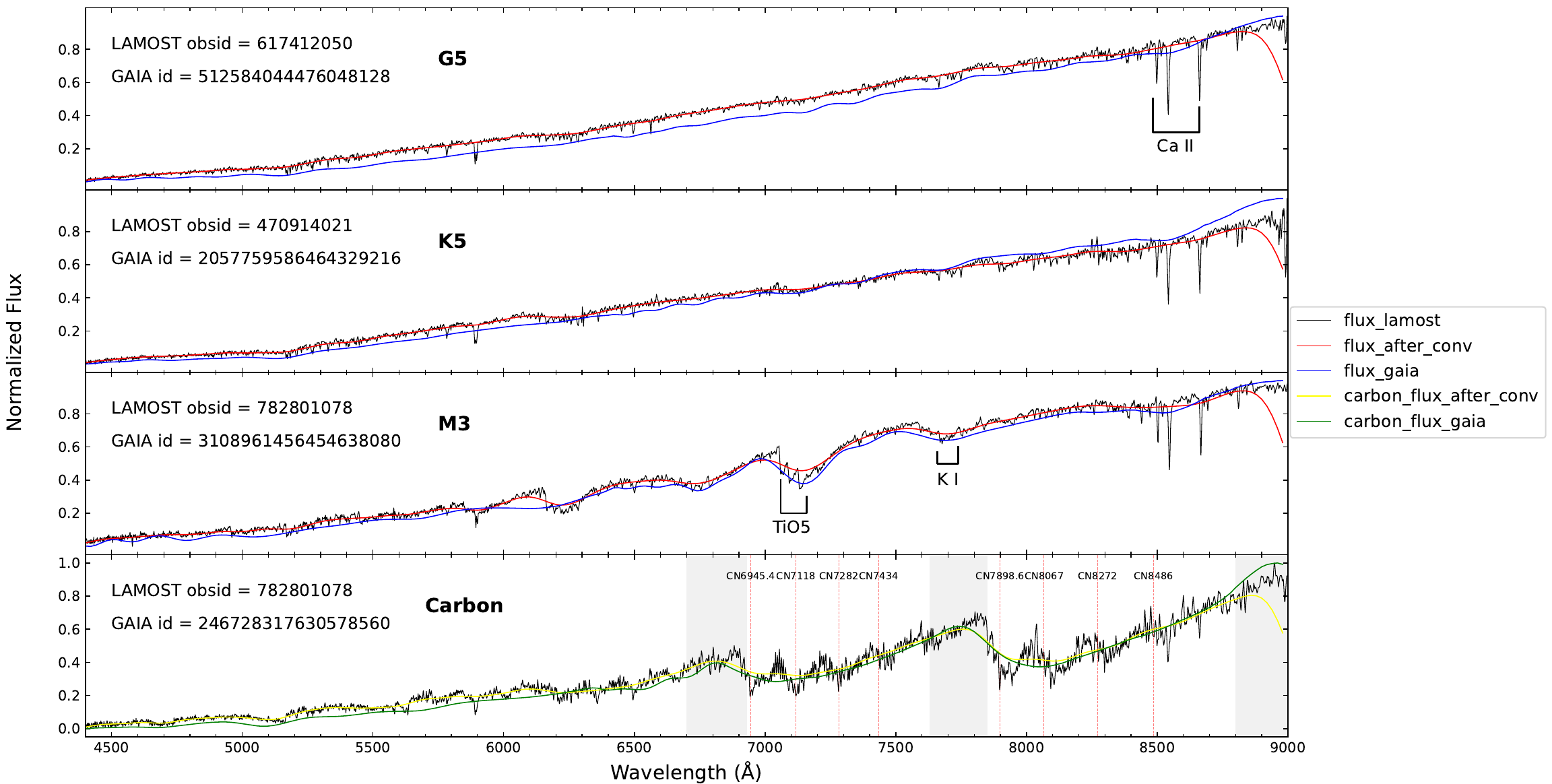}
        \caption{The spectra with wavelength range intercept to 4400-9000\,\AA. The LAMOST spectra of some common sources are plotted with black lines. The red and yellow lines are the result of their convolution to 50 resolution. The blue and green lines are the $Gaia$ spectra. In the bottom panel, the light grey shaded areas pick out the peaks and the red dashed lines in the bottom panels mark several obvious CN absorption line locations.}
        \label{fig:conv_sp}
    \end{figure*}

    To better understand the formation of characteristic troughs in the $Gaia$ XP spectra of carbon stars, the LAMOST spectrum of the same source was drawn in Figure \ref{fig:conv_sp} to determine the CN absorption position. Using the fast Fourier transform and interpolation method of \citet{2021ApJS..256...14Z}, the LAMOST spectrum was convolved to the resolution of the approximate $Gaia$ XP spectra to show the difference in the absorption region under different resolutions. We also plotted the same figures for common sources of G, K, and M type stars as a comparison to illustrate the areas of feature differentiation. 
    
\begin{figure}[htbp]
          \centering
          \includegraphics[width=0.5\textwidth]{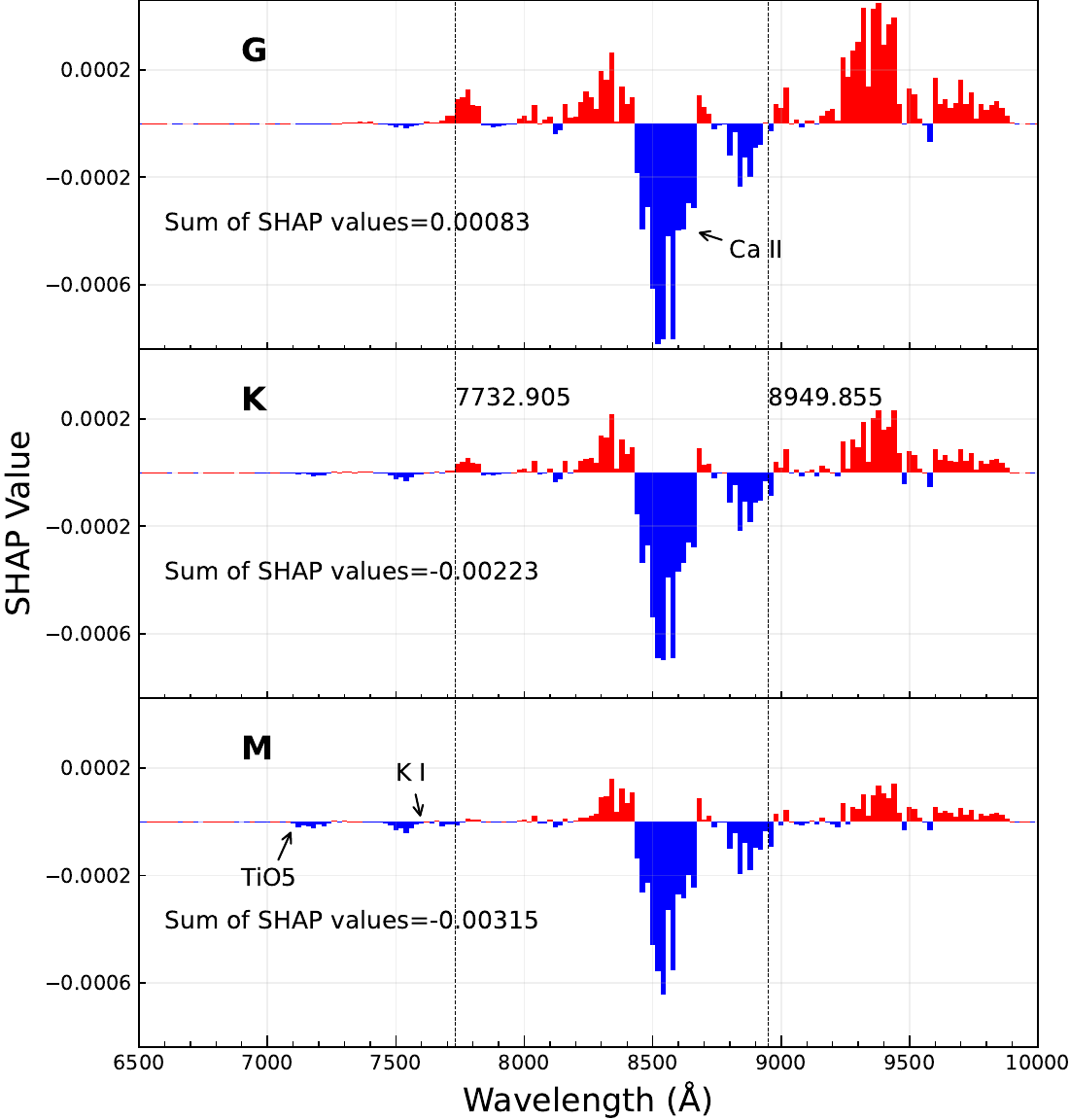}
          \caption{The wavelength range is intercepted from 6500 to 10000\,\AA. The average distribution of SHAP values per features for these three types was plotted respectively.}
          \label{fig:gkm_shap}
    \end{figure}

    We found that the most significant feature shared by the negative sample of G, K, and M type stars compared with the positive sample of carbon stars is the Ca\,II line absorption feature in the range of 8484-8662\,\AA\ \citep{2023ApJS..266....4L}, which can also be identified in the $Gaia$ XP spectra. When the temperature is low, near the M type, the prominent TiO5 (7126-7135\,\AA; \citeauthor{2023ApJS..266....4L} \citeyear{2023ApJS..266....4L}) and K\,I (7664.9-7699.0\,\AA; \citeauthor{lepine2003spectroscopy} \citeyear{lepine2003spectroscopy}) absorption line features are prominent and also visible in the LAMOST and $Gaia$ XP spectra after resolution reduction. During the change of spectral type from M to G, the absorption characteristics of TiO5 and K\,I gradually weaken in the spectra until they almost disappear. This is consistent with the Figure \ref{fig:gkm_shap} interpreted by our model. However, carbon stars exhibit broader and deeper CN absorption features at 6900-7300\,\AA, and 7900-8300\,\AA, and a set of three peak tops at 6830\,\AA, 7780\,\AA, and 9000 \,\AA\ \citep{2023MNRAS.521.2745S}. While M-type non-carbon stars also display peak structures, such as the peak tops at 7050\,\AA, and 7600\,\AA, however their positions are entirely different. Therefore, both the characteristic peaks and troughs in the spectra of carbon stars are crucial distinguishing features from other types of stars.
    
    From Figure \ref{fig:gkm_shap}, we can also see that the absorption characteristics near 8300\,\AA\ and 9400\,\AA\ are gradually enhanced and the sum of SHAP values increases with temperature increase. Therefore, we  hypothesize that in the CSTAR\_XPM\_nonG, G, and K type stars are more likely to be carbon star candidates. After a visual inspection of the 1131 stars, we estimate that G stars may contain 10\% candidates and K type stars may contain 2.6\% candidates. The reason is that if a low-temperature star has a high C abundance, its red-end molecular band characteristics will be more obvious. It will be easier to select by the method of C2023. On the contrary, if at high-temperature conditions, the carbon molecular band characteristics would be relatively weak. In other words, if a source can still show the carbon characteristics at high temperatures, its C abundance can be considerable.

\subsection{Comparison with the results of Sanders \& Matsunaga (2023)} 
    \citet{2023A&A...674A..15L} released a candidate catalog of 1,720,588  long-period variables (LPVs). After cross-matching with $Gaia$ DR3 data, we found that there are 13,513 stars in the golden sample of carbon stars and only 1289 stars in CSTAR\_XPM\_nonG are classified as LPVs. Based on the UMAP unsupervised algorithm, \citet{2023MNRAS.521.2745S} have presented 23,737 C-rich candidates from the LPV candidate catalog above. There are 40 of the above 1289 LPVs in the list of our 451 new carbon star candidates, and 26 are also identified as C-rich stars by \citet{2023MNRAS.521.2745S}. There are 20 common objects between our above 40 carbon candidates and 26 Carbon stars of \citet{2023MNRAS.521.2745S}, where six objects are not included in our results. We show the heatmaps of these six object in Figure \ref{fig:error_6}.
    
   We can see that Gaia 963568667249950592 (the first subplot of Figure \ref{fig:error_6}) is an obvious problematic spectrum, our model also assigns it a small confidence level. The spectra in subplots 2-6 of Figure \ref{fig:error_6} have relatively weak absorption features in the region around 8000\,\AA\ and 9300\,\AA, i.e., Feature 1 and 2 in Table \ref{tab:key characteristic}. Among these objects, Gaia 3586939158411536768 and 4056303080211083392 lack distinct CN molecular absorption band, then our model gives them lower confidence. Gaia 4314316918298388992 is weak in all Features 1-5, so the confidence level is also low. Gaia 4339500235641264256 has a stronger Feature 2, but a weaker Feature 1, resulting in a confidence level not exceeding 0.5. The molecular absorption bands of Gaia 5546572717982752640 are not prominent enough, so the confidence level is low.
  
   The majority of the golden sample carbon stars are TP-AGB stars, typical of long-period variable stars, so it is not surprising that they are overwhelmingly classified as LPVs by \citet{2023MNRAS.521.2745S}. However, the proportion of CSTAR\_XPM\_nonG stars labeled as LPV candidates is quite small. This may be attributed, to their low S/N ratio, a relative lack of data points in the $Gaia$ $G$ time series, small variability amplitudes, or short periods. This would lead to their non-detection as period signals by the SOS module of \citep{2023A&A...674A..15L}. On the other hand, they have not yet evolved to the TP-AGB phase with periodic variability, which is consistent with their location site on the HR diagram (Figure \ref{fig:BPRP_MG}). Regarding why only a few dozen of the 1289 LPVs in the CSTAR\_XPM\_nonG are classified as C-rich stars, this is likely due to selection effects. In CSTAR\_XPM, the majority of C-rich LPVs were selected by the algorithm of C2023 and included in the golden sample, resulting in the remaining 1289 stars being predominantly O-rich LPVs, as well as some C-rich LPVs with weak carbon spectral features that were not detected by the algorithm. Therefore, mining possible carbon stars from CSTAR\_XPM\_nonG is not a simple task. Our candidates contain the vast majority of C-rich stars identified by \citet{2023MNRAS.521.2745S}, which demonstrates the high completeness of our method. The other six stars are significantly different from the golden sample, but there is no denying that they still are possible carbon stars. Our model scores them with a not-low confidence level, just not above the 0.5 threshold (see Section \ref{sec:confidence}). Although the positive sample our model learned is from the golden sample of carbon stars with strong molecular bands, it is still able to identify quite a few carbon stars exhibiting relatively weak CN molecular band features. This demonstrates the good generalization ability of our model and encourages future systematic exploration of carbon stars across all of $Gaia$'s XP spectra.
    
\subsection{Explanation of key features}
\begin{figure*}[h]
        \centering
        \includegraphics[width=1.0\textwidth]{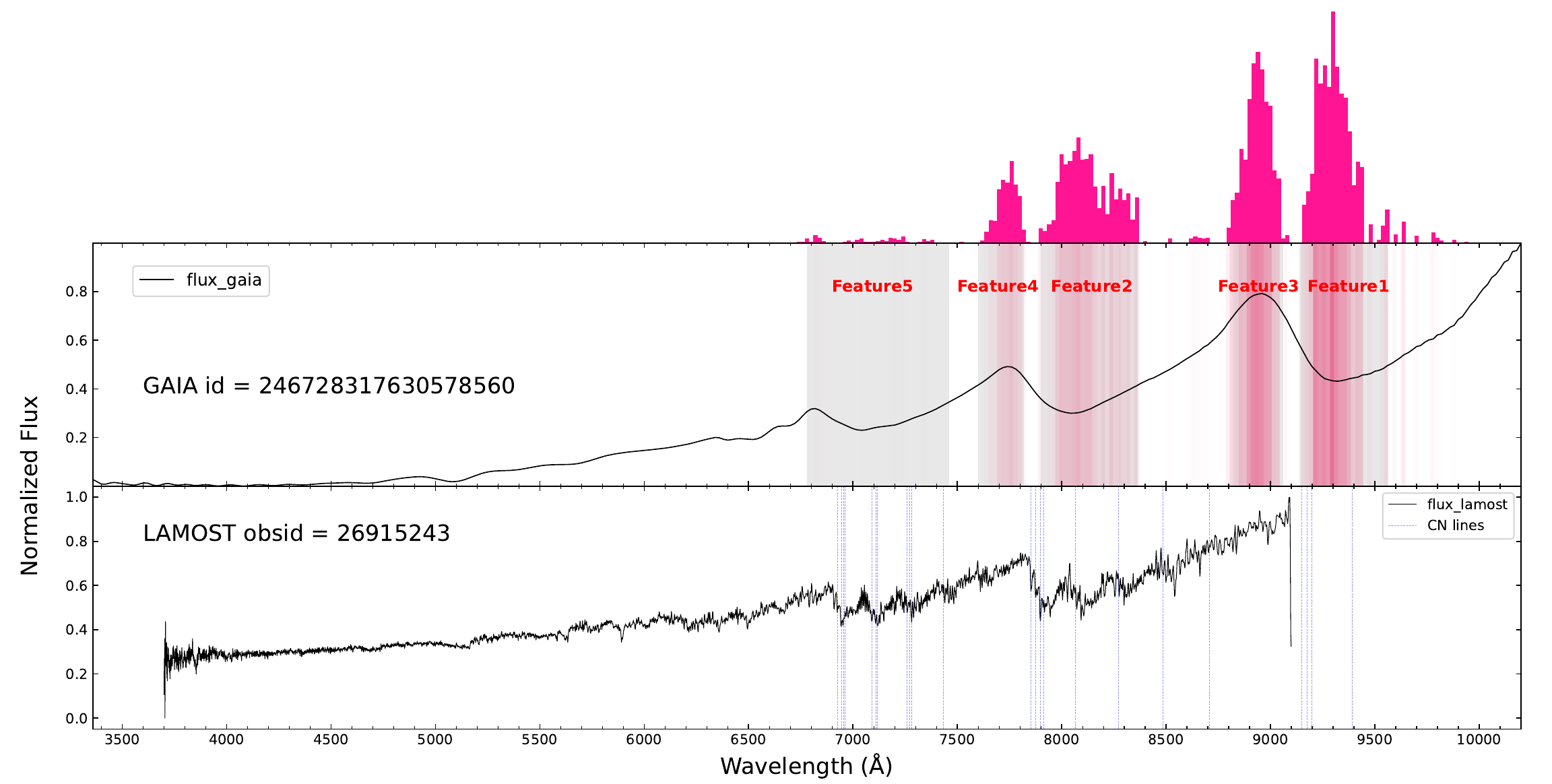}
        \caption{A Gaia/LAMOST comparison example where the model explains the distribution of the most critical features (light gray areas) in the Gaia spectrum. The red gradient against a light grey background and the deep pink energy bar at the upper panel indicate the intensity of the feature SHAP values. The corresponding CN molecular absorption line positions (light blue dotted lines) are plotted in the LAMOST spectrum.}
        \label{fig:explanations}
    \end{figure*}

    The position of the CN absorption lines with a wavelength greater than 6900\,\AA\ in the literature \citep{1962ApJS....7..165W} are marked in the Figure \ref{fig:explanations}. 
    
    From the figure, we can see that there are some differences in the shape of the molecular absorption band regions in the spectrum at different resolutions. The continuous and dense CN absorption line region in the LAMOST spectrum shows the form of wide and deep CN characteristic troughs in the Gaia spectrum, such as Feature 5 and Feature 2. We posit that this is the combined absorption effect of superimposed multiple molecular absorption lines in the spectrum at $Gaia$ spectral resolution, we call it the "mixed-molecular-absorption-band" in the Gaia spectrum\label{sec:interpretation}. It can be seen that Feature 1 and Feature 2 are consistent with the dense CN absorption line regions except for the small SHAP value assigned by the absorption region of Feature 5. which may be caused by the weaker signal flux values in the spectra compared to Feature 1 and Feature 2.
    
    As for the peaks observed in the spectra, they are a reflection of the relative decrease of opacity. When the molecular absorption on both sides of the peak becomes stronger, it will increase opacity and thus form the characteristic trough. Consequently, the region with less opacity stands out as a peak in the spectrum. Hence, the characteristic peaks of Feature 3 and Feature 4 in the shadow region are are also meaningful for distinguishing the positive and negative samples.
    
\subsection{Reliability analysis of new candidates}
    To verify the reliability of the new candidates, we mainly analyzed the relationship between the molecular band head strengths (R773.3 and R895.0) and the confidence levels, as well as the relationship between the strengths and the SHAP values. We also calculated the average sum of SHAP values for the new candidates.
    
    \begin{figure}[ht]
        \centering
        \begin{minipage}{0.15\textwidth}
            \centering
            \includegraphics[width=\linewidth]{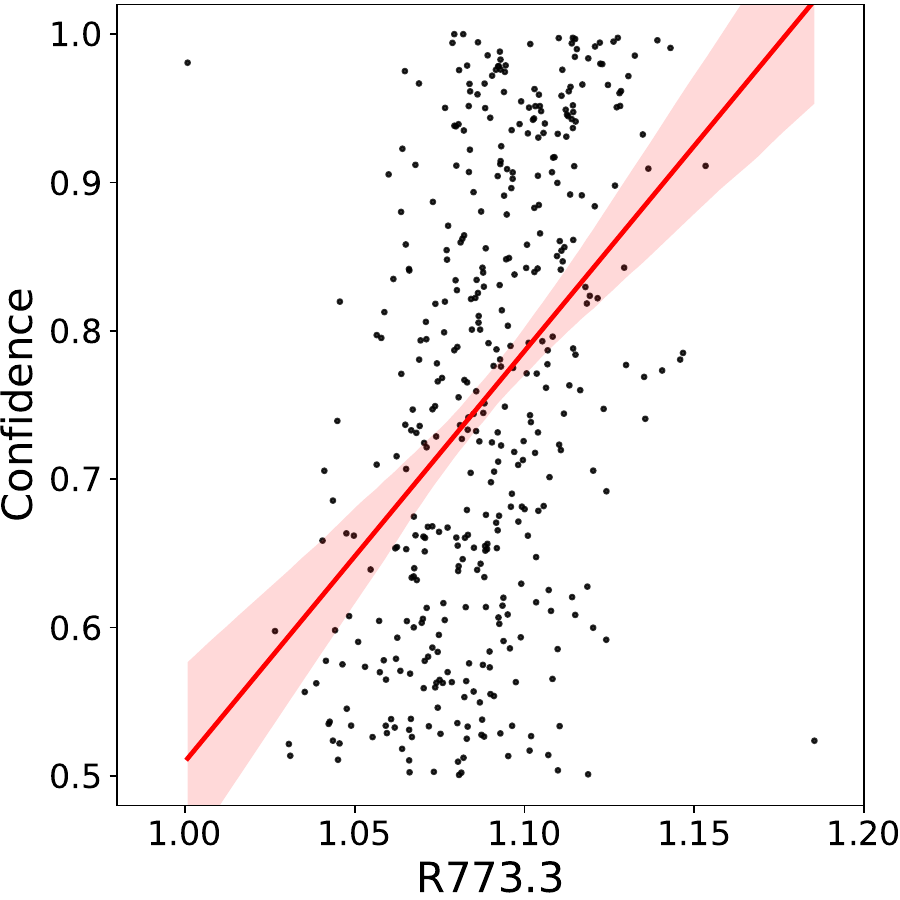}
        \end{minipage}%
        \hfill
        \begin{minipage}{0.15\textwidth}
            \centering
            \includegraphics[width=\linewidth]{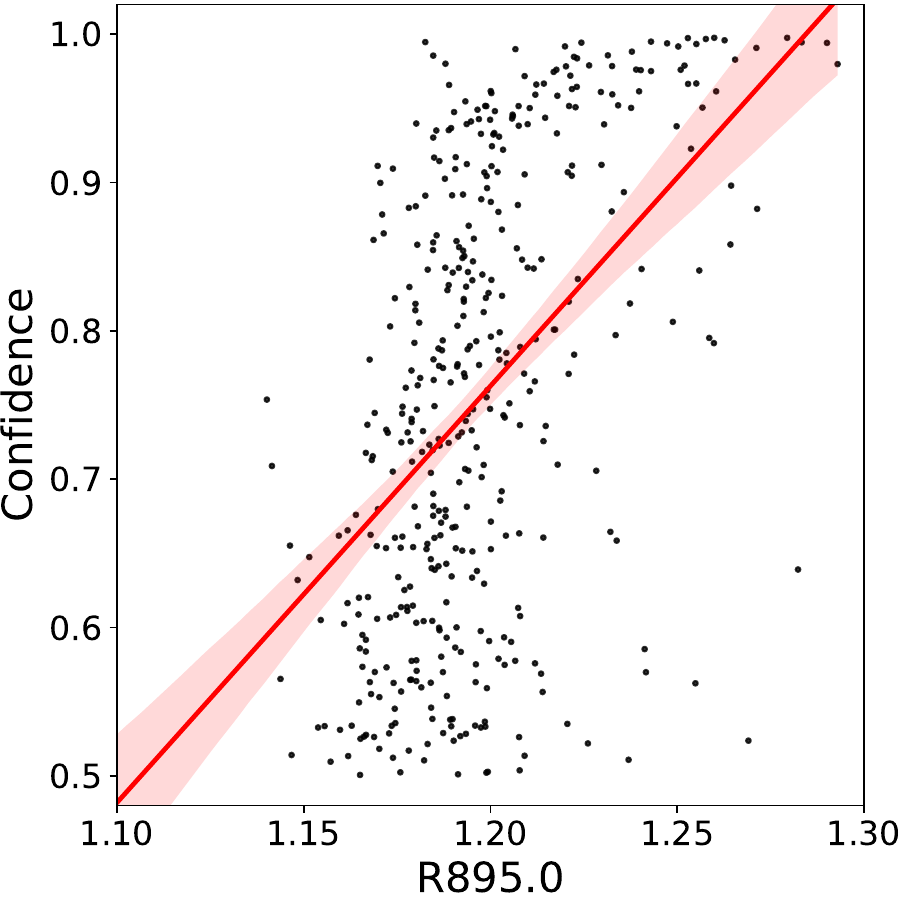}
        \end{minipage}%
        \hfill
        \begin{minipage}{0.185\textwidth}
            \centering
            \includegraphics[width=\linewidth]{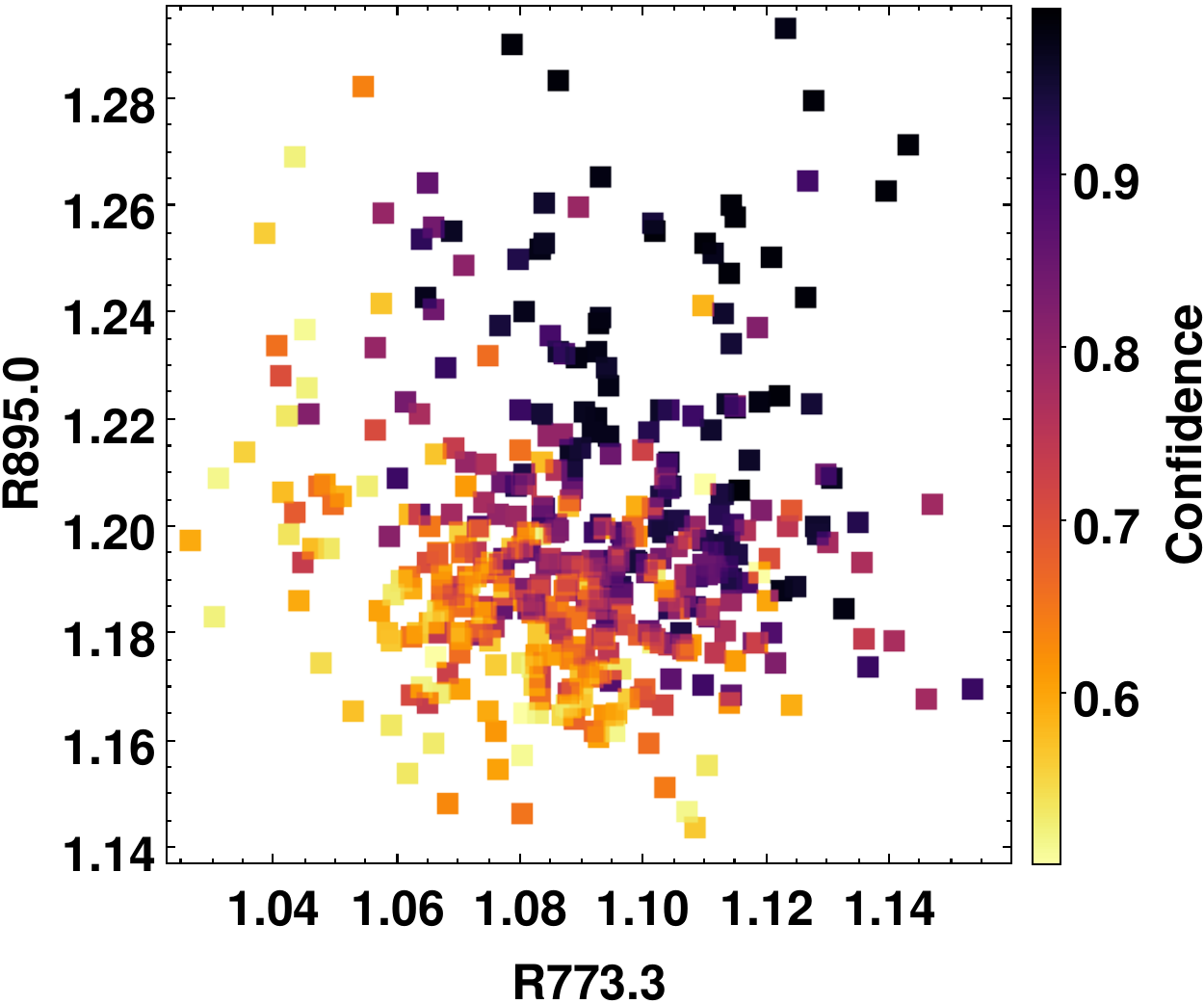}
        \end{minipage}%

        \vspace{1em}
        
        \begin{minipage}{0.15\textwidth}
            \centering
            \includegraphics[width=\linewidth]{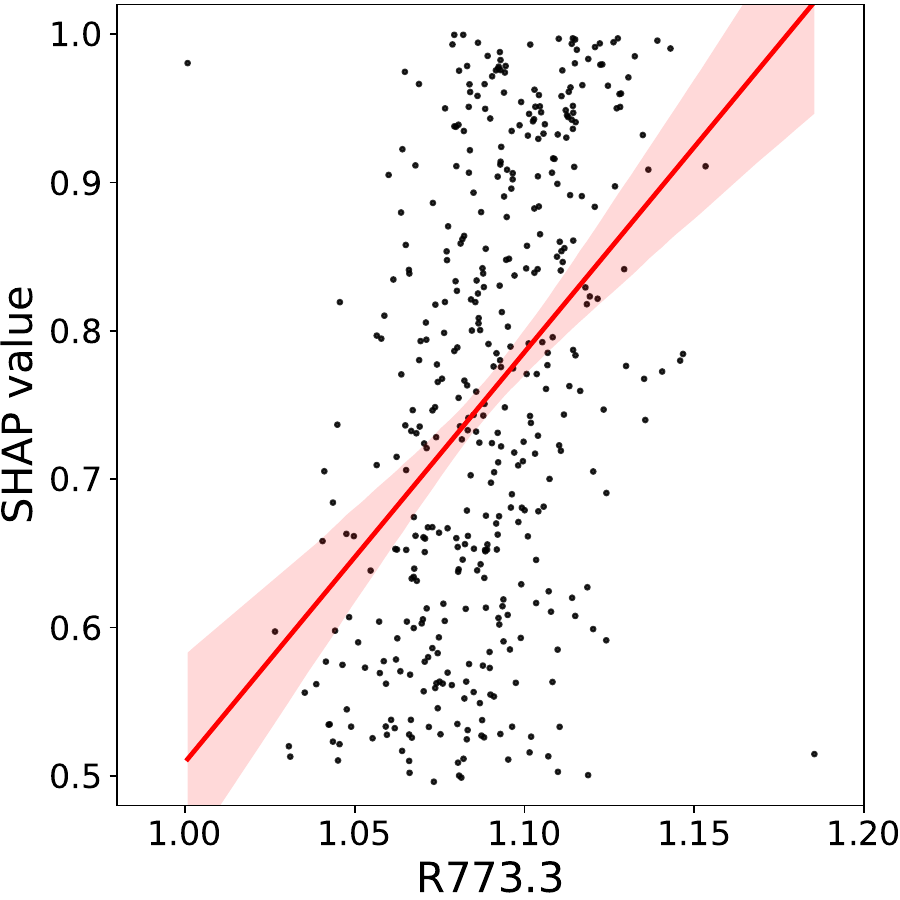}
        \end{minipage}%
        \hfill
        \begin{minipage}{0.15\textwidth}
            \centering
            \includegraphics[width=\linewidth]{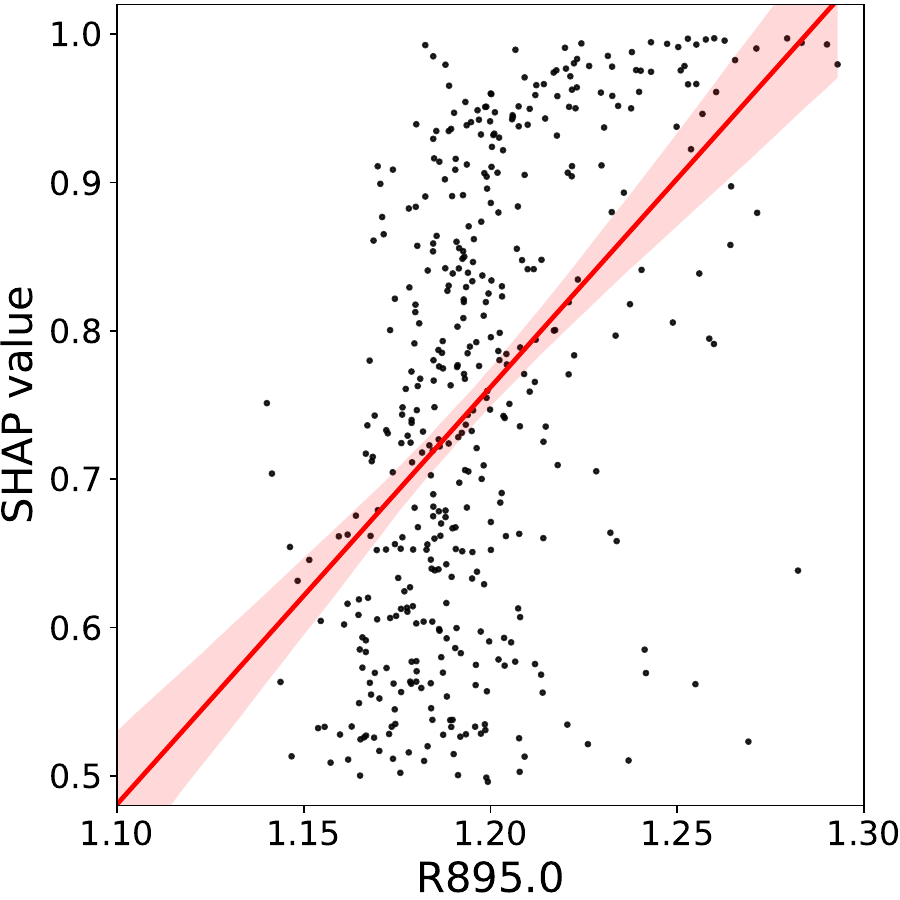}
        \end{minipage}%
        \hfill
        \begin{minipage}{0.185\textwidth}
            \centering
            \includegraphics[width=\linewidth]{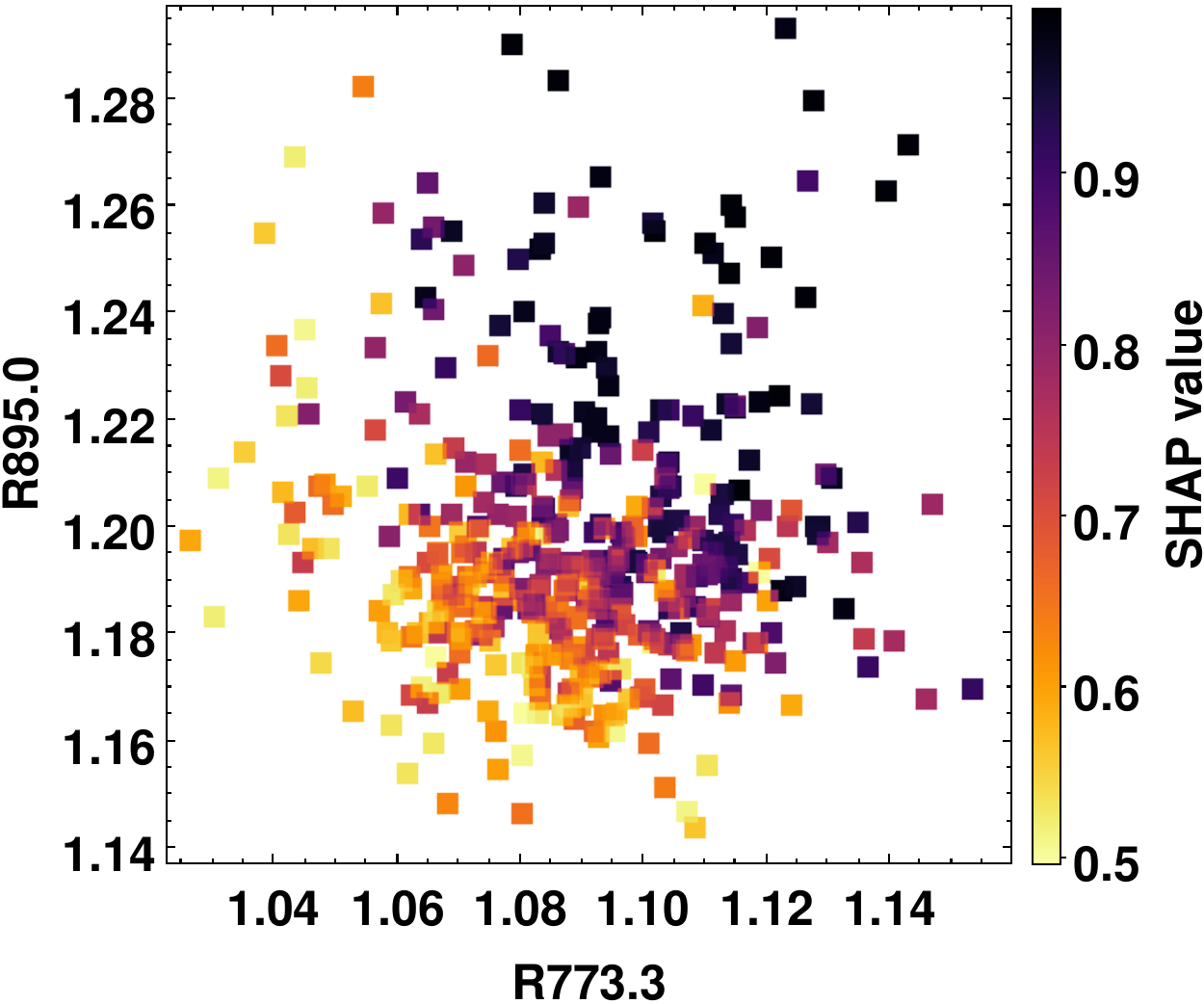}
        \end{minipage}%
        \caption{The upper panels show the relationship between the confidence given by the classification model and R773.3, R895.0. The bottom panels show the relationship between the sum of SHAP values interpreted by the SHAP model for each spectrum and R773.3, R895.0.}
        \label{fig:R_confidence_shape}
    \end{figure}
    From Figure \ref{fig:R_confidence_shape}, it is clear that there is an approximately linear relationship between the strengths and the confidence, SHAP value given by the models. Moreover, it is evident that smaller values of R773.3 and R895.0 correspond to lower confidence and SHAP values from the color-coded scatter plots. Conversely, as the values of R773.3 and R895.0 increase, so do their confidence and SHAP scores (dark color). Such strong positive correlations prove that our candidates should be reliable.

\begin{figure}[htbp]
          \centering
          \includegraphics[width=0.5\textwidth]{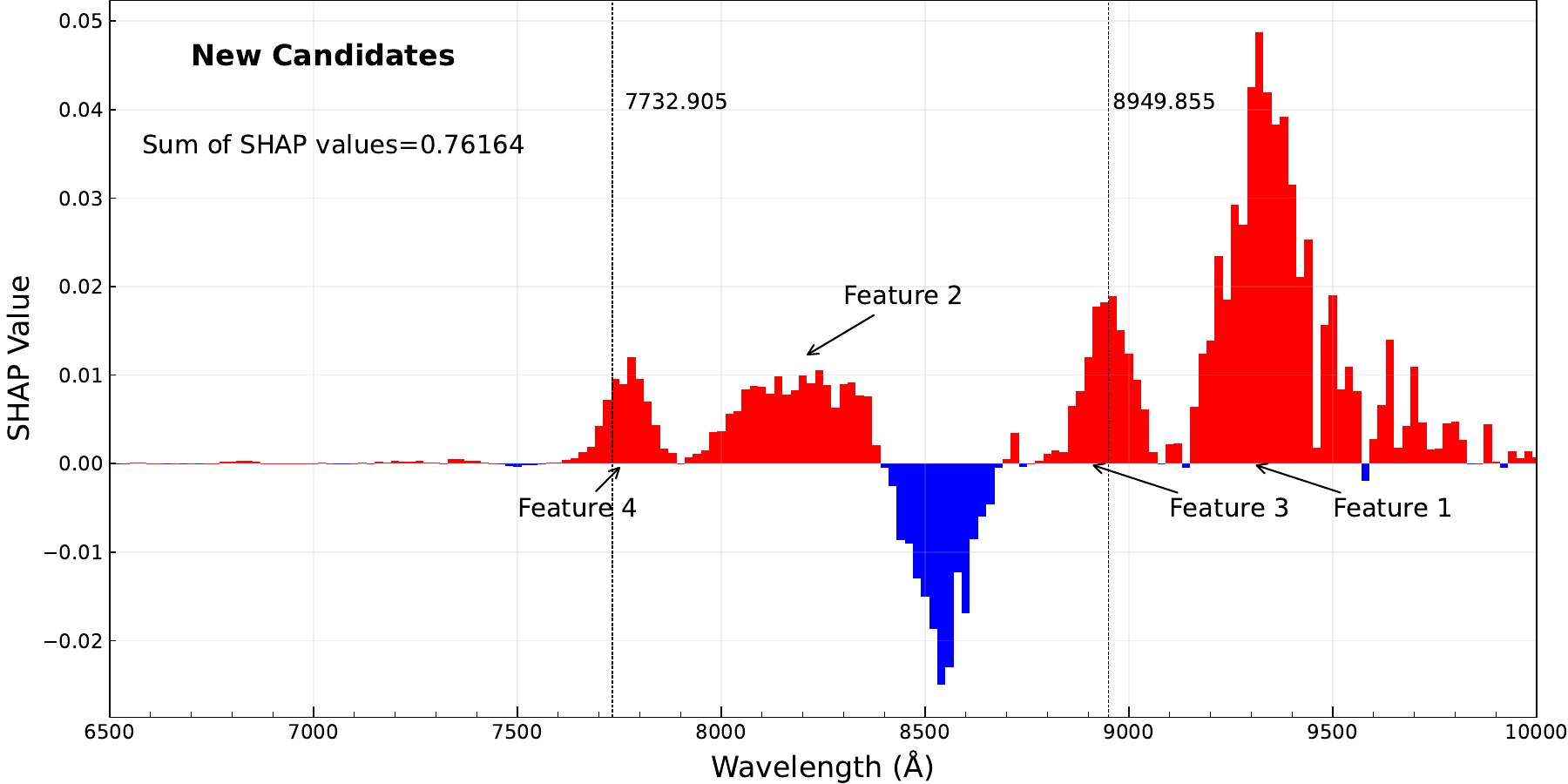}
          \caption{The same wavelength ranges from 6500 to 10000\,\AA. The average SHAP values across feature distributions are shown for the 451 new carbon candidates.}
          \label{fig:car_candidates_shap}
    \end{figure}

    The average SHAP values distribution of features is shown in Figure \ref{fig:car_candidates_shap}. As can be seen from this figure, the sum of average SHAP values (0.76164) of the 451 carbon star candidates is slightly lower than that of the positive sample (0.99894; see Figure \ref{fig:car_sum}), but much larger than that of the negative sample (0.00056), which also confirm the reliability of our candidates. We also see that the contribution of Feature 5 of the candidates almost disappears and Feature 2 becomes weaker compared to the positive sample, which is consistent with the relatively weaker carbon signatures of their spectra. While the negative feature near 8500\,\AA\ (Ca\,II absorption lines) becomes stronger, this seems to imply that the candidates have greater metallicity (see Figure 8; \citeauthor{1997A&AS..124..359M} \citeyear{1997A&AS..124..359M}). It makes sense that our model treats this as a negative contribution to carbon star identification since metallicity tends to be positively correlated with the oxygen levels of a star's atmosphere \citep{2023MNRAS.521.2745S}. Another possible reason is that the neighboring molecular bands become weaker due to temperature or C/O ratio effect, thereby highlighting the Ca\,II absorption line features. Although there is a slight difference in the statistical distribution of SHAP values between the new candidates and positive sample, the overall characteristics suggest that the model is more inclined to classify them as carbon stars, indicating that these candidates are more similar to the golden sample carbon stars.
    
\subsection{Origins of the new carbon star candidates}
    Since the absolute magnitudes of our new carbon star candidates are all less than 3, according to the dividing line at $(M_G)_{0}$ = 5.0 mag between carbon dwarfs and carbon giants proposed by \citet{2024ApJS..271...12L}, we have selected carbon giants. Additionally, only 40 of these candidates have been reported as LPV candidates by \citet{2023A&A...674A..15L}. This implies that many of them might not have evolved to the TP-AGB phase with intense TDU episodes, or alternatively, they could be TP-AGB stars but whose S/N ratio or observed amplitudes are insufficient for detection by the algorithm. The smaller amplitudes might not only be related to the evolutionary state of the stars but could also be influenced by observations in crowded regions, where the light curves of variable stars may be compressed, leading to measured amplitudes smaller than their true values \citep{2023A&A...674A..15L}, and potentially resulting in a poor S/N ratio. 
    
    Extinction correction in crowded and highly reddened sky regions must be treated with caution. If the color index corrections for the candidates are accurate, their bluer colors ($(G_{\text{BP}}-G_{\text{RP}})_{0}$ < 2) compared to the positive sample likely indicate that they are extrinsic carbon stars, i.e., products of binary evolution. Nevertheless, it cannot be excluded that some of these stars are in the TP-AGB phase. Their weaker carbon spectral features may result from temperature and metallicity effects. A higher temperature promotes the dissociation of carbon-bearing molecules in stellar atmosphere, while a higher metallicity can directly influence the C/O ratio, which determines the global spectral shape of carbon stars \citep{2002ApJ...579..817A}. Additionally, they might be newly formed carbon stars that have undergone limited evolution (hence warmer) with few thermal pulses (TPs). Fewer TDU episodes imply that the dredged-up carbon in their atmospheres remains at relatively low levels but is already sufficient to establish a C/O ratio larger than unity (\citeauthor{2002ApJ...579..817A} \citeyear{2002ApJ...579..817A}; \citeauthor{2005ARA&A..43..435H} \citeyear{2005ARA&A..43..435H}). In contrast to the evolved N-type carbon stars (C/O ratio significantly exceeding unity), that have a more pronounced mass loss \citep{1993ApJ...413..641V} and become optically obscured due to dust formation, these 'young' carbon stars appear to be more valuable for abundance analysis \citep{2002ApJ...579..817A}. Furthermore, the limited amount of dredged-up carbon could also be due to the weak TDU strength in low-mass stars \citep{2023MNRAS.521.2745S}. Finally, it is worth noting that their weak spectral carbon features may also be attributed to the lower S/N ratio or to observations in crowded or highly reddened sky regions, rather than a genuinely low carbon abundance in their atmospheres.

\section{Conclusions}\label{sec:conclusion}  

    In this work, we have demonstrated the capability of the proposed `GaiaNet' model in classifying carbon stars within the CSTAR\_XPM sample. Additionally, we have provided compelling feature attributional interpretations for the model's outputs using the SHAP method. This "magic" empowers the originally incomprehensible black-box model with the ability to proactively "speak up" and make its discriminations traceable. Our main results and conclusions are as follows:
    
    \begin{enumerate}
        \item[(i)] The method presented in C2023 for calculating the strength of molecular bands has proven effective in selecting carbon stars with strong CN features. However, it lacks effectiveness in identifying hotter carbon stars with weaker CN features and may introduce contamination. In contrast, the proposed `GaiaNet' model demonstrates superior capability in distinguishing between carbon and non-carbon stars robustly from the training datasets, particularly in identifying carbon stars with weak CN features from CSTAR\_XPM\_nonG. Compared to four conventional machine learning methods, the `GaiaNet' model exhibits an average accuracy improvement of approximately 0.3\% on the validation set, with the highest accuracy reaching up to 100\%. The flexible parameter architecture provides the model with enhanced fitting capability and stability achieved through the incorporation of operations such as 1×1 Convolution, Global Average Pooling, Batch Normalization (BN), and Dropout layers. Furthermore, Optimised training methods including momentum and weight decay were employed. By utilizing deep learning models with varying convolutional kernels during the 1D convolution process, the `GaiaNet' model effectively captures both local and global features from low-resolution smoothed spectra. Coupled with the nonlinear capabilities provided by an appropriate model's depth, the model possesses natural advantages in handling $Gaia$'s XP spectra.

        \item[(ii)] We utilized SHAP for feature attribution interpretation with the `GaiaNet' classification model based on positive sample. We presented clear feature heatmaps for each spectrum, which serve as a basis for judgment and identification of each spectrum. Based on the statistical analysis of SHAP values from the spectra, we found that the C$_{\text{2}}$ features provide only a  weak signal at the blue end of low flux values and have a minimal presence in many cold carbon stars. Therefore, they were not selected by our model. We summarised five key spectral feature regions: 6780-7460\,\AA, 7600-7820\,\AA, 7900-8360\,\AA, 8820-9060\,\AA, and 9140-9560\,\AA. Among them, the 7900-8360\,\AA\ and 9140-9560\,\AA\ regions are the most obvious trough areas. These two regions contribute over 80\% of the SHAP value due to CN molecular absorption. The 7600-7820\,\AA\ and 8820-9060\,\AA\ regions contribute about 16\%, which are are also important distinguishing features due to the peak areas from the strong CN molecular absorption on both sides. The 6780-7460\,\AA\ region contains relatively weak peaks and troughs, making the smallest contribution less than 4$\%$. We hypothesize that this is related to the spectral values, which seems to be addressable through improved spectral normalization methods. These results suggest that our model can effectively learn and capture the key distinguishing features, which are consistent with CN molecular absorption lines and visual identifications by astronomers in the classification of carbon star spectra. These five key feature regions will be essential for future systematic mining of carbon stars from $Gaia$'s XP spectra.

        \item[(iii)] We identified 451 new carbon star candidates from CSTAR\_XPM\_nonG using the trained classification model. Compared to the band head strength calculation method, our approach effectively detects more carbon star candidates with weaker CN features than those found in the golden sample of carbon stars. These candidates align with the weak CN feature carbon stars depicted in Fig. 21 of C2023. The weaker CN features might be a function of temperature and metallicity effects. Because many of these stars become bluer according to their corrected colors ($(G_{\text{BP}}-G_{\text{RP}})_{0}$ < 2), and show stronger negative contributions caused by Ca\,II absorption features near 8500\,\AA\ based on the statistical analysis of SHAP values compared to the positive sample. Moreover, most of them are not classified as LPVs by \citet{2023A&A...674A..15L}, indicating that they could be extrinsic carbon stars. However, the possibility that they are AGB carbon stars with less evolution (limited TPs or mild TDU events) can not be ruled out. Their observed weak carbon spectral feature may suggest a small amount of dredged-up carbon, with C/O ratio not significantly larger than 1. However, caution must be taken when analyzing these candidates, as they are mainly located in the crowded region of Galactic inner disc.

       \item[(iv)] We cross-matched the 451 new candidates with LAMOST DR10, resulting in a total of 4 common sources. Among them, 2 were identified as carbon stars by the LAMOST 1D pipeline. By cross-matching with the SIMBAD database, we obtained 73 common sources. Among these, 28 had already been reported as carbon stars or carbon star candidates in previous work. A reference sample of 1,289 LPVs was constructed to compare the results in \citet{2023MNRAS.521.2745S}. In this work, we have discovered 40 potentially carbon stars from it, and they identified 26 C-rich stars including only 6 stars that are not in our candidates. This is due to the significantly different spectral features of the six stars from the positive sample, which results in their confidence lower than 0.5.  Furthermore, the assigned confidence and SHAP values for each spectrum were used to quantify the model's output, revealing a strong correlation with the band head strengths of R773.3 and R895.0. This suggests that the classification model primarily focuses on the crucial CN features, and the learned features make sense. Hence, our model can efficiently learn the key features between carbon and non-carbon stars in the CSTAR\_XPM sample, and the model's reliability is further validated.  
    \end{enumerate}

    Due to the limited diversity of the training positive sample, there are still some carbon stars with weaker CN features that are not recognized by our algorithm. Their probabilities given by the model's output do not exceed the threshold of 0.5 but significantly larger than 0, thus the proportion of candidates in CSTAR\_XPM\_nonG could be higher, especially the G and K type stars, predictably if we set a lower threshold. we roughly estimate that the loss percentage is at least about 5\%. In future work, it will be feasible to make reasonable improvements to the classification model and incorporate more carbon stars with weaker CN bands into the positive sample to search for and identify more potential warm candidates from CSTAR. We will then expand the algorithm to the entire $Gaia$ XP spectral dataset, thereby significantly enriching the number of known carbon stars. Indeed it would be interesting to perform further classification and meaningful analysis. Given the resolution of $Gaia$'s XP spectra, our method can be directly transferred to the upcoming China Space Station Telescope (CSST), especially for the interpretation of different types of stars whose spectral properties are still unclear.

\begin{acknowledgments}
    We thank Bo Zhang for his effective spectral smoothing method and helpful discussions. We thank Xiao Kong for his help with the data. This study is supported by the National Natural Science Foundation of China (NSFC) under grant Nos. 12173013, 12273078, and U1931209; the National Key Basic R$\&$D Program of China (No. 2019YFA0405500); the project of Hebei provincial department of science and technology under the grant number 226Z7604G, and the Hebei NSF (No. A2021205006). W.Y. thanks the science research grants from the China Manned Space Project, and we also acknowledge the grants from the China Manned Space Project under Nos. CMS-CSST-2021-A10. This work has made use of data from the European Space Agency (ESA) mission $Gaia$ (\url{https://www.cosmos.esa.int/gaia}), processed by the $Gaia$ Data Processing and Analysis Consortium (DPAC; \url{https://www.cosmos.esa.int/web/gaia/dpac/consortium}). This research has used the SIMBAD database, operated at CDS, Strasbourg, France. This research also used the low-resolution spectral data of LAMOST DR10 (the Large Sky Area Multi-Object Fiber Spectroscopic Telescope DATA RELEASE 10; \url{https://www.lamost.org/dr10/v1.0}). LAMOST is operated and managed by the National Astronomical Observatories, Chinese Academy of Sciences.
\end{acknowledgments}

\bibliographystyle{aa} 
\bibliography{reference}

\newpage
\onecolumn
\appendix
\section{Heatmaps of six different stars}\label{sec:error_6}
    \begin{figure*}[htbp]
      \centering
      \subfigure{
        \includegraphics[width=0.48\linewidth]{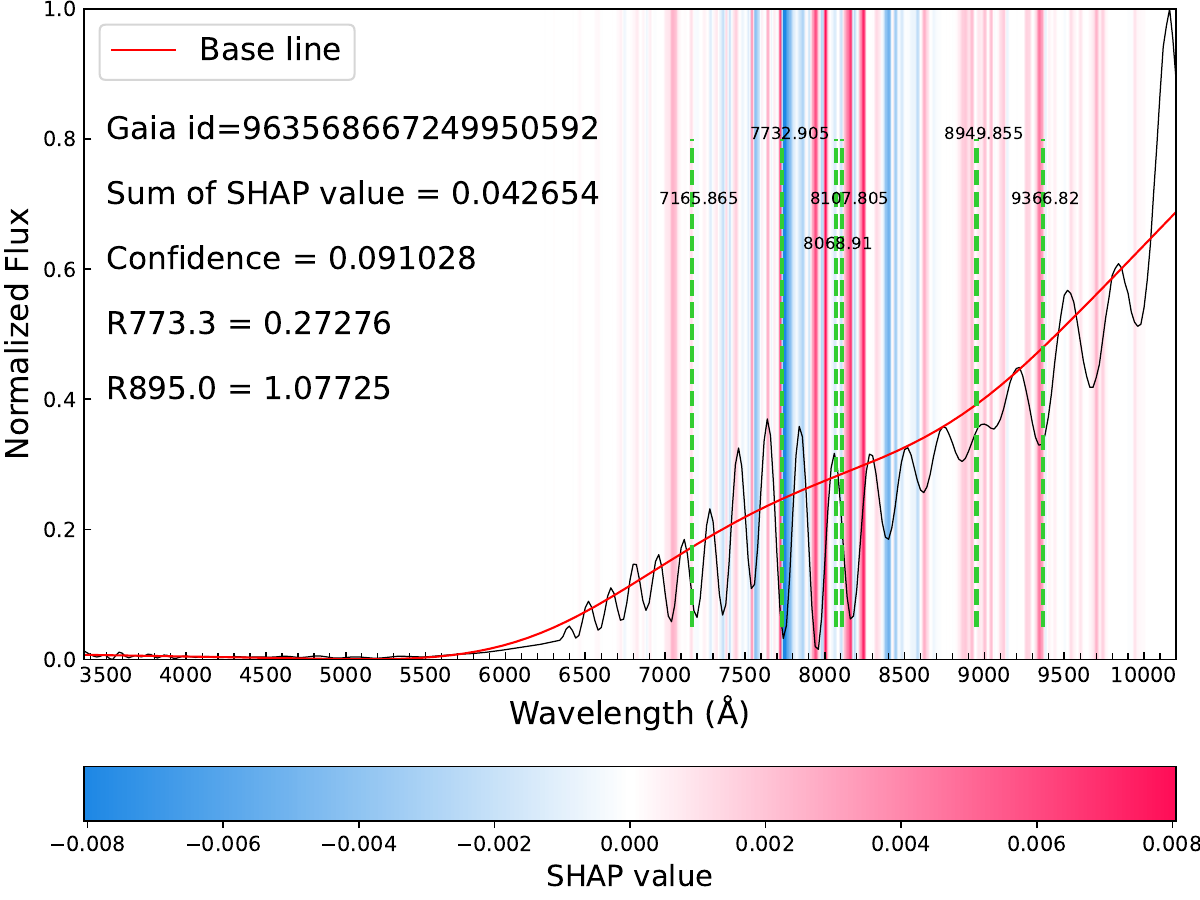}
      }
      \hfill
      \subfigure{
        \includegraphics[width=0.48\linewidth]{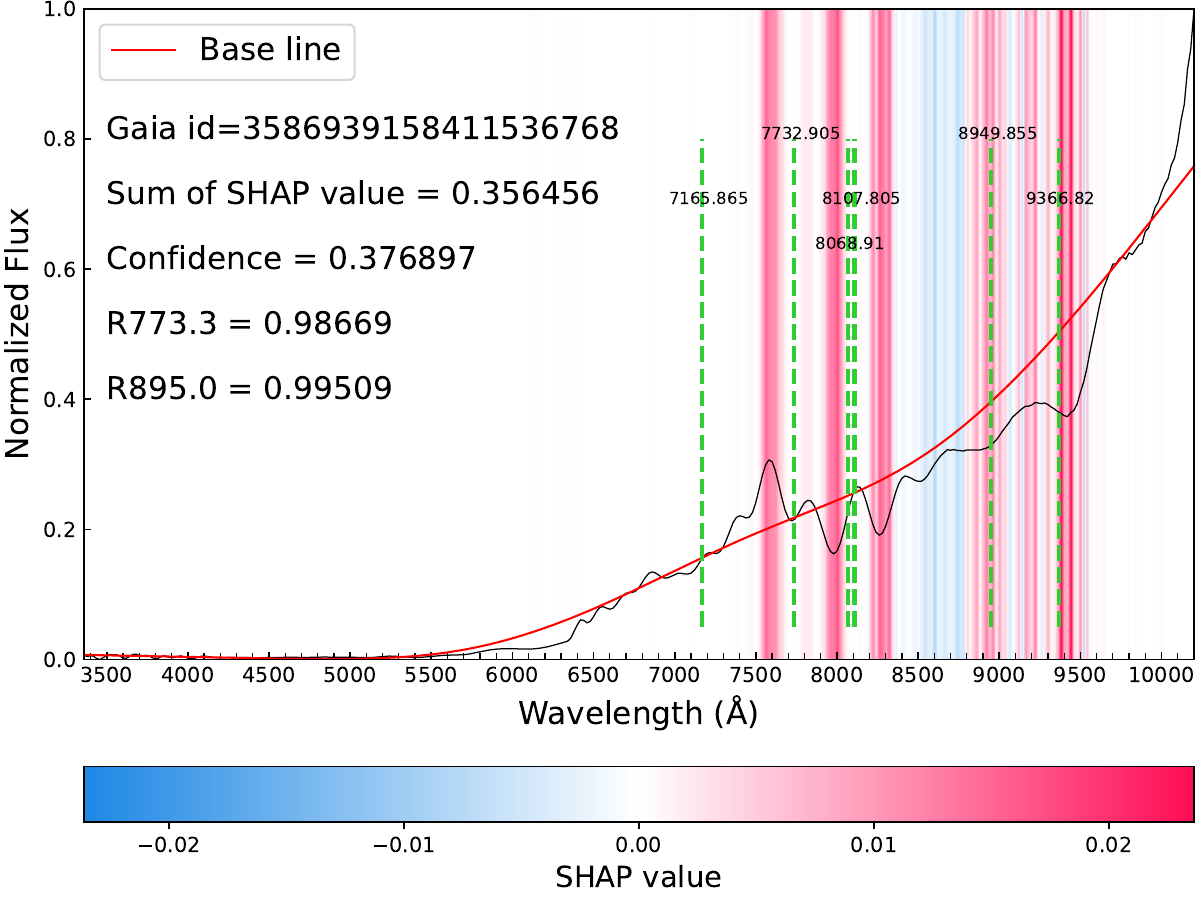}
      }

      \vspace{1em}
      
      \subfigure{
        \includegraphics[width=0.48\linewidth]{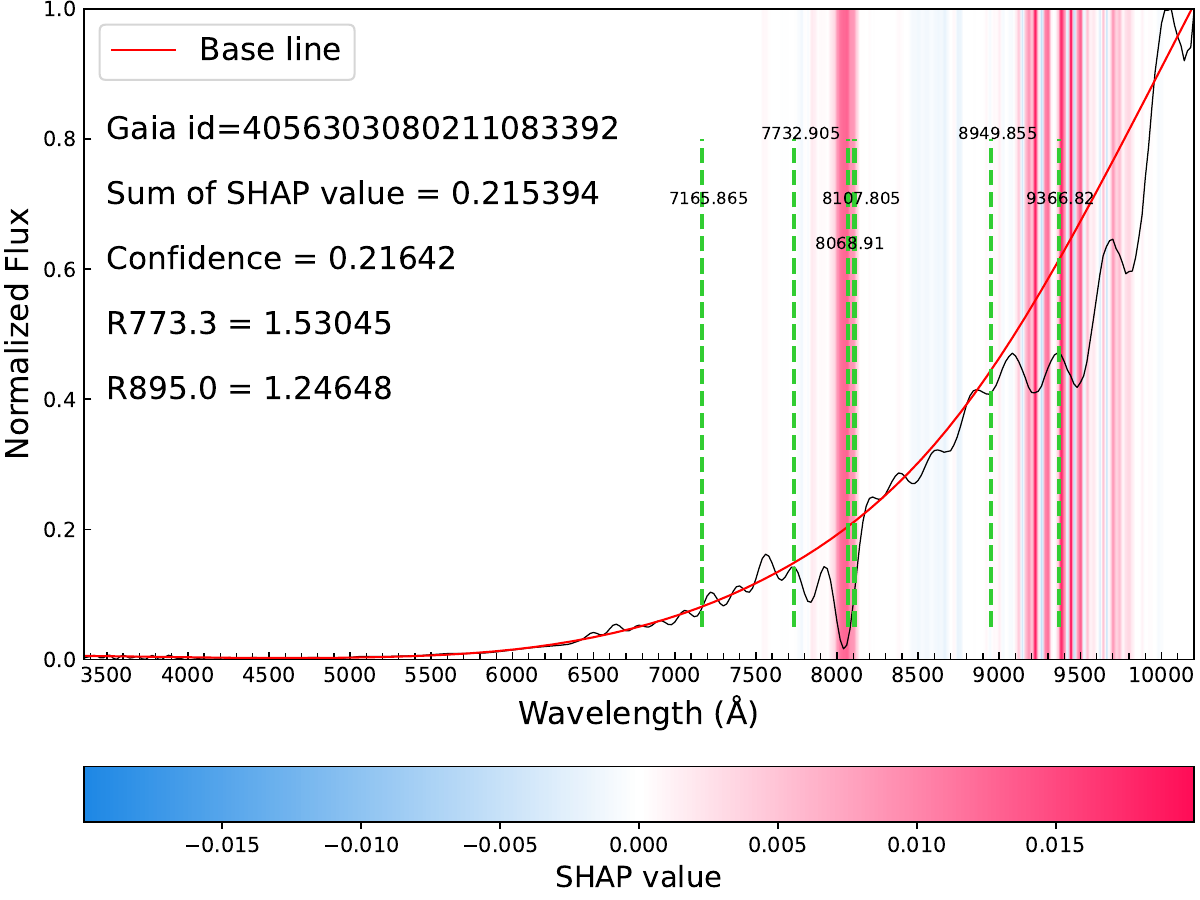}
      }
    \hfill
      \subfigure{
        \includegraphics[width=0.48\linewidth]{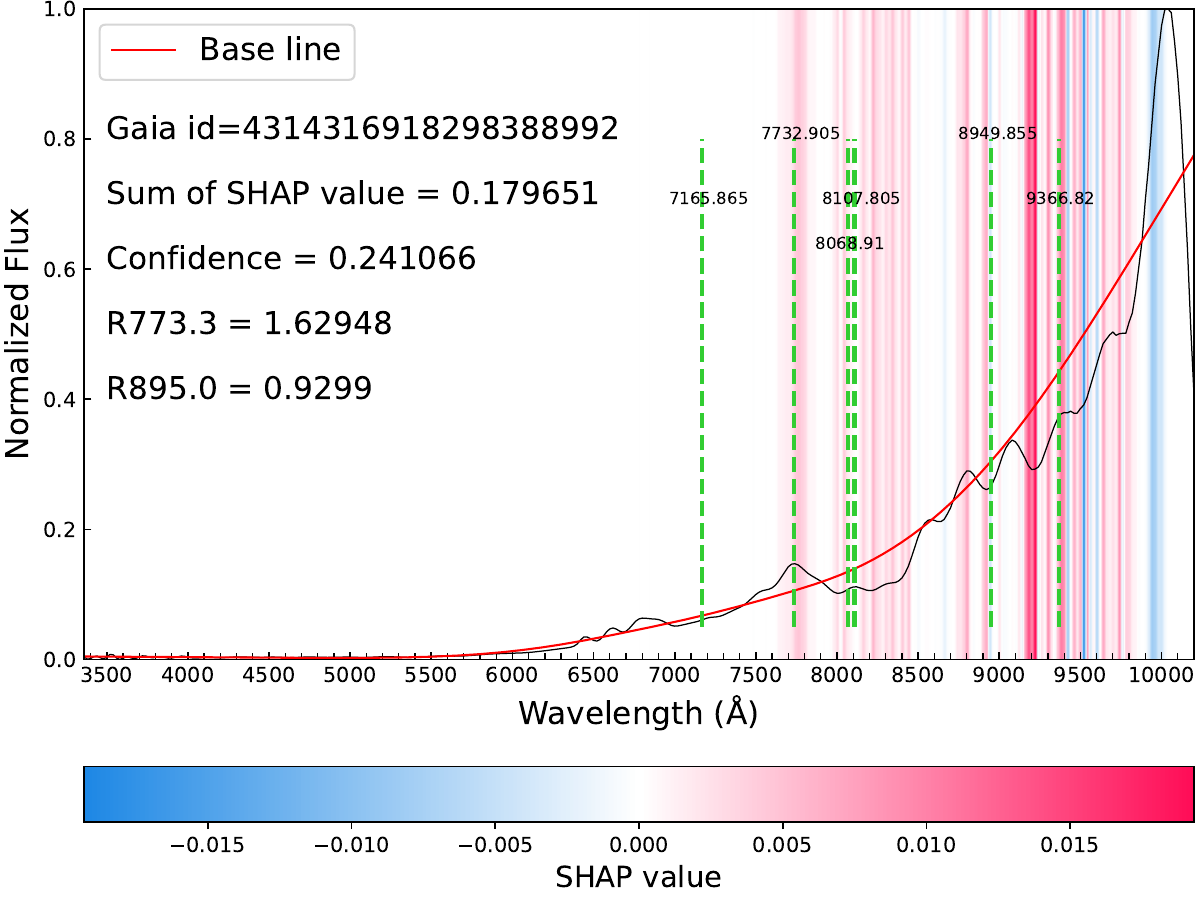}
      }
      \vspace{1em}
      
      \subfigure{
        \includegraphics[width=0.48\linewidth]{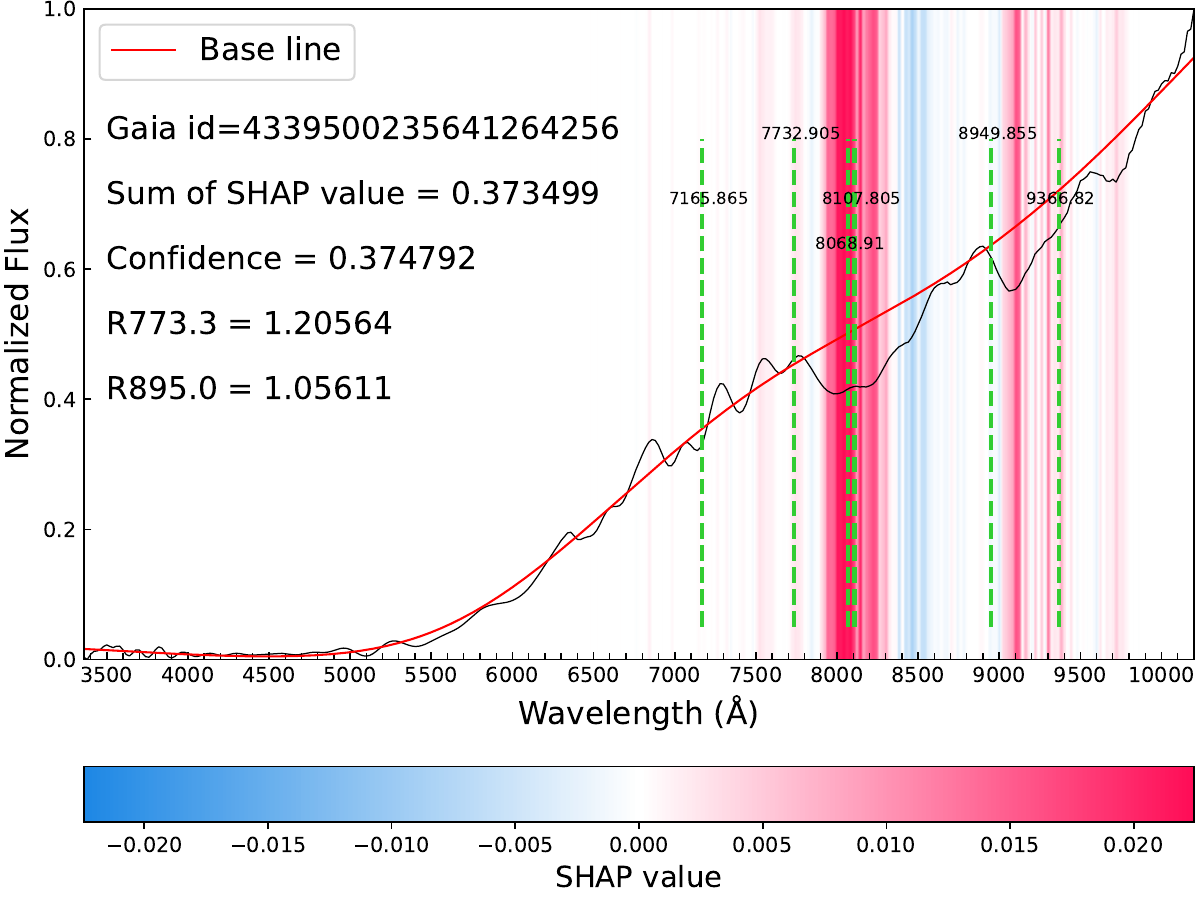}
      }
      \hfill
      \subfigure{
        \includegraphics[width=0.48\linewidth]{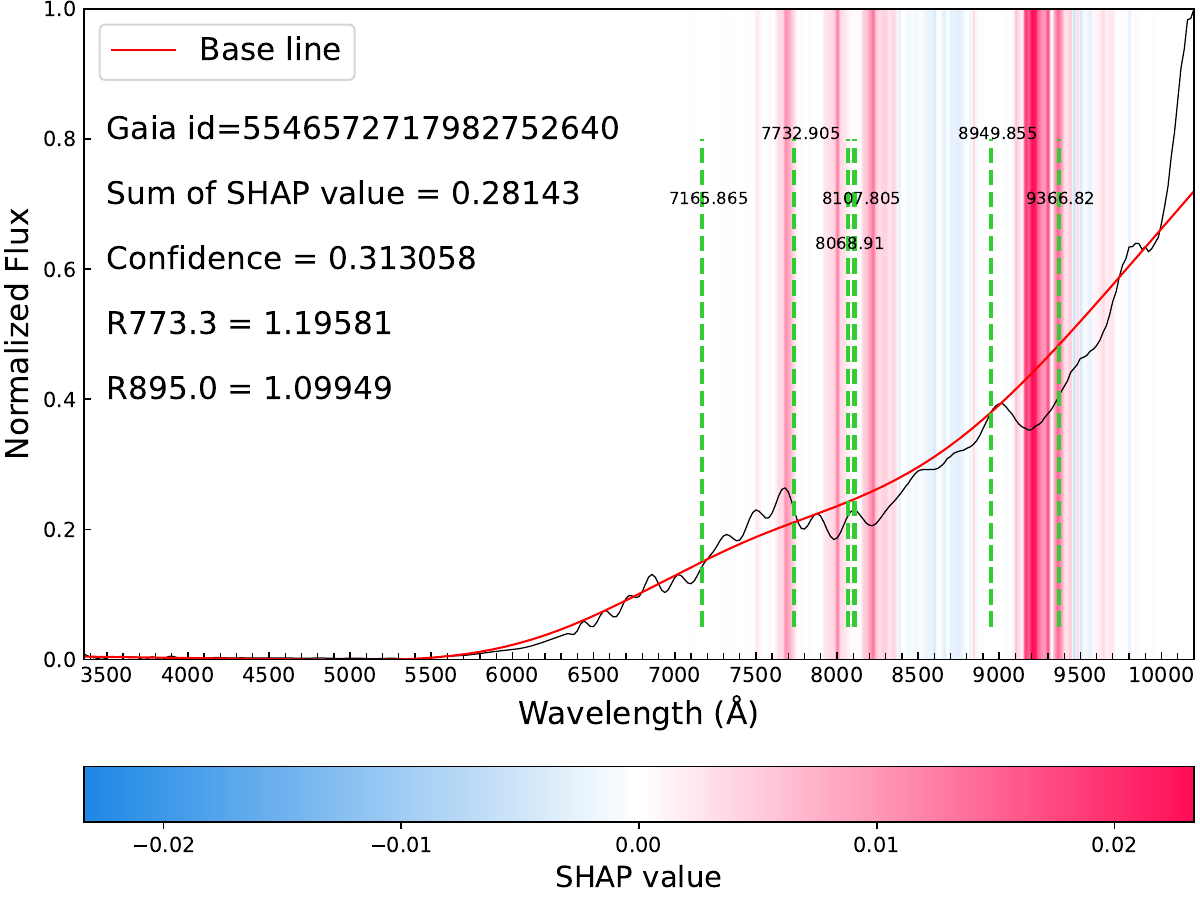}
      }
      \caption{From top to bottom, left to right are sub-figures 1-6. The green dashed lines mark the positions used to calculate the CN band head strengths. 7732.9\,\AA\ and 8950.0\,\AA\ are the top of the band head, others are used to identify the location of band troughs.}
      \label{fig:error_6}
      \end{figure*}

\section{New Carbon star candidates from CSTAR\_XPM\_nonG} \label{sec:appendix}
     The catalog has 451 entries, which are all the most likely carbon star candidates selected by our model. The model obtains this result by a training positive sample selected from CSTAR\_XPM\_G and negative sample selected from CSTAR\_XPM\_non\_G. All candidates are sorted in descending order of confidence, and we show the top 100 candidates.  
    
\centering
\begin{longtable}{lllccll}
        \caption{The catalog of carbon star candidates in CSTAR\_XPM\_nonG}
    
        \label{star_table} \\

        \hline
        \hline
        \multicolumn{1}{c}{\makecell{Source ID \\ (GAIA)}} &
        \multicolumn{1}{c}{\makecell {R.A \\ (deg)}} &
        \multicolumn{1}{c}{\makecell {Dec. \\ (deg)}} &
        \multicolumn{1}{c}{\makecell{Main type \\ (SIMBAD)}} &
        \multicolumn{1}{c}{\makecell{Other types \\ (SIMBAD)}} &
        \multicolumn{1}{c}{Confidence} &
        \multicolumn{1}{c}{SHAP value} \\
        \hline
        \endfirsthead

        \multicolumn{7}{c}{{Continued on next page}} \\
        \endfoot
        \hline
        \endlastfoot
        5938465054747237248\textsuperscript{\hyperlink{tb2:sample-notes}{($S$)}} & 255.244434 & -48.950652 & LPV* & *|C*?|LP*|LP?|MIR|NIR & 1.0 & 0.99955 \\
        5835730639669683968\textsuperscript{\hyperlink{tb2:sample-notes}{($S$)}} & 241.084489 & -57.364918 & LPV* & *|C*?|LP*|LP?|MIR|NIR & 0.99999 & 0.99957 \\
        5313143689277626880\textsuperscript{\hyperlink{tb2:sample-notes}{($S$)}} & 140.040299 & -52.485428 & C* & *|C*|C*?|LP?|MIR|NIR & 0.99994 & 0.99955 \\
        5489622585506231552 & 116.27471 & -53.090801 & Star & *|IR|NIR & 0.99991 & 0.9703 \\
        656271591242470528 & 125.428401 & 17.285119 & S* & C*|IR|LP*|MIR|Mi*|NIR & 0.99972 & 0.99313 \\
        5941474349332334976 & 245.800041 & -48.712454 &   &   & 0.99755 & 0.9972 \\
        5854048743845734528 & 212.221097 & -62.928023 & Star & *|IR & 0.99754 & 0.99715 \\
        5854243902825896064 & 215.206128 & -62.274869 &   &   & 0.99737 & 0.99694 \\
        5980488835879061888 & 259.325419 & -30.927344 &   &   & 0.99676 & 0.99641 \\
        5604270350862378752\textsuperscript{\hyperlink{tb2:sample-notes}{($S$)}} & 106.535192 & -31.87378 & LPV* & *|C*?|IR|LP*|LP?|NIR & 0.99597 & 0.99524 \\
        4061196422397873920 & 262.590675 & -28.135145 &   &   & 0.99586 & 0.99566 \\
        4062631113240876672 & 268.808798 & -28.434779 &   &   & 0.99498 & 0.99456 \\
        3047772534862552320\textsuperscript{\hyperlink{tb2:sample-notes}{($S$)}} & 110.696578 & -8.815217 & Mira & *|C*|C*?|IR|LP*|LP?|Mi* & 0.99465 & 0.99262 \\
        4094930504172297984 & 271.316536 & -20.239457 &   &   & 0.99451 & 0.99423 \\
        5865171467859742848 & 200.505129 & -63.584571 &   &   & 0.99428 & 0.99374 \\
        4059056016839338752 & 261.375858 & -30.012123 & LP*\textunderscore Candidate & *|LP?|NIR & 0.99413 & 0.99307 \\
        514263548487890176 & 36.376851 & 63.054981 &   &   & 0.9938 & 0.99351 \\
        5862198598937747456 & 199.179033 & -63.386326 &   &   & 0.99335 & 0.993 \\
        354762791118252288\textsuperscript{\hyperlink{tb2:sample-notes}{($S$)}} & 33.533472 & 47.662447 & C* & *|C*|C*?|IR|LP*|MIR|Mi* & 0.99306 & 0.99121 \\
        5716413183908004224\textsuperscript{\hyperlink{tb2:sample-notes}{($S$)}} & 114.258845 & -19.548336 & C* & *|C*|C*?|IR|LP*|LP?|Mi* & 0.99174 & 0.99082 \\
        4070093150664753792 & 271.441386 & -21.348017 &   &   & 0.9917 & 0.99131 \\
        2014593550230928896 & 342.495667 & 60.299062 & RedSG* & *|IR|LP*|MIR|Mas|NIR|pA? & 0.99072 & 0.99034 \\
        4045821465839438976 & 274.444504 & -32.26074 & Mi*\textunderscore Candidate & *|IR|MIR|Mi?|NIR|V* & 0.99022 & 0.98829 \\
        5990083449153819392 & 243.082181 & -46.963369 &   &   & 0.98988 & 0.98948 \\
        5937319015683955968 & 250.688189 & -50.815719 &   &   & 0.98822 & 0.98793 \\
        5255092670818771968 & 153.907807 & -59.55129 & deltaCep & *|MIR|NIR|V*|cC* & 0.98572 & 0.98533 \\
        2178053614615673856 & 324.821403 & 55.932396 &   &   & 0.98554 & 0.9851 \\
        4058721456082780544 & 262.301994 & -29.531504 &   &   & 0.98474 & 0.98036 \\
        2029261378964662272 & 300.317391 & 28.786557 & Star & *|IR|MIR|NIR & 0.98369 & 0.98328 \\
        6057834256157778432 & 182.445267 & -61.527859 &   &   & 0.98285 & 0.98253 \\
        5258718997589107712 & 154.820297 & -57.32384 & post-AGB* & *|AB?|IR|LP*|LP?|MIR|NIR & 0.98075 & 0.98049 \\
        4056545208914430720 & 268.940437 & -29.089513 &   &   & 0.98008 & 0.97941 \\
        4065758154277969408 & 271.411365 & -24.856418 &   &   & 0.97982 & 0.97961 \\
        4054558494476007296 & 264.153127 & -32.681742 &   &   & 0.97892 & 0.97856 \\
        2026948006494690048 & 298.042068 & 25.93909 &   &   & 0.97884 & 0.97853 \\
        6054537473621712128 & 185.729092 & -63.137329 &   &   & 0.97847 & 0.9781 \\
        4056219684743385600 & 269.089299 & -30.304209 &   &   & 0.97829 & 0.97679 \\
        4104914825724879360 & 278.259201 & -12.901917 &   &   & 0.97615 & 0.9758 \\
        5932639326833555328 & 242.63607 & -53.942337 &   &   & 0.97602 & 0.97562 \\
        2026983981142199296 & 297.299261 & 25.949449 &   &   & 0.97599 & 0.97557 \\
        4259206577970407936 & 282.889751 & -1.597829 &   &   & 0.97575 & 0.97536 \\
        5960473527976734208 & 261.260894 & -39.265847 &   &   & 0.97508 & 0.97463 \\
        4044123652438371456 & 269.878542 & -30.884273 &   &   & 0.97454 & 0.97414 \\
        5862947950437062656 & 191.303003 & -63.410645 &   &   & 0.97201 & 0.9716 \\
        4062586784782656896 & 269.371786 & -28.626903 &   &   & 0.97174 & 0.97085 \\
        4057240787453141376 & 267.31741 & -29.262486 &   &   & 0.96676 & 0.9664 \\
        4067075609721316992 & 266.855157 & -25.982856 &   &   & 0.96669 & 0.9663 \\
        5972266168138911488 & 258.397097 & -39.766671 & Star & * & 0.96648 & 0.96618 \\
        4066701402179072640 & 272.329522 & -22.653578 &   &   & 0.96605 & 0.96562 \\
        4062660525189492608 & 269.324272 & -28.23621 &   &   & 0.96576 & 0.96525 \\
        3451032236956436096 & 89.311802 & 32.377569 & S* & *|C*|C*?|IR|LP*|LP?|MIR & 0.96564 & 0.96266 \\
        448608341933318656 & 52.575367 & 56.563998 &   &   & 0.96447 & 0.96409 \\
        4068826998652116480 & 266.535475 & -23.246602 &   &   & 0.96299 & 0.96256 \\
        486835887330942464 & 54.455116 & 62.415504 &   &   & 0.96165 & 0.95998 \\
        5942937868055138304 & 249.816206 & -46.113176 &   &   & 0.96146 & 0.96111 \\
        4060674841494267904 & 266.021275 & -27.667819 &   &   & 0.96144 & 0.961 \\
        5886865420714514944 & 230.322616 & -54.987715 &   &   & 0.96101 & 0.96063 \\
        4062342693202139136 & 269.461274 & -29.236824 &   &   & 0.96024 & 0.95963 \\
        5311959480914027392 & 138.009931 & -51.782806 &   &   & 0.95941 & 0.95839 \\
        4064870814038165504 & 271.674298 & -25.950435 &   &   & 0.9592 & 0.95895 \\
        5972008611885974016 & 258.9437 & -40.182199 &   &   & 0.95847 & 0.95822 \\
        5338251170445346560 & 162.625494 & -60.525464 &   &   & 0.95468 & 0.95429 \\
        5972260022101577216 & 259.209972 & -39.244572 &   &   & 0.95202 & 0.95162 \\
        5618457383755161216\textsuperscript{\hyperlink{tb2:sample-notes}{($S$)}} & 112.738987 & -23.464088 & C* & *|C*|C*?|IR|LP*|LP?|MIR & 0.95178 & 0.95018 \\
        5942980542846977408 & 249.414567 & -45.784273 &   &   & 0.9516 & 0.95094 \\
        2030353670740758656 & 298.576264 & 29.883676 &   &   & 0.95156 & 0.95101 \\
        5968203992396813056 & 250.512108 & -41.840297 &   &   & 0.95154 & 0.95131 \\
        5974841263083171456 & 263.673464 & -35.429062 &   &   & 0.95146 & 0.95108 \\
        526430675440161536 & 15.541814 & 66.977695 &   &   & 0.95074 & 0.95003 \\
        4063963721314562304 & 269.949861 & -26.829915 &   &   & 0.9505 & 0.94625 \\
        4322164334588338816 & 293.372593 & 17.214312 & LPV* & *|LP*|MIR|NIR & 0.95031 & 0.94997 \\
        5878424087663723904 & 217.675214 & -60.971208 & Star & * & 0.95018 & 0.94967 \\
        4069866513830121856 & 270.629017 & -22.40119 &   &   & 0.94908 & 0.94872 \\
        4256920826432562944 & 279.872356 & -4.237485 &   &   & 0.94807 & 0.94733 \\
        4118583194001521920 & 267.444473 & -21.508462 & SB* & *|NIR|SB* & 0.94748 & 0.947 \\
        2201585087589197696 & 335.620434 & 60.982812 &   &   & 0.94573 & 0.94533 \\
        5334981845684641280 & 177.503771 & -61.224682 &   &   & 0.94472 & 0.94413 \\
        2031872508602623360 & 297.895438 & 29.923054 &   &   & 0.94363 & 0.94314 \\
        5884051972285306880 & 236.587553 & -56.50199 &   &   & 0.94315 & 0.94265 \\
        4319411299229958784 & 290.048676 & 13.672668 &   &   & 0.94273 & 0.94232 \\
        4056588678371817600 & 268.51291 & -29.083316 &   &   & 0.94228 & 0.94124 \\
        4255740779922180608 & 283.449916 & -3.247059 &   &   & 0.94114 & 0.94066 \\
        4117181282312558208 & 265.719645 & -22.047087 &   &   & 0.93976 & 0.93926 \\
        5940367789632743040 & 248.633299 & -50.116494 &   &   & 0.93939 & 0.9386 \\
        5972190026958636416 & 258.096814 & -39.644554 &   &   & 0.9393 & 0.93895 \\
        2137619448150914432\textsuperscript{\hyperlink{tb2:sample-notes}{($S$)}} & 297.93139 & 53.692079 & S* & *|C*|C*?|Em*|IR|LP*|MIR & 0.93918 & 0.93707 \\
        5878466526231182208 & 219.454802 & -61.054658 &   &   & 0.93825 & 0.93786 \\
        4104919842247578240 & 278.542311 & -12.778434 &   &   & 0.93789 & 0.93758 \\
        887462018563120128\textsuperscript{\hyperlink{tbl:sample-notes}{($L$)}}\textsuperscript{\hyperlink{tb2:sample-notes}{($S$)}} & 104.758731 & 28.814589 & C* & *|C*|C*?|IR|LP*|NIR & 0.93704 & 0.93209 \\
        4255321483686501504 & 282.459084 & -3.991967 &   &   & 0.93671 & 0.93612 \\
        5972269432314624256 & 258.456518 & -39.618216 &   &   & 0.93528 & 0.93473 \\
        4110745157943191936 & 259.92529 & -25.261413 &   &   & 0.93506 & 0.93471 \\
        5964551444805061632 & 253.839017 & -44.03463 &   &   & 0.93337 & 0.93295 \\
        5969733756664102912 & 253.454426 & -40.520653 &   &   & 0.93309 & 0.93164 \\
        4266241042905220352 & 283.816115 & 0.179313 &   &   & 0.9328 & 0.93237 \\
        4068727046189164416 & 266.400872 & -23.76739 &   &   & 0.93238 & 0.93201 \\
        5961670105863745792 & 263.017942 & -39.635199 &   &   & 0.93092 & 0.93027 \\
        4040149880018961920 & 266.545071 & -36.492316 &   &   & 0.9303 & 0.92944 \\
        5934333021361329792 & 247.918148 & -50.699238 &   &   & 0.92446 & 0.92401 \\
        5932728936967577216 & 240.743333 & -54.404695 &   &   & 0.92275 & 0.92243 \\   
    \end{longtable}
    
    \begin{tablenotes}
        \item {\textbf{Notes.}} \\
        \textsuperscript{\hypertarget{tbl:sample-notes}{($L$)}} represents a carbon star labeled by LAMOST's pipeline. \\
        \textsuperscript{\hypertarget{tb2:sample-notes}{($S$)}} represents a carbon star identified by \citet{2023MNRAS.521.2745S}. \\
        Other types are given a maximum of 6 types due to table width limitations.
    \end{tablenotes}
    
\clearpage

\end{document}